\documentclass[useAMS,usenatbib]{mnras}
\usepackage{graphicx}
\usepackage[section]{placeins}

\usepackage{yfonts}
\usepackage{times} 
\usepackage{amsmath}
\usepackage{amssymb}
\usepackage{tikz}
\usepackage{xcolor}

\def\icarus{Icarus}
\def\aaps{A\&AS}
\def\aj{Astron.\ J. }

\def\apjl{Astrophys.\ J. }
\def\apjs{Astrophys.\ J.\ (Supp.) }
\def\aap{Astron.\ Astrophys. }

\def\bssA_{Bull.\ Seismol.\ Soc.\ Am. }

\def\gcA_{Geochim.\ Cosmochim.\ Acta }

\def\mnras{Mon.\ Not.\ R.\ Astron.\ Soc. }

\def\planss{Plan.\ Space Sci. }
\def\prA_{Phys.\ Rev.\ A }

\begin{document}

\title[Resonance dynamics at arbitrary inclination]{Resonance libration and width at arbitrary inclination}
\author[F. Namouni and M. H. M. Morais]{F. Namouni$^{1}$\thanks{E-mail:
namouni@obs-nice.fr (FN) ; helena.morais@rc.unesp.br (MHMM)} and  M. H. M. Morais
$^{2}$\footnotemark[1]\\
$^{1}$Universit\'e C\^ote d'Azur, CNRS, Observatoire de la C\^ote d'Azur, CS 34229, 06304 Nice, France\\
$^{2}$Instituto de Geoci\^encias e Ci\^encias Exatas, Universidade Estadual Paulista (UNESP), Av. 24-A, 1515 13506-900 Rio Claro, SP, Brazil}

\date{Accepted 2020 February 4. Received 2020 February 4; in original form 2019 December 30.}

\maketitle

\begin{abstract} We apply the analytical  disturbing function for arbitrary inclination derived in our previous work to characterize resonant width and libration of mean motion resonances at arbitrary inclination obtained from direct numerical simulations of the three-body problem.  We examine the 2:1 and 3:1 inner Jupiter  and 1:2 and 1:3 outer Neptune resonances {and their possible asymmetric librations} using a new  analytical pendulum model of resonance that includes the simultaneous libration of multiple arguments and their second harmonics. The numerically-derived resonance separatrices are obtained using the Mean Exponential Growth factor of Nearby Orbits ({\sc megno} chaos indicator).  We find that the analytical and numerical estimates are in agreement {and that resonance width is determined by the first few fundamental resonance modes that librate simultaneously on the resonant timescale. Our results demonstrate} that the new pendulum model may be used to ascertain resonance width analytically, and more generally, that the disturbing function for arbitrary inclination is a powerful analytical tool that describes resonance dynamics of low as well as high inclination asteroids in the solar system.
\end{abstract}

\begin{keywords}
celestial mechanics--comets: general--Kuiper belt: general--minor planets, asteroids: general -- Oort Cloud.
\end{keywords}

\section{Introduction}
The origin of the solar system's Centaurs and transneptunian objects on highly-inclined orbits  is an outstanding challenge for solar system formation theory. It has recently been found that some objects on high inclinations must have been captured in the Sun's birth cluster and made their way towards the inner solar system under the gravitational influence of the planets \citep{Namouni18b,kaib19}.  Jupiter's coorbital asteroid Ka`epaoka`awela is one such object. Its orbit is Sun-bound and has a retrograde inclination of  $163^\circ$ relative to the Solar system's invariable plane.  Ka`epaoka`awela's motion is stable over the age of the solar system  thanks to Jupiter's co-orbital resonance. Understanding the process that led to Ka`epaoka`awela's capture and more generally its relationship to other high inclination asteroids in the Solar system relies on understanding precisely how such bodies cross the outer planets region and interact with their mean motion resonances at arbitrary inclination. 

Numerical simulations that study capture at arbitrary inclination \citep{NamouniMorais15,MoraisNamouni16,NamouniMorais17,NamouniMorais18c} are useful tools in this respect but they are not able to cover large portions of parameter space as would an analytical approach that quantifies the dynamics of resonance and capture. To this end, we have recently developed a disturbing function (\cite{Namouni18a}, Paper I) that generalizes the classical one for low inclination prograde motion \citep{ssdbook} as well as that of low inclination retrograde motion \citep{MoraisNamouni13a} and that of polar motion (\cite{NamouniMorais17b}, Paper II).  In this work, we validate the use of the disturbing function by comparing the analytical resonance widths it yields to those inferred from the {\sc megno} chaos indicator itself based on direct numerical integrations of the full equation of motion \citep{cincotta00,goz03}. We further examine the resonance libration centers obtained from the direct numerical integrations and discuss the underlying mechanism that controls what resonant argument is likely to librate. 

Resonance dynamics at arbitrary inclination may also be studied using semi-analytical methods where the unexpanded disturbing function is averaged numerically. This has been done for the co-orbital resonance at high  inclination \citep{Namouni99,Namounietal99}, for coorbital resonance with retrograde motion \citep{MoraisNamouni13a} and for mean motion resonances in the Solar system \citep{gallardo06,gallardo19a}. The numerically-averaged disturbing function has the advantage of automatically including all possible harmonics as well as arguments at a given mean motion location defined by nominal resonance. However, this method can only yield approximate estimates of resonance width as the underlying theoretical hamiltonian is too complex because of all the arguments and their harmonics requiring the use of  pendulum model approximations \citep{gallardo20}. Recently, the semi-analytical method was compared to our first analytical estimates of pure eccentricity resonance widths in Paper I \citep{gallardo19a} and some doubt was cast on the ability of the disturbing function for arbitrary inclination to describe resonance dynamics. To dispel this doubt, we expand the analytical pendulum models we used in Paper I by including the possible presence of multiple librating arguments to the one-harmonic and two-harmonics pendulum models and apply {them} to Jupiter's inner 2:1 and 3:1 resonances as  well as Neptune's outer 1:2 and 1:3 resonances and their asymmetric librations. We show through these models that  the disturbing function reproduces accurately resonance structure including the domains of asymmetric librations of outer resonances. 
 We also discuss the recent work of \cite{lei19} who developed a model based on rewriting the disturbing function of Paper I in terms of harmonics of the pure inclination resonance whose amplitudes depend on the argument of perihelion and applied it to Jupiter's 2:1 and 3:1 {inner}  resonances. In particular, we explain why their conclusion, about the crucial role of the pure inclination resonant argument for all types of resonance, is erroneous. 

The paper is organized as follows. In Section 2, we recall the general expression of the disturbing function of Paper I and its main properties. In Section 3, we recall how the pendulum model works with one and two harmonics as well as the main underlying assumption {for} its validity. In Section 4, we expand the previous pendulum models to include the presence of multiple librating arguments and derive the corresponding analytical widths. Section 5 is devoted to the comparison of our analytical estimates with the {\sc megno} maps and the discussion of how analytical modeling help us understand resonance {dynamics}.  In Section 6, we compare our {results} to the works of \cite{gallardo19a} and  \cite{lei19}. Section 7 contains our conclusions. 

\section{Disturbing function for arbitrary inclinations}
The motion of  an asteroid  under the gravitational influence of the sun  of mass $M_{\star}$ and a planet, of mass $m^\prime\ll M_{\star}$, on a circular orbit of semi-major axis  $a^\prime$ is governed by the potential:
\begin{equation}
R= {G\,m^\prime}{a^{\prime-1}} (\Delta^{-1}-r\,\cos{\psi})\equiv {G\,m^\prime}{a^{\prime-1}}\bar{R},
\end{equation}
where the asteroid's osculating Keplerian orbit with respect to the sun has semi-major axis $a$, eccentricity $e$, inclination $I$, true anomaly $f$, argument of perihelion $\omega$, and  longitude of ascending node $\Omega$.  The asteroid's orbit radius $r=\alpha (1-e^2)/(1+e \cos f)$ and $\alpha=a/a^\prime$ is its normalized semimajor axis. The planet-asteroid distance $\Delta^2=1+r^2-2\,r\,\cos{\psi}$ and  the angle between the radius vectors of  the planet and asteroid, $\psi$,  is given as:
\begin{equation}
\cos\psi=\cos(\Omega-\lambda^\prime)\cos(f+\omega )-\sin(\Omega-\lambda^\prime)\sin(f+\omega )\cos I,  \label{cospsi}
\end{equation}
where $\lambda^\prime$ is the planet's longitude, and inclination is measured with respect to the planet's orbital plane. 

The first term of $\bar{R}$  is  the direct perturbation that we denote $\bar{R}_d$ and  the second term, that we denote $\bar{R}_i$, is the indirect perturbation that comes from the reflex motion of the sun under the influence of the planet as the standard coordinate system is chosen to be  centered on the sun. 

The disturbing function for arbitrary inclination is an expansion of $\bar{R}$ with respect to eccentricity and inclination in the vicinity of circular motion on a plane inclined by the arbitrary inclination $I_r$. Paper I is devoted to the expansion derivation. Here we note the salient properties of {that} expansion. The direct and main part of the perturbation may be written as:
\begin{eqnarray}
\bar{R}_d&=&\!\!\!\!\!\!\!\!\sum_{
{\scriptsize\begin{array}{c}
-N\leq k\leq N \\
|k|\leq m\leq N\\
0\leq n\leq N\\ 
m+n=N 
\end{array}}}\!\!\!\!\!\!\!\!
c^{k}_{mn}(p,q,\alpha,I_r)\, e^m s^n \ \cos \phi^{p:q}_k, \label{RY}\\
\phi^{p:q}_k&=&q \lambda -p \lambda^\prime +(p-q) \Omega-k\omega. \nonumber
\end{eqnarray}
where $p$, $q$ and $k$ are integers and $s=\sin(I-I_r)$.  The force coefficients $c^{k}_{mn}(p,q,\alpha,I_r)$ have an  important property related to the resonance order $|p-q|$.  For an odd resonance order, $c^{k}_{mn}(p,q,\alpha,I_r)=0$ when $k$ is even, whereas for an even resonance order, $c^{k}_{mn}(p,q,\alpha,I_r)=0$ when $k$ is odd. This property guarantees that the integer coefficient of the longitude of ascending node, $\Omega$, that reads $p-q+k$ is always even. The property is familiar from the classical disturbing function \citep{ssdbook} and will be used particularly in Section 6 when we discuss the models and results of \cite{lei19}.  The force coefficients satisfy another important relationship that relates positive and negative $k$ terms, and may be written as:
\begin{equation}
 c^{-k}_{mn}(p,q,\alpha,I_r)=c^{k}_{mn}(p,-q,\alpha,I_r+180^\circ \mbox{modulo}\ 180^\circ). \label{cmkpk}
 \end{equation}
An expansion of order $N$ in terms of eccentricity $e$ and inclination $s=\sin(I-I_r)$ models all resonant angles $\phi^{p:q}_k=q\lambda-p\lambda^\prime-(q-p)\Omega -k \omega$  with $|k|\leq N$ regardless of the values of $p$ and $q$. {Consequently}, an expansion of order $N$ does not limit the type of resonance that can be modeled. For instance, an expansion of order 2 may be used to study the 1:6 resonance. This property is not found in the classical disturbing function (for prograde low inclination motion) {as} an expansion of order $N$ may model only resonances with $q=p\pm N$. For the 1:6 resonance, an expansion order of at least 5 is required to get the relevant force terms.  

Another property of the disturbing function for arbitrary inclination that differs significantly from the classical disturbing function is the fact that all even (odd) order resonances have force coefficients that are quadratic (linear) in eccentricity and inclination to lowest order. In the classical disturbing function, however, to lowest order, force coefficient powers are determined {strictly} by the resonance order. For the 1:6 resonance, the lowest order force terms are of the type $e^m\sin (I/2)^{2n}$ where $m+2n=5$ is the resonance order. 

The presence of the reference inclination $I_r$ gives the disturbing function for arbitrary inclination two possible interpretations. The first is that of a double expansion with respect to eccentricity $e$ and inclination sine $s=\sin(I-I_r)$ measured with respect to the reference plane defined by $I_r$. The second interpretation is found when $s$ is set to zero identically. Then the reference inclination becomes the inclination of the asteroid and the disturbing function is {an expansion with respect to eccentricity only}. This interpretation is useful for the determination of resonance width as shown in Paper I and will be used in this work.

The disturbing function for arbitrary inclination may be used to study prograde as well as retrograde motion. In this respect, there are two particularly important resonant arguments at nominal resonance: the prograde and retrograde pure eccentricity arguments. The former is obtained by choosing $k=q-p$ and the latter with $k=q+p$ (see Paper I for details).  When we apply the analytical models in Section 5 to Jupiter's 2:1 and 3:1 inner resonances by including among other modes $k$ the retrograde ones with $k=3$ and $k=4$ respectively, the expansion of the disturbing function given explicitly to order $N=4$ in Paper I is sufficient. This however is not true of Neptune's 1:2 and 1:3 outer resonances as the pendulum model that describes them requires the second harmonic of every argument. Consequently, we expand the disturbing function to order 8 so as to include the effects of retrograde motion for outer resonances. In order to avoid overcrowding the article with the full tables of the 8th-order expansion, we only give in Appendix A, the expressions of the force amplitudes $f^{p:q}_k$ of the argument  $\phi^{p:q}_k=q \lambda -p \lambda^\prime +(p-q) \Omega-k\omega$ that we study in this work namely: those of the 2:1, 3:1, 1:2 and 1:3 resonances.

\section{Single-argument pendulum models of resonance}
The pendulum model of resonance is a standard tool to determine the width of a mean motion resonance in the three-body problem. Its basic approximation is that among the three angles that describe the orbit of the massless asteroid, $\lambda$, $\omega$ and $\Omega$, only the mean longitude has by far the fastest frequency variations implying that all other frequency changes should be neglected. Neglecting secular accelerations of the remaining angles is only justified for orbits with moderate to large eccentricity. For the objects of interest {in our work}, namely Centaurs and TNOs, the pendulum approximation is satisfied. For further details about the validity of the pendulum approximation and how to remedy {its shortcoming for circular orbits}, we refer the reader to \cite{ssdbook}. In the following, we review briefly the classical pendulum model and the two-harmonics pendulum model that describe the {dynamics} of a {single} resonant argument then generalize it to the possible presence and simultaneous libration of different arguments at the same resonant location. 

\subsection{Classical pendulum model}
For an asteroid in a $p$:$q$ resonance with the planet of  resonant argument $\phi_k^{p:q}=q\lambda-p\lambda^\prime-(q-p)\Omega-k\omega$, the {relevant part in} the disturbing function with its direct and  indirect parts may be written as $\bar R=f_{k}^{p:q}\cos \phi_k^{p:q}$. The force amplitude  $f_{k}^{p:q}$  includes the eccentricity and inclination dependence of the resonant term. The secular part of the disturbing function is absent in accordance with the pendulum approximation. The time evolution of the resonant argument satisfies the pendulum equation:
\begin{equation}
\ddot \phi_k^{p:q}= \frac{3 n^2 q^2 m^\prime  \alpha}{M_\star}  f_{k}^{p:q} \ \sin \phi_k^{p:q},\label{oneharmonic}
\end{equation}
where the dot indicates the derivative with respect to time. {For the equation's  derivation, see Paper I}.  The pendulum's angle, $\phi_k^{p:q}$, librates stably at the natural frequency $|3 n^2 q^2 m^\prime M_\star^{-1}\alpha  f_{k}^{p:q}  |^{1/2}$   around $\phi_k^{p:q}=0^\circ$ if the sign of $f_{k}^{p:q} $ is negative and around $\phi_k^{p:q}=180^\circ$  if the sign is positive. The resonance's width in terms of semi-major axis is given as:
\begin{equation}
\Delta_0 a_k^{p:q} = \left[\frac{16\, \alpha\, m^\prime f_{k}^{p:q}}{3M_\star}\right]^\frac{1}{2} a_k^{p:q}\label{reswidth0}
\end{equation}
where  $a_k^{p:q}=\alpha a^\prime$ is the semi-major axis of nominal resonance. 

\subsection{Two-harmonics pendulum model}
The classical pendulum model describes accurately the global  dynamics of inner resonances for moderate to large eccentricity. It is, however, not adequate for outer resonances because the topology of phase space  includes asymmetric librations, that is, librations of $\phi_k^{p:q}$ around arbitrary values other than $0^\circ$ and $180^\circ$ \citep{Bruno,Malhotra96,WinterMurray97}. In Paper I, we were inspired by the Andoyer Hamiltonian model \citep{Andoyer02,Beauge94} to solve this problem {analytically}  by developing a new pendulum model that includes the effect of the second harmonic $\phi_{2k}^{2p:2q}=2\phi_k^{p:q}$. The two-harmonics pendulum equation is  given as:
\begin{eqnarray}
\ddot \phi_k^{p:q}&=& \frac{3 n^2 q^2 m^\prime  \alpha}{M_\star} \left( f_{k}^{p:q}\ \sin \phi_k^{p:q}+ 2 f_{2k}^{2p:2q}\ \sin 2 \phi_k^{p:q}\right).\label{twoharmonic}
\end{eqnarray}
It was found in Paper I that the behavior of this one-dimensional dynamical system depends on the parameter:
\begin{equation}\beta=\frac{4f_{2k}^{2p:2q}}{|f_{k}^{p:q}|},\label{beta}\end{equation} 
that determines the possible formation of critical points other than classical ones around $0^\circ$ and $180^\circ$ at
\begin{equation}
\phi_{\rm asymmetric}=\pm\cos^{-1}(-\beta^{-1}) \label{alib}
\end{equation} 
if $f_{k}^{p:q}\geq0$ otherwise the asymmetric critical points are shifted by $180^\circ$ with respect to (\ref{alib}). 
The two-harmonics pendulum equation can in effect be simplified by considering only positive $f_{k}^{p:q} $. If the latter quantity is negative, the variable change $\phi=180^\circ+\bar\phi$ produces a two-harmonics pendulum equation for  $\bar \phi$ identical to that of $\phi$ with a positive  $f_{k}^{p:q} $.  
 
It was found in Paper I, assuming a positive $f_{k}^{p:q} $, that when $|\beta|<1$, stable libration occurs around $180^\circ$  and the semimajor axis resonance width is given by the classical pendulum model resonance width  (\ref{reswidth0}). 

When,  $\beta\leq-1$, stable librations may occur around $0^\circ$ and $180^\circ$ whose semimajor axis widths are respectively:
\begin{eqnarray}
\Delta_1 a_k^{p:q} \!\!\!\!\!\!&=& \!\!\!\!\!\!\!\left[\frac{ \alpha\, m^\prime}{3M_\star}\right]^\frac{1}{2}  \!\!\frac{|4f_{2k}^{2p:2q}+|f_{k}^{p:q}||}{|f_{2k}^{2p:2q} |^{1/2}}a_k^{p:q}, \label{reswidth1}\\
\Delta_2 a_k^{p:q} \!\!\!\!\!\!&=& \!\!\!\!\!\!\!\left[\frac{ \alpha\, m^\prime}{3M_\star}\right]^\frac{1}{2}  \!\!\frac{|4f_{2k}^{2p:2q} -|f_{k}^{p:q} ||}{|f_{2k}^{2p:2q} |^{1/2}}a_k^{p:q}.\label{reswidth2}
\end{eqnarray}

When $\beta\geq1$, asymmetric librations at either asymmetric points (\ref{alib}) are possible and their resonance's semimajor axis width is given by $\Delta_2 a_k^{p:q}$. A larger amplitude libration may also occur around both asymmetric points. Centered around $\phi=180^\circ$, its resonance width given by $\Delta_1 a_k^{p:q}$. 

\section{Multiple-argument pendulum models of resonance}
In this section, we build on the previous two models and develop a pendulum model of resonance that describes the situation where different {arguments} pertaining to the same resonance may librate simultaneously.  Such situations are not exceptional. In effect, the pendulum approximation itself looks at evolution intervals where the secular accelerations are negligible. In that case, one can, for instance, subtract from the resonant argument $\phi_k^{p:q}$ any multiple of the argument of perihelion $\omega$, as it is mostly constant, and form new arguments  $\phi_{k+l}^{p:q}=\phi_k^{p:q}-l\omega$ that can librate within the same time interval where the pendulum approximation is valid. This implies that resonance width and libration can be influenced by different modes $k$ depending on their force amplitudes. However, argument libration in the pendulum approximation does not mean that any argument can librate at nominal resonance indefinitely. The reason is that the timescale of the pendulum approximation that is larger than the resonant timescale is also smaller than the secular timescale.  Secular evolution of the angles neglected by the pendulum approximation introduces drifts in the resonant arguments so that usually only one argument may librate at resonance on the secular timescale unless the argument of perihelion is locked in one of the equilibria of the Kozai-Lidov secular resonance \citep{Kozai62,Lidov62}. 

\begin{figure}
\begin{center}
{ 
\hspace*{-3mm}\includegraphics[width=85mm]{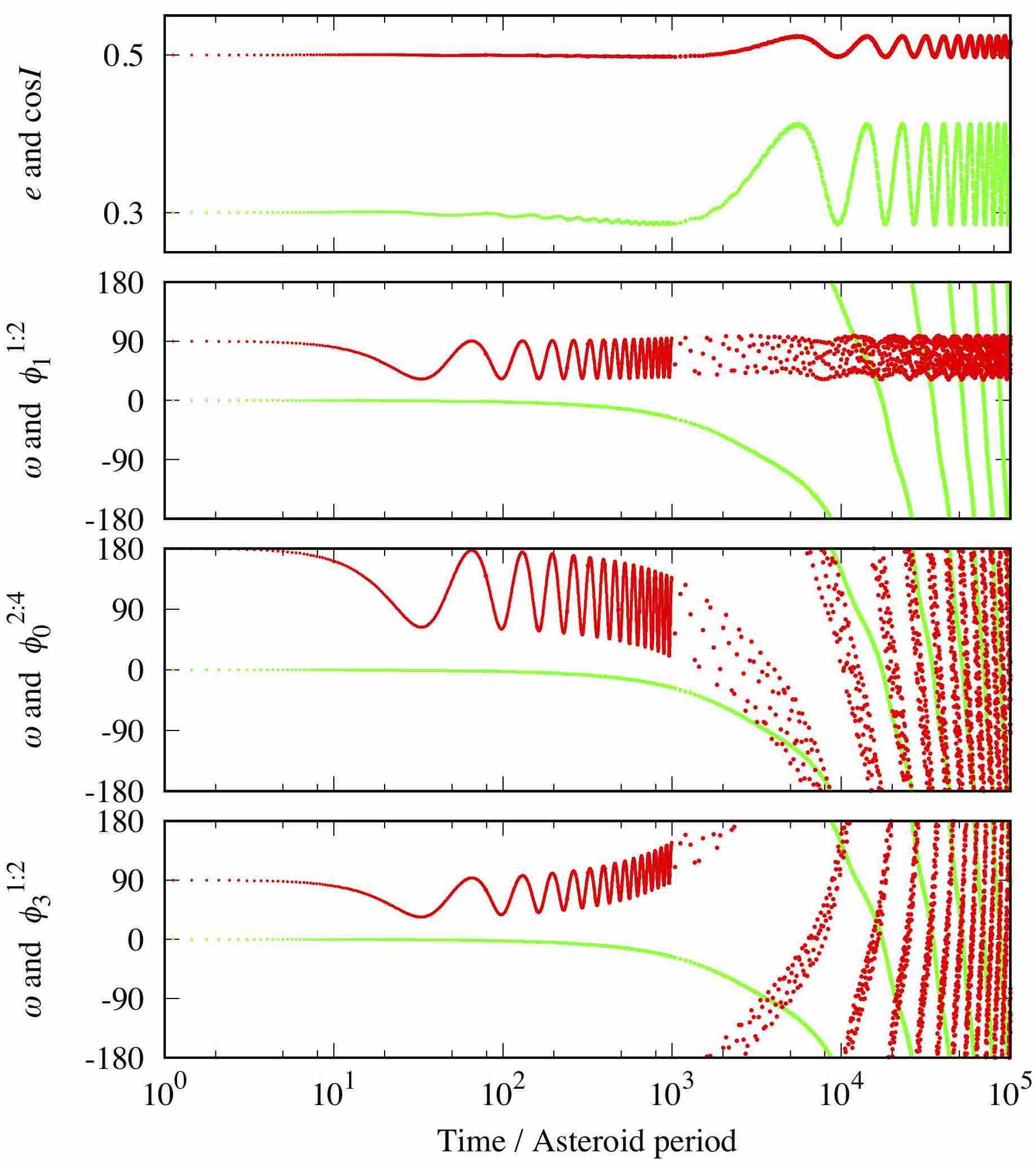}}
\caption{Time evolution of an asteroid in Neptune's 1:2 outer resonance. In the top panel, the top red (bottom green) curve denotes the  inclination's cosine (eccentricity). In the argument panels,  the top red (bottom green) curve in the first $10^3$ periods denotes the resonant argument (argument of perihelion).  After $10^3$ periods, the sampling of time evolution data was reduced to make the plots easily readable.}\label{orbits}
\end{center}
\end{figure}

To illustrate this concept, we show in Figure \ref{orbits}, the evolution of an asteroid in Neptune's outer 1:2  resonance over $10^5$ asteroid orbits. The asteroid's orbit has initial eccentricity $e=0.3$,  inclination $I=60^\circ$, $\phi^{1:2}_1=90^\circ$, and $\omega=0^\circ$. A logarithmic scale is used to show the evolution contrast between the resonant and secular timescales.  The modes shown define the prograde pure eccentricity argument $\phi^{1:2}_{1}=2\lambda- \lambda^\prime-\varpi$ where $\varpi$ is the longitude of perihelion, the retrograde pure eccentricity resonance $\phi^{1:2}_{3}=2\lambda^\star-\lambda^\prime-3\varpi^\star$ where  $\lambda^\star=\lambda-2\Omega$ and  $\varpi^\star=\varpi-2\Omega$ are the physical angles used to study retrograde motion. We also plot the pure inclination mode $k=0$ that appears only as a second harmonic $\phi^{2:4}_0$ because its force amplitude affects odd order outer  resonances as their dynamics depend on second harmonics as well as first harmonics contrary to odd order inner resonances whose evolution is governed by first harmonics as explained in Section 3.  

The initial conditions result in asymmetric libration around $\phi^{1:2}_1=60^\circ$ of amplitude $30^\circ$ and a resonant period of 65 asteroid periods. The argument of perihelion $\omega$ regresses secularly with a period of $17500$ asteroid periods. On the smaller timescale of $10^3$ asteroid periods, the argument of perihelion $\omega$ is mostly constant, the pendulum approximation applies, and all three modes $k=1$, 3 and 0 librate. Therefore, the pendulum model that determines resonance width should take into account these {modes} and others if necessary. On the secular timescale, however, only the mode $k=1$ retains its asymmetric libration. The 65-year oscillation is always present in $k=3$ and $k=0$ but, {on the secular timescale}, the modes drift along with the argument of perihelion $\omega$. Further examples that illustrate this dynamics may be found in Paper II, Section 6. In the following, we expand pendulum models to include any number of librating modes on the resonant timescale. 

\subsection{One-harmonic pendulum model for simultaneous argument librations}
\subsubsection{The case of two arguments}
It is possible to modify analytically the pendulum models to account for simultaneous libration and assess resonance width. In a first step, we do this for two arguments of an inner resonance then generalize it to an arbitrary number of arguments.  The pendulum equation for two arguments $\phi_{k_1}^{p:q}$ and  $\phi_{k_2}^{p:q}$ is given as:
 \begin{eqnarray}
\ddot \phi_{0}^{p:q}&=& \frac{3 n^2 q^2 m^\prime  \alpha}{M_\star}  
\left( f_{k_1}^{p:q}  \ \sin \phi_{k_1}^{p:q} + f_{k_2}^{p:q} \ \sin \phi_{k_2}^{p:q} \right)\label{sonehar}
\end{eqnarray}
where we have chosen to write the equation for $\phi_{0}^{p:q}$ only for convenience as any multiple of $\omega$ may be added or subtracted from that side of the equation without affecting its validity. For instance, there is no $k=0$ mode for {the} inner 2:1 resonance as its resonance order is odd.  The choice of $k=0$ reflects the lack of preferential treatment of any of the two modes $k_1$ and $k_2$.   With simple algebra, the previous equation can be written as a pendulum equation as follows:
\begin{eqnarray}
\ddot \phi_{0}^{p:q}&=& \frac{3 n^2 q^2 m^\prime  \alpha}{M_\star}\  
g_{k_1,k_2}^{p:q} \ \sin \left(\phi_{0}^{p:q}-\delta \phi_{k_1,k_2}^{p:q}\right)  \label{sonehar2}
\end{eqnarray}
where the effective force amplitude and argument shift are given as
\begin{eqnarray}
g_{k_1,k_2}^{p:q} &=& \left[{f_{k_1}^{p:q}}^2+ {f_{k_2}^{p:q}}^2 +2 f_{k_1}^{p:q} f_{k_2}^{p:q} \cos([k_1-k_2]\omega)\right]^{1/2}, \label{ampgk1k2}\\
\cos\delta \phi_{k_1,k_2}^{p:q}&=& \left[ f_{k_1}^{p:q} \cos k_1\omega + f_{k_2}^{p:q} \cos k_2\omega\right] \left(g_{k_1,k_2}^{p:q}\right)^{-1},\nonumber\\
\sin\delta \phi_{k_1,k_2}^{p:q}&=& \left[f_{k_1}^{p:q}\sin k_1\omega +f_{k_2}^{p:q}\sin k_2\omega\right] \left(g_{k_1,k_2}^{p:q}\right)^{-1}. \label{ampdelta}
\end{eqnarray}
These equations imply that the resonance semi-major axis width is still given by equation (\ref{reswidth0}) because the system is described  exactly using a classical pendulum equation (\ref{sonehar2}). However this time, the force amplitude of simultaneous librations  $g_{k_1,k_2}^{p:q}$ depends on the force amplitudes of both modes and crucially on the value of the argument of perihelion $\omega$. Furthermore, both libration arguments are shifted by $\delta \phi_{k_1,k_2}^{p:q}$ also a function of $\omega$. We call this type of asymmetric libration `displaced libration' so as not to confuse it with bona fide asymmetric librations that are related to the existence of additional critical points in the dynamical system. We emphasize that this derivation and its consequences are valid only on the resonant timescale and do not mean that both $k_1$ and $k_2$ will librate indefinitely. The secular evolution of $\omega$ in particular will define what mode if any may librate on the secular timescale.

 \subsubsection{The case of multiple arguments}
When $N_k$ different arguments librate simultaneously at the nominal resonant location on the resonance timescale, the pendulum equation is similar to (\ref{sonehar}) but with $N_k$ terms and may be reduced to classical pendulum equation similar to (\ref{sonehar2}) where the corresponding effective force amplitude and argument shift are given as:
\begin{eqnarray}
{{g}^{p:q}}^2 &=& 
\!\!\!\!\!\!\!\sum_{
{\scriptsize\begin{array}{c}
0\leq \mu,\nu\leq N_k\\ 
\mu\neq \nu
\end{array}}} \!\!\!f_{k_\nu}^{p:q}  f_{k_\mu}^{p:q} \cos([k_\mu-k_\nu]\omega)+\nonumber\\
&&\ \ \ \ \ \ \ \ \ +\sum_{\mu=1}^{N_k}{f_{k_\mu}^{p:q}}^2,  \label{gmultiple}\\ 
\cos\delta \phi^{p:q}&=& \left[\sum_{\mu=1}^{N_k}f_{k_\mu}^{p:q} \cos k_\mu\omega \right]\left(g^{p:q}\right)^{-1},\label{deltamultiple}\\
\sin\delta \phi^{p:q}&=& \left[\sum_{\mu=1}^{N_k}f_{k_\mu}^{p:q}\sin k_\mu\omega \right]\left(g^{p:q}\right)^{-1},\nonumber
\end{eqnarray}
where $\{k_\mu,1\leq\mu\leq N_k\}$ define the different arguments at the nominal location and $f_{k_\mu}^{p:q}$ are the corresponding force amplitudes.

\subsection{Two-harmonics pendulum model for simultaneous multiple argument librations}
\subsubsection{Pendulum Hamiltonian for two librating arguments}
The presence of more than one resonant argument for outer resonances modifies not only resonance width through the contribution of the various mode amplitudes and the value of the argument of perihelion,  but also the nature of librations as the relative strength of the perturbing terms would affect the effective $\beta$-parameter of Section 3.2 that will be called $b$ in this section in order to avoid confusion. We first consider the two-arguments two-harmonics pendulum equation that may be written as follows: 
\begin{eqnarray}
\ddot \phi_0^{p:q}&=& \frac{3 n^2 q^2 m^\prime  \alpha}{M_\star} \left( f_{k_1}^{p:q} \ \sin \phi_{k_1}^{p:q}+ 2  f_{2k_1}^{2p:2q} \ \sin 2 \phi_{k_1}^{p:q}\right.\nonumber\\
&& + \left.  f_{k_2}^{p:q} \ \sin \phi_{k_2}^{p:q} +2  f_{2k_2}^{2p:2q} \ \sin 2 \phi_{k_2}^{p:q}\right).
\label{twoargtwoharmonic}
\end{eqnarray}
We again use the angle $\phi_0$  for convenience as explained previously.  Next, we apply the same algebraic transformation in Section 4.1. to the first and third terms, and  to the second and fourth terms. The pendulum equation now reads:
\begin{eqnarray}
\ddot \phi_{0}^{p:q}&=& \frac{3 n^2 q^2 m^\prime  \alpha}{M_\star}  \left[
  g_{k_1,k_2}^{p:q} \ \sin \left(\phi_{0}^{p:q}-\delta \phi_{k_1,k_2}^{p:q}\right) \right.\\ \nonumber
 &&+  \left. 2 g_{2k_1,2k_2}^{2p:2q} \ \sin \left(2\phi_{0}^{p:q}-\delta \phi_{2k_1,2k_2}^{2p:2q}\right)\right].\label{redtwotwohar}
\end{eqnarray}
Using the angle $\psi=\phi_{0}^{p:q}-\delta \phi_{2k_1,2k_2}^{2p:2q}/2$ and the new time variable $t^\prime = t |3 n^2 q^2 m^\prime  \alpha g_{k_1,k_2}^{p:q} /M_\star)|^{1/2}$, the pendulum equation is rewritten as $\psi^{\prime\prime}=\sin (\psi-\delta) +b\ \sin \psi\cos\psi
$ where $\psi^{\prime\prime}$ is the second derivative of $\psi$ with respect to $t^\prime$ and the parameters $b$ and $\delta$  are given as:
\begin{equation}
b=\frac{4g_{2k_1,2k_2}^{2p:2q}}{|g_{k_1,k_2}^{p:q}|},\ \ \mbox{and}\ \ \delta=\delta \phi_{k_1,k_2}^{p:q}-\delta \phi_{2k_1,2k_2}^{2p:2q}/2.
\end{equation}
In writing the pendulum equation, we assumed that $g_{k_1,k_2}^{p:q}$ is positive. If it is not then a simple phase shift of $\psi$ by $180^\circ$ makes the coefficient of the first harmonic positive (see Paper I).  The dynamical system derives from   the Hamiltonian: 
\begin{equation}H=\frac{p^2}{2} +\cos(\psi -\delta)+\frac{b}{4}\, \cos(2\psi),\label{hamil}\end{equation}
where $p=\psi^\prime$ is the momentum. The Hamiltonian is similar to that of two harmonics (Paper I) except for the presence of the argument shift $\delta$. The system is invariant under the following two transformations: ($\psi\rightarrow \psi+90^\circ$, $\delta\rightarrow \delta+90^\circ$, $b\rightarrow -b$) and  ($\psi\rightarrow \psi-90^\circ$, $\delta\rightarrow \delta-90^\circ$, $b\rightarrow -b$). This allows us to limit our study to $b\geq0$.  We will later restrict the  argument shift $\delta$ to the interval [$0^\circ:90^\circ$] as the Hamiltonian structure corresponding to all other values may be recovered by the following transformations that bring $\delta$ back to the interval [$0^\circ:90^\circ$]:
\begin{eqnarray}
-180^\circ\leq \delta\leq -90^\circ&& \delta \rightarrow 180^\circ+\delta,\  \psi\rightarrow 180^\circ+\psi,\label{tras1}\\
-90^\circ\leq \delta\leq 0^\circ&& \delta \rightarrow -\delta,\  \psi\rightarrow -\psi,\label{tras2}\\
90^\circ\leq \delta\leq 180^\circ&& \delta \rightarrow 180^\circ-\delta,\ \psi\rightarrow 180^\circ-\psi.\label{tras3}
\end{eqnarray}  
These relationships are useful for the determination of the resonance widths as the critical points are given by different expressions depending on the value of the argument shift $\delta$ as we explain in the following.

\subsubsection{Critical points}
\begin{figure*}
\begin{center}
{ 
\includegraphics[width=40mm]{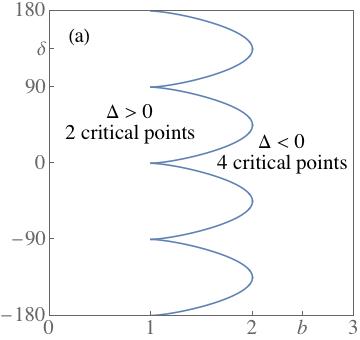}\hspace{3mm}\includegraphics[width=40mm]{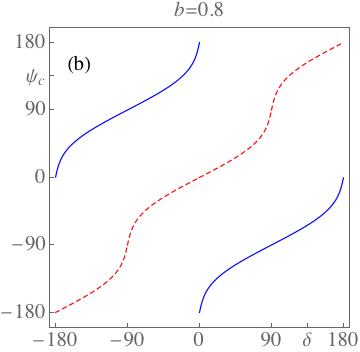}\hspace{3mm}\includegraphics[width=40mm]{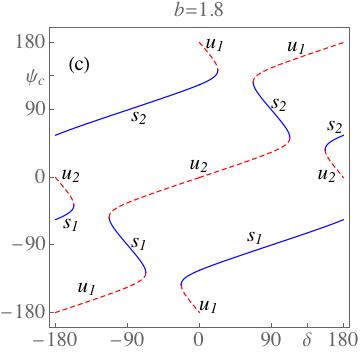}\hspace{3mm}\includegraphics[width=40mm]{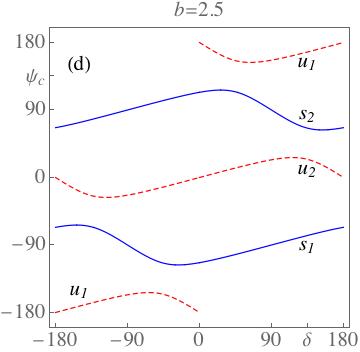}
\\[1mm]
\hspace*{3mm}\includegraphics[width=38mm]{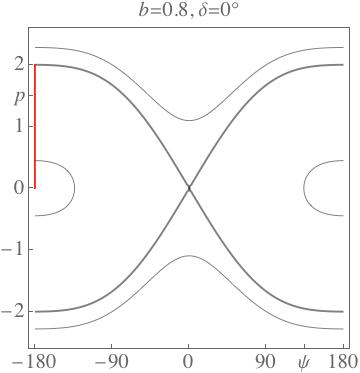}\hspace{5mm}\includegraphics[width=38mm]{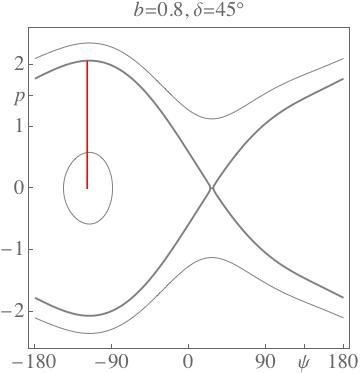}\hspace{5mm}\includegraphics[width=38mm]{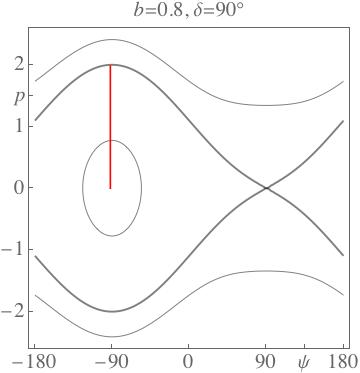}\hspace{5mm}\includegraphics[width=38mm]{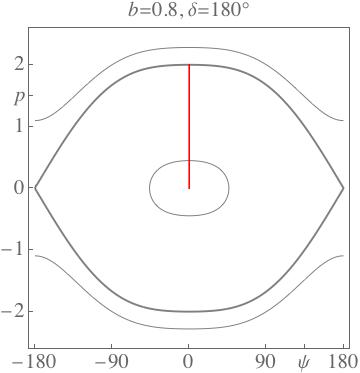}
\\[1mm]
\hspace*{3mm}\includegraphics[width=38mm]{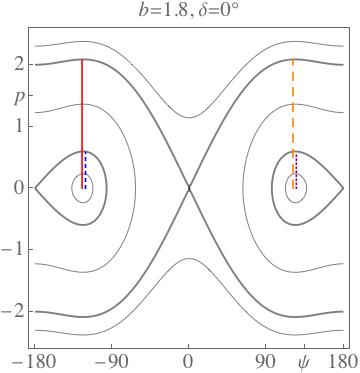}\hspace{5mm}\includegraphics[width=38mm]{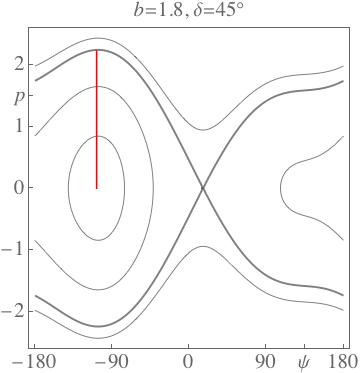}\hspace{5mm}\includegraphics[width=38mm]{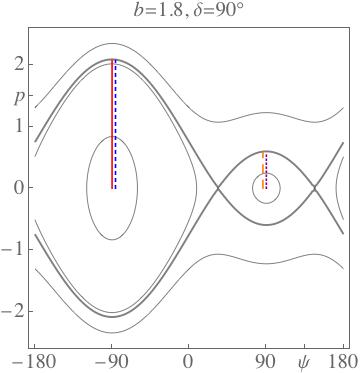}\hspace{5mm}\includegraphics[width=38mm]{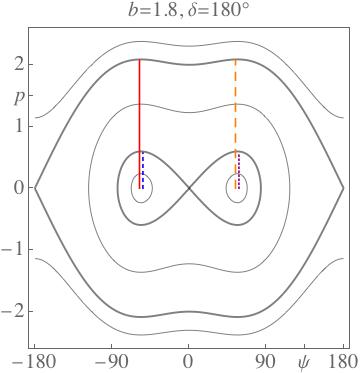}
\\[1mm]
\hspace*{3mm}\includegraphics[width=38mm]{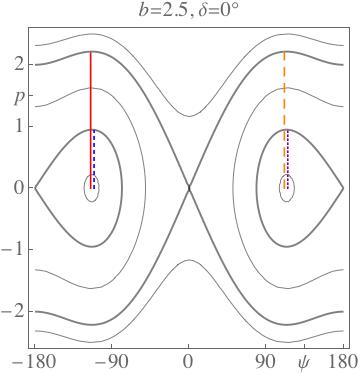}\hspace{5mm}\includegraphics[width=38mm]{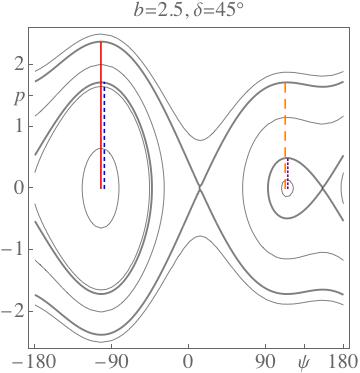}\hspace{5mm}\includegraphics[width=38mm]{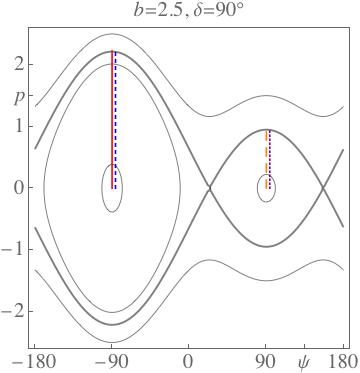}\hspace{5mm}\includegraphics[width=38mm]{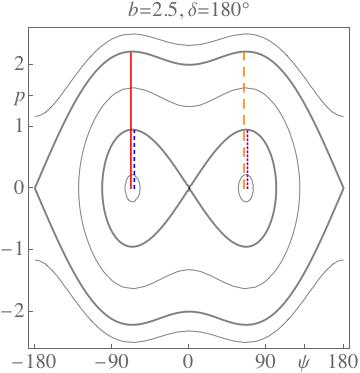}
}
\caption{Topology of the two-arguments two-harmonics pendulum model. First row: (a) Number of critical points in the $b\delta$-plane, (b,c,d) Positions of the stable (unstable) critical points are shown with solid blue lines (dashed red lines) for $b=0.8$, 1.8 and 2.5.  Second to fourth rows: Level curves of Hamiltonian (\ref{hamil}) for  $b=0.8$, 1.8 and 2.5 respectively and the four values of $\delta=0^\circ, \ 45^\circ, \ 90^\circ$ and $180^\circ$. The vertical lines starting from the location of the stable equilibrium points denote the resonance widths whose line-types and colors are defined in Figure \ref{figamp}.}\label{fd}
\end{center}
\end{figure*}

In order to  determine the nature of librations and estimate the resonance width {analytically}, we write the equations for the critical points of (\ref{hamil}) as $\partial_pH=0$ and $\partial_\psi H=0$.  The first yields $p=0$ and the second may be written as:
\begin{equation}
\sin\delta\, v^4- 2(b-  \cos\delta) \, v^3 +2(b+\cos\delta)\, v -\sin\delta=0
\end{equation}
where $v=\tan(\psi_c/2)$ {and} $\psi_c$ is the value of $\psi$ at the critical points. The solutions of the quartic polynomial depend on its discriminant $\Delta=32 [8b^6 - 24 b^4  -3 (1-9 \cos4\delta)b^2 -8]$. 
There are two real solutions if $\Delta<0$ and four (including multiple ones)  if $\Delta\geq0$. The regions of two and four solutions are shown in Figure \ref{fd}, panel (a) in the $b\delta$-plane for $b\geq 0$.  The existence of four equilibrium points for  $1\leq b\leq 2$ is no longer guaranteed like the one-argument two-harmonics pendulum model of Section 3.2 but depends on the argument shift $\delta$. 

We determine analytically the location of the critical points using the four roots of the quartic:
\begin{eqnarray}
v_1&=&\frac{ b - \cos\delta}{2\sin\delta}- S -R_{-},\\
v_2&=&\frac{ b - \cos\delta}{2\sin\delta}- S +R_{-},\\
v_3&=&\frac{ b - \cos\delta}{2\sin\delta}+S -R_{+},\\
v_4&=&\frac{ b - \cos\delta}{2\sin\delta}+S +R_{+}, 
\end{eqnarray} 
where 
\begin{eqnarray}
S&=&\frac{1}{2}\left[\left(\frac{b - \cos\delta}{\sin\delta}\right)^2+\frac{2^\frac{1}{3}S_0}{9^\frac{1}{3}}   +\frac{32^\frac{1}{3} ( b^2-1) }{
 3^\frac{1}{3} S_0\sin^2\delta}\right]^\frac{1}{2} \nonumber,\\
 S_0&=& \left[\frac{36 b \cos\delta}{\sin^{2}\delta}- \frac{(-3\Delta)^\frac{1}{2}}{4 \sin^{3}\delta}\right]^\frac{1}{3}, \nonumber\\
 R_{\pm}&=&\frac{1}{2}\left[\frac{3(b-\cos\delta)^2}{\sin^2\delta}-4S^2\right. \nonumber \\
 && \ \ \ \ \ \  \left. \pm \frac{(b-\cos\delta)^3-2(b+\cos\delta)\sin^2\delta}{S \sin^3\delta}\right].^\frac{1}{2}
 \end{eqnarray} 
 
A given root does not correspond to a single critical point as the former is not a continuous function of $\delta$. The critical points $\psi_c(b,\delta)$ themselves, however, are continuous with respect $\delta$ and may be expressed in terms of the roots as follows. For $0\leq b \leq 1$, there are only two critical points shown in Figure \ref{fd} top  panels (a) and (b) with $b=0.8$. Their expressions are given as:
\begin{eqnarray}
&\mbox{Unstable point}&\label{ubl1}\\
&  -180^\circ\leq \delta\leq -90^\circ,\   \psi_u=2 \arctan v_1,&\nonumber\\
&   -90^\circ\leq \delta\leq 90^\circ, \   \psi_u=2 \arctan v_3,&\nonumber\\
 &   90^\circ\leq \delta\leq 180^\circ,\   \psi_u=2 \arctan v_4,&\nonumber\\
&\mbox{Stable point}&\label{sbl1}\\ 
&  -180^\circ\leq \delta\leq -90^\circ,\   \psi_s=2 \arctan v_2,&\nonumber\\
&   -90^\circ\leq \delta\leq 0^\circ, \   \psi_s=2 \arctan v_4,&\nonumber\\
 &   0^\circ\leq \delta\leq 180^\circ,\   \psi_s=2 \arctan v_1.&\nonumber
\end{eqnarray}
For $1\leq b\leq 2$, the existence of four critical points depends on $\delta$ as explained above.  There is a range of halfwidth:
\begin{equation}
 \delta_c=\arccos[(8+3b^2+24b^4-8b^6)/(27b^2)]/4
\end{equation}  
around each of $\delta=-90^\circ$, $0^\circ$, $90^\circ$ and $180^\circ$ where four critical points exist. For all other values of $\delta$, there are only two critical points. An example is shown in Figure \ref{fd} top panels (a) and (c) with $b=1.8$ corresponding to $\delta_c\simeq 23^\circ$. The expressions of the critical points are given as:
\begin{eqnarray}
&\mbox{Unstable point}&\\
&  -180^\circ\leq \delta\leq -90^\circ+\delta_c\  \mbox{and} \nonumber\\
&   -\delta_c\leq \delta\leq 0^\circ : \   \psi_{u1}=2 \arctan v_1,&\nonumber\\
 &   0^\circ\leq \delta\leq \delta_c \   \mbox{and}\nonumber \\
 &   90^\circ-\delta_c\leq \delta\leq 180^\circ: \  \psi_{u1}=2 \arctan v_4.&\nonumber\\
&\mbox{Stable point}&\\ 
&  -180^\circ\leq \delta\leq -180^\circ+\delta_c,\  -90^\circ -\delta_c\leq\delta\leq -90^\circ+\delta_c &\nonumber\\
&  \mbox{and}\  -\delta_c\leq \delta\leq 0^\circ :  \   \psi_{s1}=2 \arctan v_2,&\nonumber\\
 &   0^\circ\leq \delta\leq 180^\circ,\   \psi_{s1}=2 \arctan v_1.&\nonumber\\
 &\mbox{Unstable point}&\\
&  -180^\circ\leq \delta\leq -180^\circ+\delta_c\  \mbox{and} \nonumber\\
&   -90^\circ-\delta_c\leq \delta\leq 0^\circ : \   \psi_{u2}=2 \arctan v_3,&\nonumber\\
 &   0^\circ\leq \delta\leq 90^\circ+\delta_c \   \mbox{and}\nonumber \\
 &   180^\circ-\delta_c\leq \delta\leq 180^\circ: \  \psi_{u2}=2 \arctan v_2.&\nonumber\\
 &\mbox{Stable point}&\\ 
&  -180^\circ\leq \delta\leq 0^\circ :  \   \psi_{s2}=2 \arctan v_4,&\nonumber\\
 &   0^\circ\leq \delta\leq \delta_c,\   90^\circ-\delta_c\leq \delta\leq 90^\circ+\delta_c &\nonumber \\
 &\mbox{and}\ 180^\circ-\delta_c\leq \delta\leq 180^\circ : \psi_{s2}=2 \arctan v_3.&\nonumber
\end{eqnarray}

For $b \geq 2 $, there are four critical points regardless of $\delta$ as shown in Figure \ref{fd}  top panels (a) and (d) with $b=2.5$. The expressions of the critical points are in order of increasing $\psi$-value.
\begin{eqnarray}
&\mbox{Unstable point}&\\
&  -180^\circ\leq \delta\leq 0^\circ,\   \psi_{u1}=2 \arctan v_1,&\nonumber\\
 &   0^\circ\leq \delta\leq 180^\circ,\   \psi_{u1}=2 \arctan v_4,&\nonumber\\
&\mbox{Stable point}&\\ 
&  -180^\circ\leq \delta\leq 0^\circ,\   \psi_{s1}=2 \arctan v_2,&\nonumber\\
&   0^\circ\leq \delta\leq 180^\circ, \   \psi_{s1}=2 \arctan v_1,&\nonumber\\
&\mbox{Unstable point}&\\ 
&  -180^\circ\leq \delta\leq 0^\circ,\   \psi_{u2}=2 \arctan v_3,&\nonumber\\
&   0^\circ\leq \delta\leq 180^\circ, \   \psi_{u2}=2 \arctan v_2,&\nonumber\\
&\mbox{Stable point}&\\ 
&  -180^\circ\leq \delta\leq 0^\circ,\   \psi_{s2}=2 \arctan v_4,&\nonumber\\
&   0^\circ\leq \delta\leq 180^\circ, \   \psi_{s2}=2 \arctan v_3.&\nonumber
\end{eqnarray}
\subsubsection{Libration widths}
The level curves of the two-arguments two-harmonics pendulum model are shown in the bottom three rows of Figure \ref{fd}. There are a number of differences with respect to the one-argument two-harmonics pendulum model of Section 3.2 depending on the value of the parameter $b$. 

\begin{figure*}
\begin{center}
{ 
\includegraphics[width=40mm]{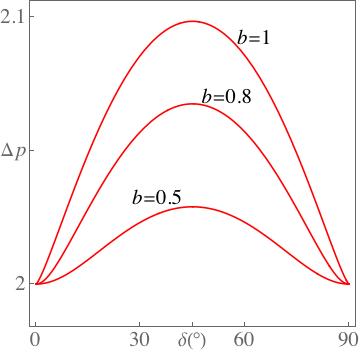}\hspace{3mm}\includegraphics[width=40mm]{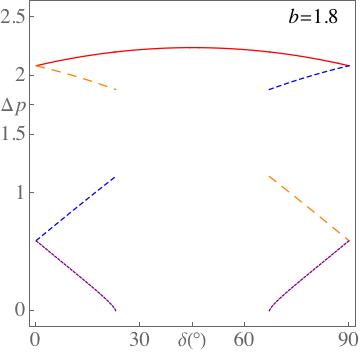}\hspace{3mm}\includegraphics[width=40mm]{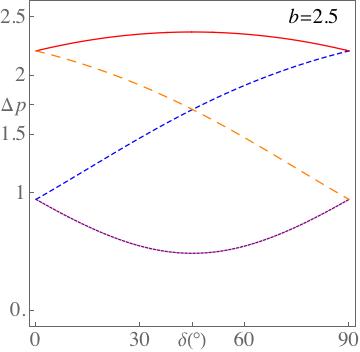}
}
\caption{Resonance widths of the two-arguments two-harmonics pendulum as a function of the argument shift $\delta$ for different values of the parameter $b$. The left panel shows the single resonance width (\ref{reswidl1}) for three values of $b\leq 1$. The middle and right panels show the widths  for $b=1.8$ and 2 given by  equations (\ref{reswidasg21},\ref{reswidasg22},\ref{reswidGg21},\ref{reswidGg22}) respectively with a dashed blue line, a dotted purple line, a solid red line, and a long dashed orange line. }\label{figamp}
\end{center}
\end{figure*}

For $0\leq b\leq 1$, libration in the one-argument two-harmonics model of Section 3.2 occurs at $180^\circ$  for  $f_{k_1,k_2}^{p:q}\geq0$ and   displaced libration does not exist. The presence of the argument shift $\delta$ in the two-arguments two-harmonics model literally shifts the argument libration continuously as a function of $\delta$ as shown in Figure \ref{fd} (b) giving rise to displaced librations. In particular, libration may occur near $0^\circ$ when  $\delta$ is near $180^\circ$.  To derive the resonance width for $0\leq b\leq 1$, we use the fact that the separatrix passes through the unstable point ($\psi=\psi_u$, $p=0$) thus defining the Hamiltonian value for maximum libration, then infer the value of $p$ at maximum libration with $\psi=\psi_s$. Writing the Hamiltonian (\ref{hamil}) as: $H=p^2/2 +V(\psi)$, the resonance width is therefore  given as:
\begin{equation}
\Delta p =  \left(2[V(\psi_u)-V(\psi_s)]\right)^\frac{1}{2},\label{reswidl1}
\end{equation}
where $\psi_u$ and $\psi_s$ are given by the expressions (\ref{ubl1},\ref{sbl1}). 
For $0\leq b\leq 1$, the one-argument two-harmonics pendulum model of Section 3.2 gives $\Delta p =2$ the same width as that of the classical one-argument one-harmonic pendulum model of Section 3.1. The addition of a second argument makes the width dependent on $\delta$ as shown in Figure \ref{figamp}. The relative difference is maximal for $b=1$.  It amounts to 5 per cent of the base width and is reached for $\delta=\pm45^\circ\pm90^\circ$. 

For $b\geq 2$, the one-argument two-harmonics pendulum of Section 3.2 has four critical points with no global asymmetry as that introduced by a finite argument shift $\delta$ displayed in the bottom two rows of Figure \ref{fd}. For that reason, only two resonance widths were defined in Paper I and explained in Section 3.2: one for asymmetric librations and another for librations around both asymmetric points or equivalently around $\psi=180^\circ$ for  $f_{k_1,k_2}^{p:q}\geq0$ (see for instance the $\delta=0^\circ$ panels of $b=2$ in Figure \ref{fd}). When the parameter  $\delta$ has a finite value, the width of asymmetric libration around each of the stable points is different. Similarly the maximal libration width about both points no longer has a single maximum but two. We therefore define two asymmetric libration  widths associated with each stable equilibrium point and given as 
\begin{eqnarray}
\Delta p_{\rm asym,1} &=&  \left(2[V(\psi_{u1})-V(\psi_{s1})]\right)^\frac{1}{2},\label{reswidasg21}\\ 
\Delta p_{\rm asym,2} &=&  \left(2[V(\psi_{u1})-V(\psi_{s2})]\right)^\frac{1}{2}, \label{reswidasg22}
\end{eqnarray}
The separatrix associated with {libration} around both stable points has two maxima  given as :
\begin{eqnarray}
\Delta p_{\rm both,l} &=&  \left(2[V(\psi_{u2})-V(\psi_{s1})]\right)^\frac{1}{2},\label{reswidGg21}\\ 
\Delta p_{\rm both,s} &=&  \left(2[V(\psi_{u2})-V(\psi_{s2})]\right)^\frac{1}{2},\label{reswidGg22}
\end{eqnarray}
with $\Delta p_{\rm both,l}\geq\Delta p_{\rm both,s}$. The expressions (\ref{reswidasg21},\ref{reswidasg22},\ref{reswidGg21},\ref{reswidGg22}) are valid for $0^\circ\leq\delta\leq90^\circ$ and are $90^\circ$-periodic.  For any other value of $\delta$, the trasformations (\ref{tras1},\ref{tras2},\ref{tras3}) can be used to bring back $\delta$ in the [$0^\circ:90^\circ$] range. The various resonance widths  are shown in Figure \ref{figamp}.  To define a single libration width around  both asymmetric points, one may use only $\Delta p_{\rm both,l}$ or a combination of the two analytical widths. The appropriate choice will depend on the numerical method used to determine the libration widths in dynamics simulations.

For $1\leq b\leq2$, the widths given by   (\ref{reswidasg21},\ref{reswidasg22},\ref{reswidGg22}) exist only for  $0^\circ\leq \delta\leq \delta_c$ and $90^\circ-\delta_c\leq \delta\leq 90^\circ$ whereas the maximal libration width  (\ref{reswidGg21}) exists regardless of $\delta$. In the range  
$\delta_c^\circ\leq\delta\leq 90^\circ-\delta_c$, the latter  becomes the maximum libration width around the only stable point present (Figure \ref{figamp}, middle panel).

Regarding the various definitions of resonance width, it should be noted that the numerical simulations presented in the next section for outer resonances rely upon initial conditions with a fixed value of the main resonant argument thus requiring additional definitions of resonance width. To understand this aspect, we consider, for instance, the Neptune outer 1:2 resonance simulations. The initial value of the argument $\phi_{1}^{1:2}$ is set to $90^\circ$ or $180^\circ$ with the former value aimed at detecting asymmetric librations and the latter aimed at librations around both asymmetric centers. However, the resonance widths defined previously, do not measure the extent of the resonance island at $\psi=180^\circ$, and since the asymmetric critical points are not located {exactly} at the fixed values $\psi=\pm 90^\circ$, it is possible that sampling with such initial conditions does not occupy the full width of resonance given by the previous estimates (\ref{reswidasg21},\ref{reswidasg22},\ref{reswidGg22}).  Three additional definitions of resonance widths are required to reflect these two choices of initial conditions. The first gives the resonance width at $\psi=180^\circ$ corresponding to libration around both critical points and is given as:
\begin{eqnarray}
\Delta p_{\rm 180^\circ} &=&  \left(2[V(\psi_{u2})+\cos\delta-b/4]\right)^\frac{1}{2}\label{reswidGRg21},
\end{eqnarray}
where the second and third terms on the right hand side are the expression of  $-V(180^\circ)$. For librations set exactly at $90^\circ$, the relevant widths for asymmetric librations and those around both critical points are written respectively as:
\begin{eqnarray}
\Delta p_{\rm asym,90^\circ} &=&  \left(2[V(\psi_{u1})-\sin\delta+b/4]\right)^\frac{1}{2}\label{reswidasR21},\\
\Delta p_{\rm both,90^\circ} &=&  \left(2[V(\psi_{u2})-\sin\delta+b/4]\right)^\frac{1}{2}\label{reswidasR22},
\end{eqnarray}
where the second and third terms on the right hand side are $-V(90^\circ)$.

All momentum resonance width formulas $\Delta p$ (\ref{reswidl1}--\ref{reswidasR22})  can be {transformed to yield resonance widths  in terms of the resonance's semi-major axis using} the relationship (see Paper I for a derivation): 
\begin{equation}
\Delta a_k^{p:q} = 2\left[\frac{\alpha\, m^\prime |g_{k_1,k_2}^{p:q}|}{3M_\star}\right]^\frac{1}{2}  \Delta p \ \ a_k^{p:q}. \label{areswidth}
\end{equation}

\subsubsection{The case of multiple arguments}
With more than two arguments, the two-harmonics pendulum equation may be reduced using the same transformation that led from equation (\ref{twoargtwoharmonic}) to equation (\ref{redtwotwohar}). The resulting pendulum equation would be identical to equation (\ref{redtwotwohar}) where $g_{k_1,k_2}^{p:q}$ and $\delta\phi_{k_1,k_2}^{p:q}$ are replaced by $g^{p:q}$ and $\delta\phi^{p:q}$ (\ref{gmultiple},\ref{deltamultiple}) respectively, and  $g_{2k_1,2k_2}^{2p:2q}$ and $\delta\phi_{2k_1,2k_2}^{2p:2q}$ are replaced respectively by:
\begin{eqnarray}
{{g}^{2p:2q}}^2 &=& 
\!\!\!\!\!\!\!\sum_{
{\scriptsize\begin{array}{c}
0\leq \mu,\nu\leq N_k\\ 
\mu\neq \nu
\end{array}}} \!\!\!f_{2k_\nu}^{2p:2q}  f_{2k_\mu}^{2p:2q} \cos(2[k_\mu-k_\nu]\omega)+\label{gmul2}\\
&&\ \ \ \ \ \ \ \ \ +\sum_{\mu=1}^{N_k}{f_{2k_\mu}^{2p:2q}}^2, \nonumber \\ 
\cos\delta \phi^{2p:2q}&=& \left[\sum_{\mu=1}^{N_k}f_{2k_\mu}^{2p:2q} \cos 2k_\mu\omega \right]\left(g^{2p:2q}\right)^{-1},\label{deltamul2}\\
\sin\delta \phi^{2p:2q}&=& \left[\sum_{\mu=1}^{N_k}f_{2k_\mu}^{2p:2q}\sin 2k_\mu\omega \right]\left(g^{2p:2q}\right)^{-1}.\nonumber
\end{eqnarray}
The multiple-arguments Hamiltonian is therefore identical to that of two arguments but the expressions of the parameters $b$ and $\delta$ are now given by:
\begin{equation}
b=\frac{4g^{2p:2q}}{|g^{p:q}|},\ \ \mbox{and}\ \ \delta=\delta \phi^{p:q}-\delta \phi^{2p:2q}/2.
\end{equation}

\section{Comparing analytical estimates with direct numerical integrations}
In the previous sections, we derived analytical {expressions} for resonance width and characterized {the corresponding libration types.} In this section, we compare these estimates with the direct integration of the full equations of motion of the three-body problem of the  inner 2:1 and 3:1 Jupiter resonances and the outer 1:2 and 1:3 Neptune resonances. In order to have accurate parameter space portraits of resonance at arbitrary inclination in the numerical simulations, we choose to represent them using the Mean Exponential Growth factor of Nearby Orbits also known as the  {\sc megno} chaos indicator so as to delineate precisely not only resonance width but also the chaotic domains that surround the resonances in parameter space  \citep{cincotta00,goz03}. The full equations of motion are therefore integrated along with the variational equations and {\sc megno} equations for $5\times 10^5$ orbital periods of the perturber, {a time interval larger than the typical secular timescales.}  The integrations use  the Bulirsch and Stoer method with a tolerance $10^{-14}$. The mean {\sc megno} indicator converges to 2 for regular orbits and increases at a rate inversely proportional to the Lyapunov time of chaotic orbits. The maximum mean {\sc megno} indicator value is set  at 8 for chaotic orbits in order to present stability maps with a high contrast between the regular and chaotic regions. The integration of an asteroid's orbit is stopped if collision or escape occur. In addition to the computation of the {\sc megno} indicator, we monitor the librations that occur on the integration timescale (which is larger than the secular timescale). {\sc megno} portraits are therefore associated with libration portraits that identify the modes $k$ that librate beyond the resonant timescale as well as  {the possible} asymmetric librations for the outer resonances.

\begin{figure*}
{ 
\includegraphics[width=37mm]{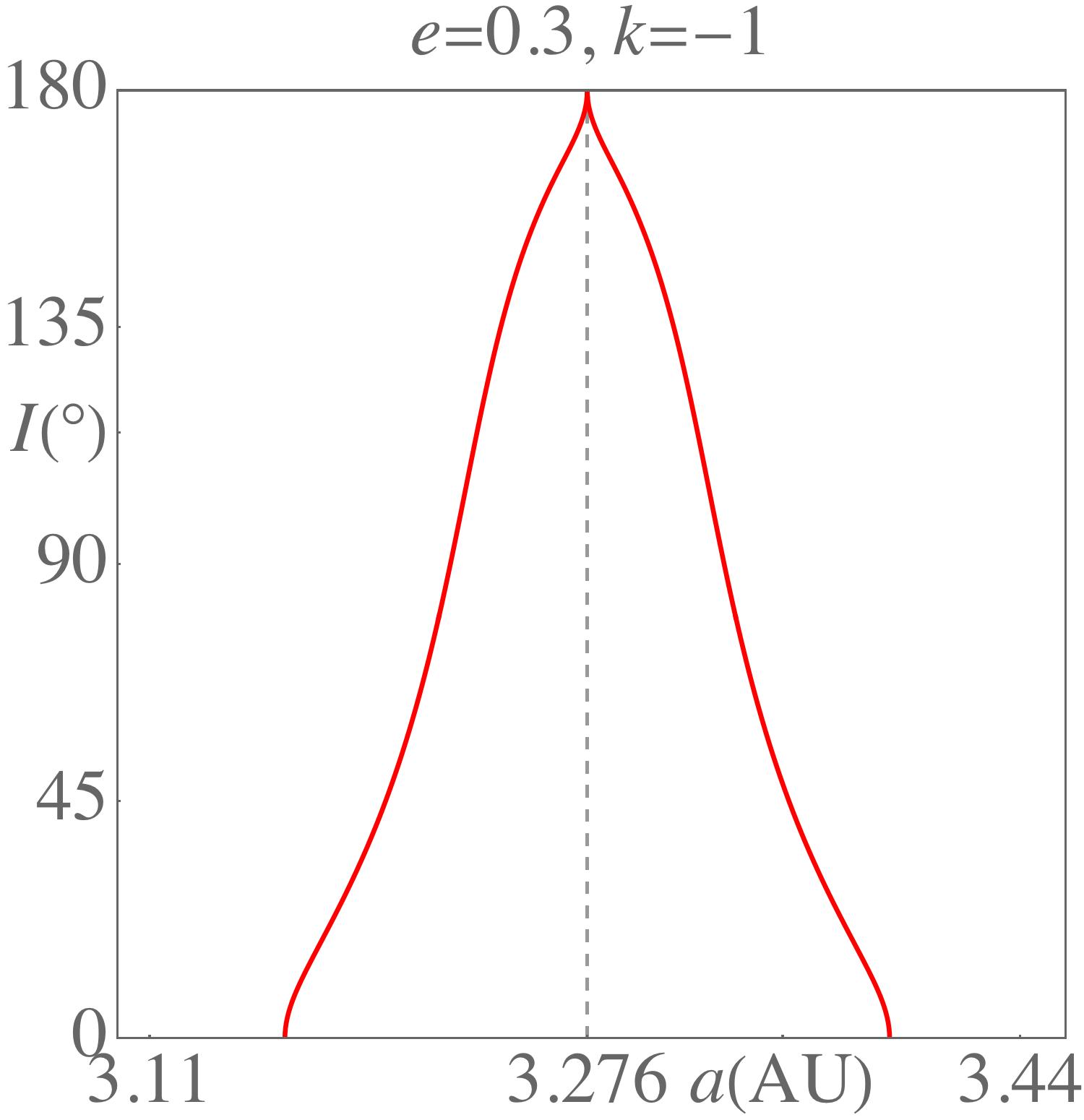}\hspace{8mm}\includegraphics[width=37mm]{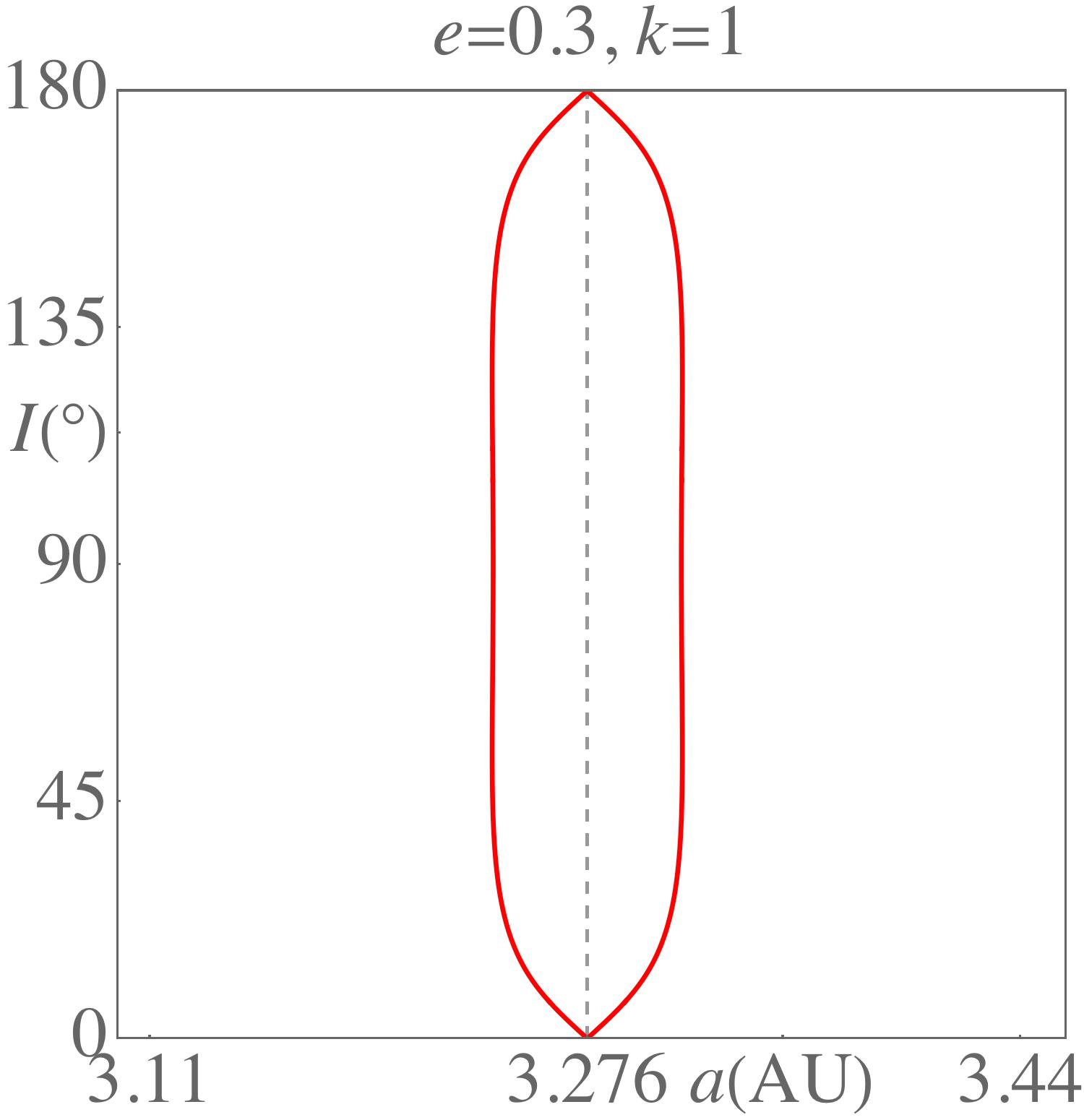}\hspace{8mm}\includegraphics[width=37mm]{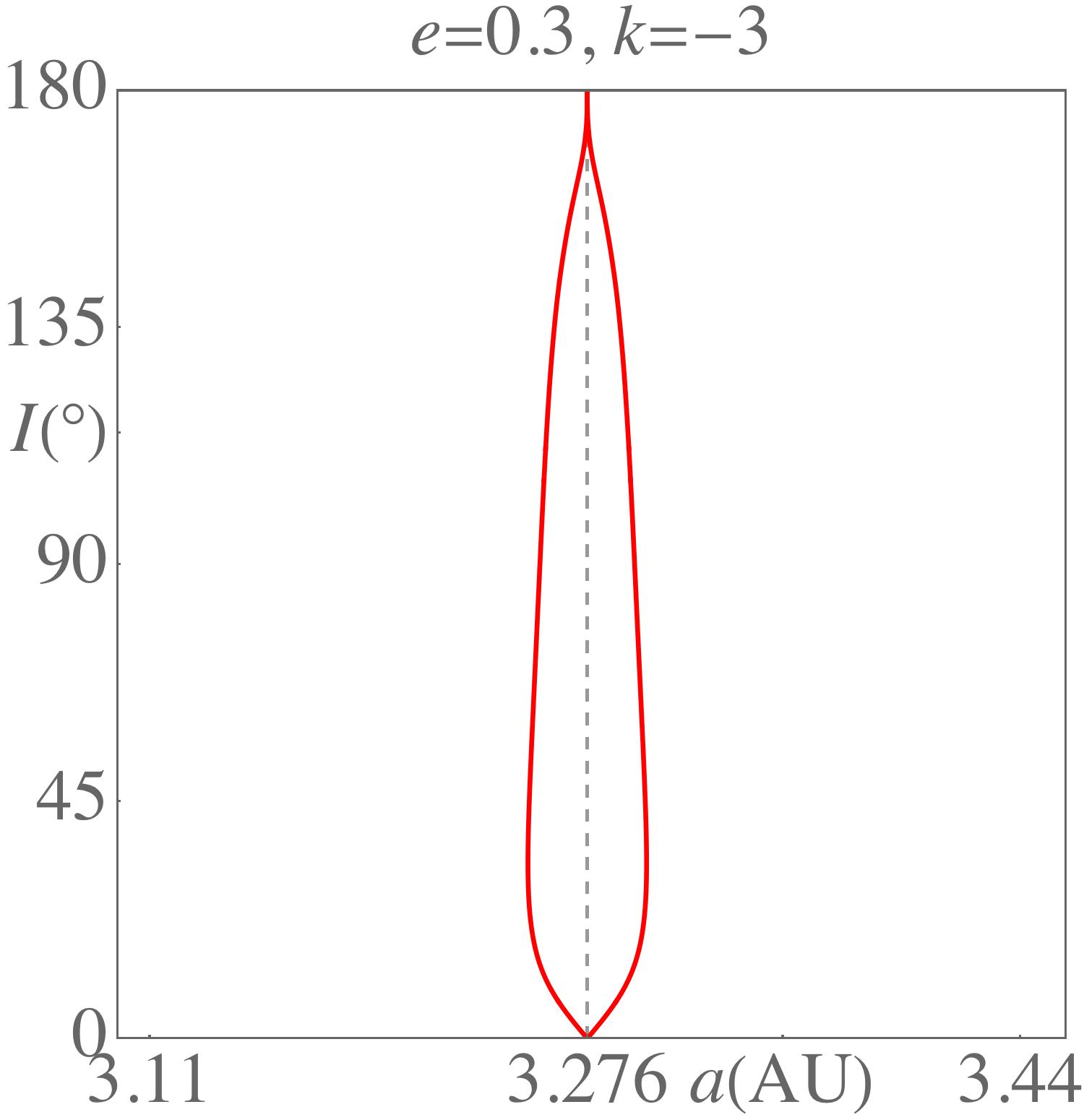}\hspace{8mm}\includegraphics[width=37mm]{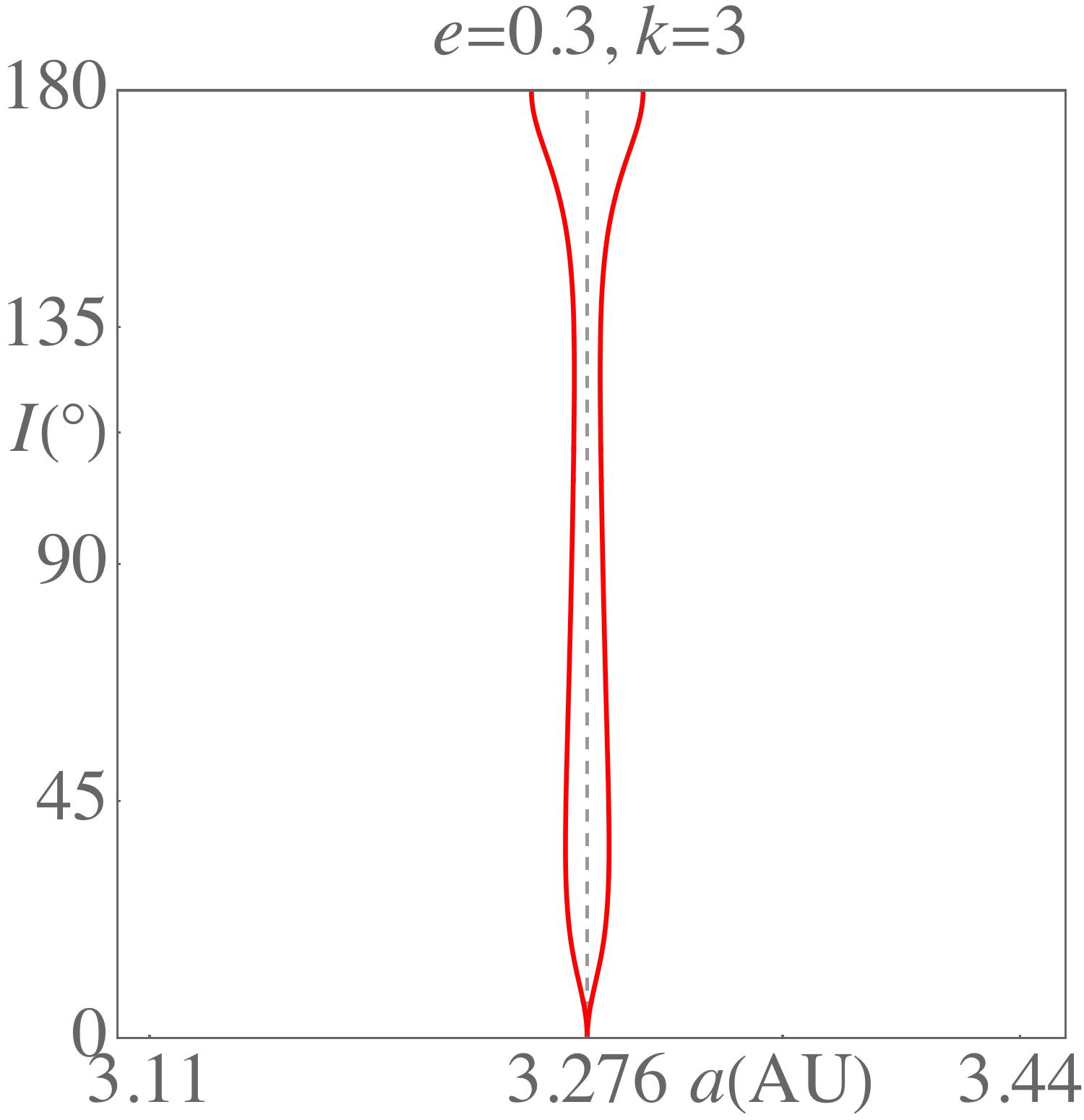}\\
\includegraphics[width=37mm]{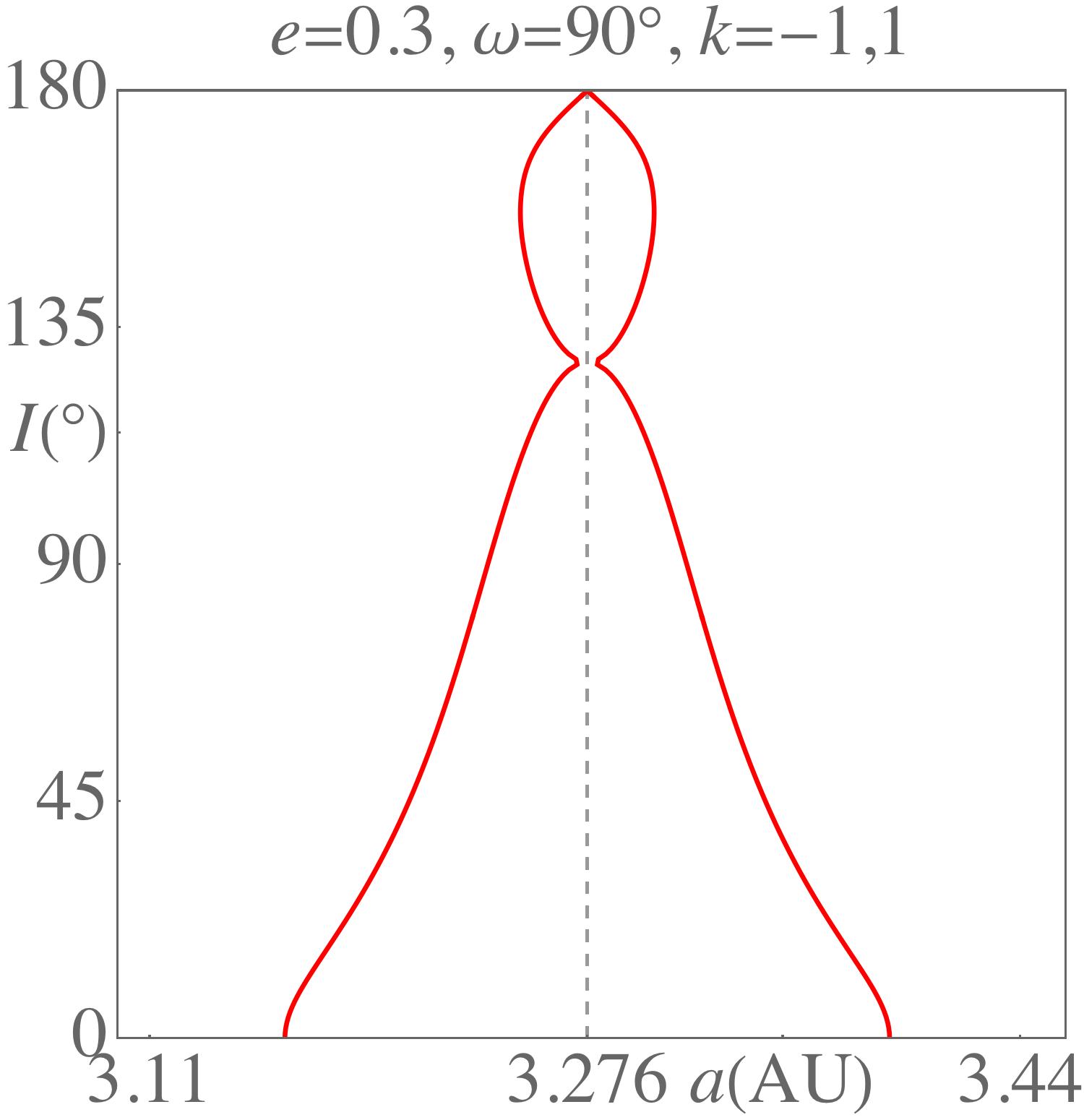}\hspace{8mm}\includegraphics[width=37mm]{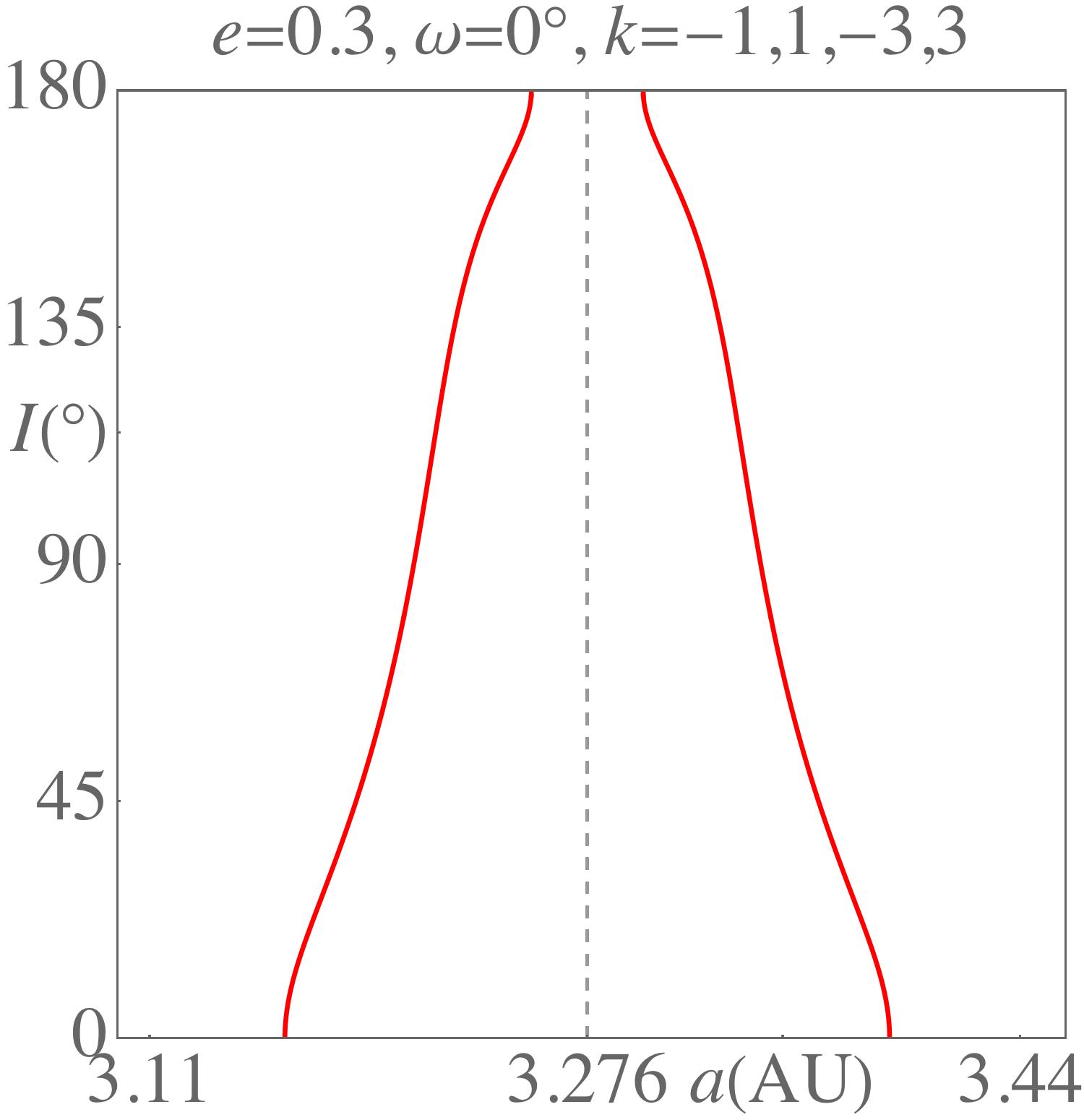}\hspace{8mm}\includegraphics[width=37mm]{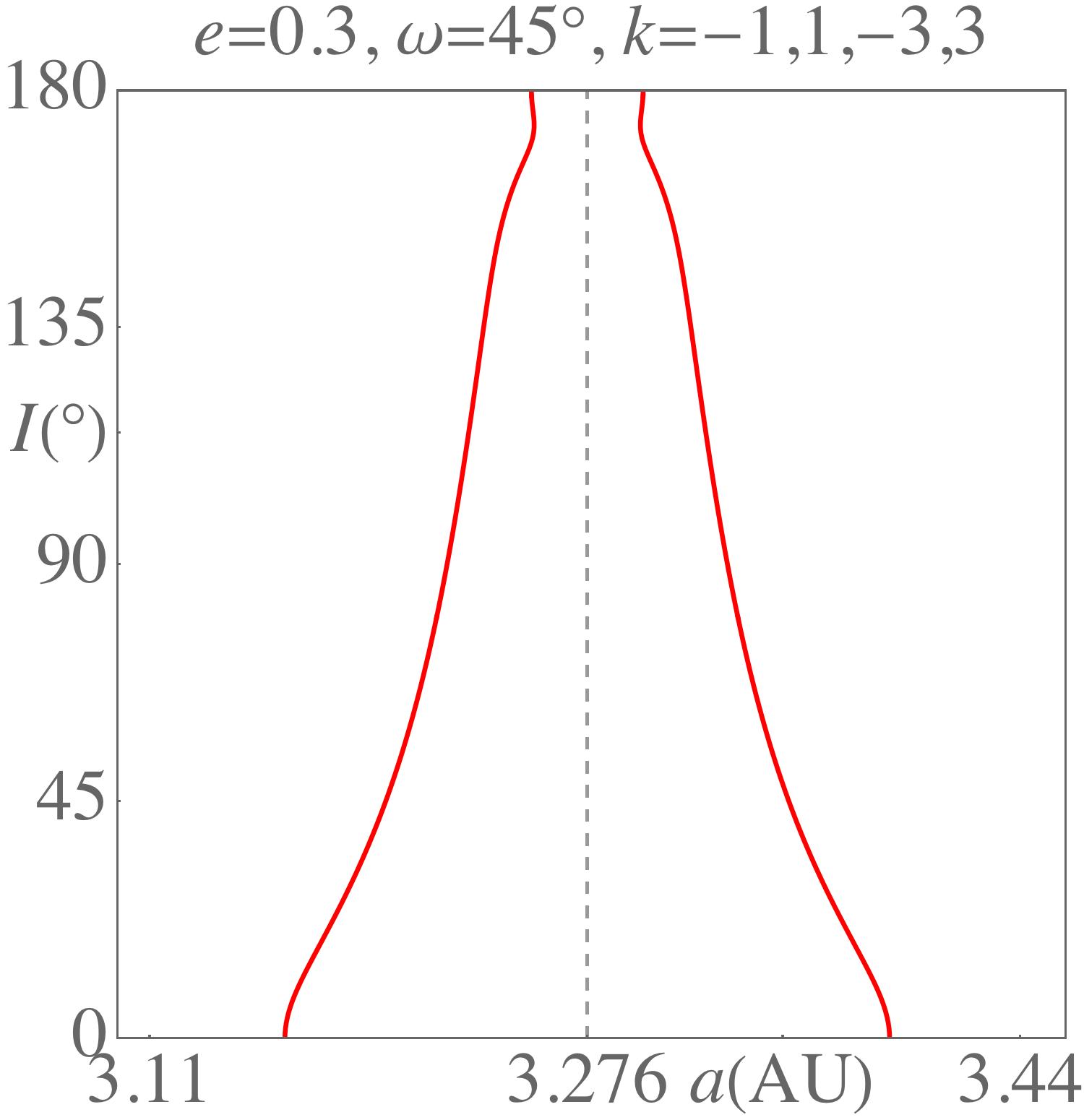}\hspace{8mm}\includegraphics[width=37mm]{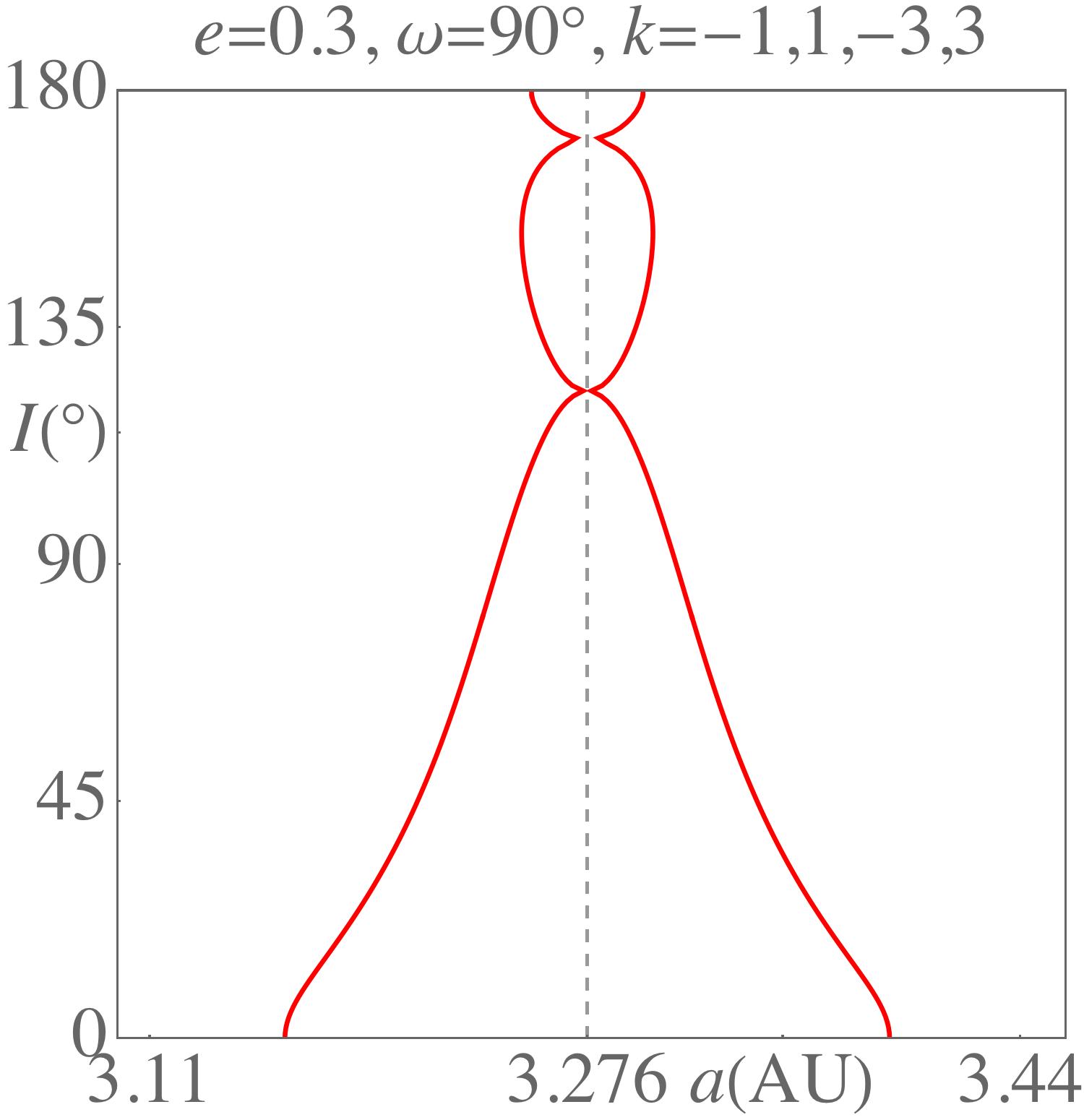}\\
\includegraphics[width=42mm]{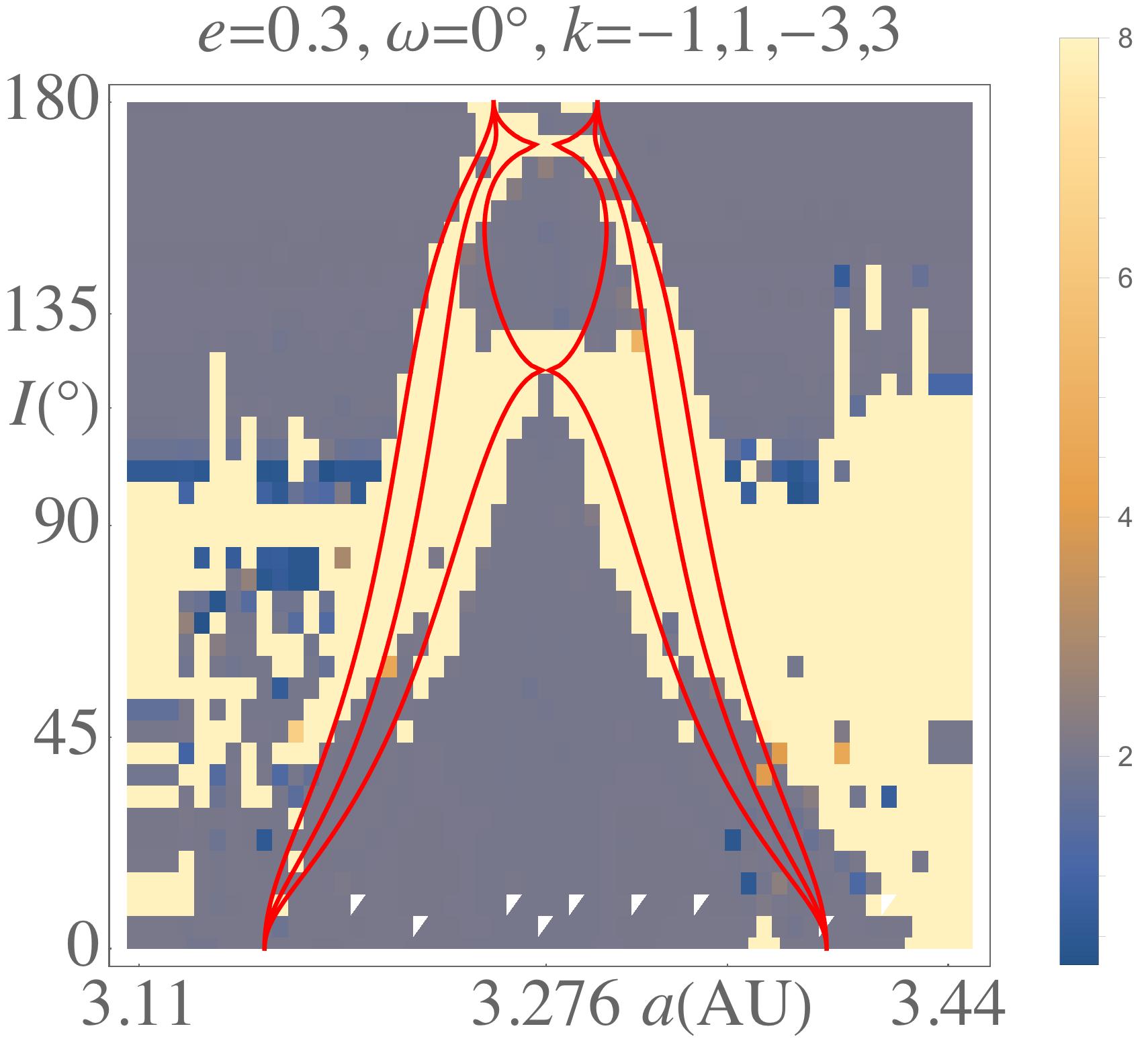}\hspace{3mm}\includegraphics[width=42mm]{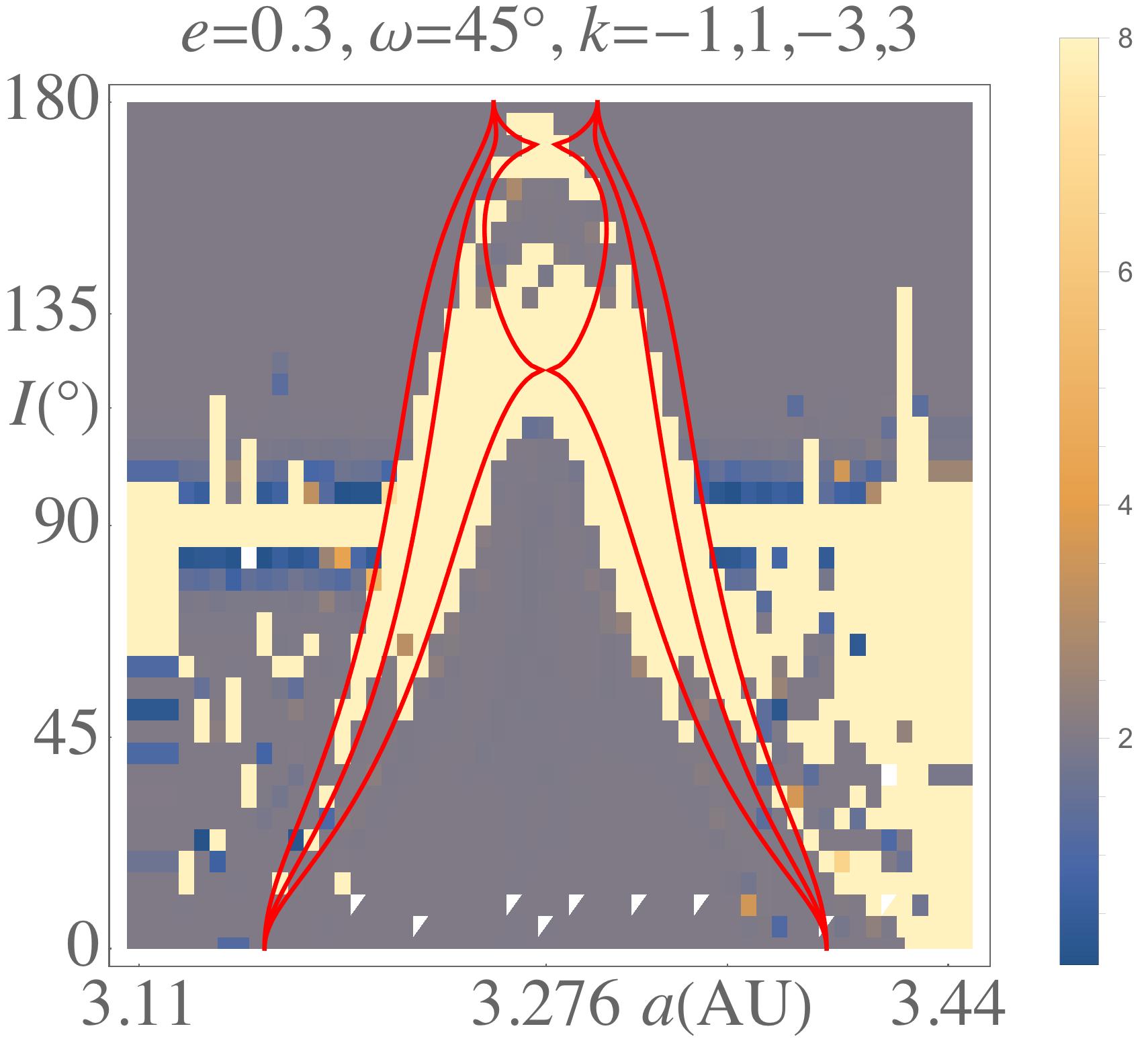}\hspace{3mm}\includegraphics[width=42mm]{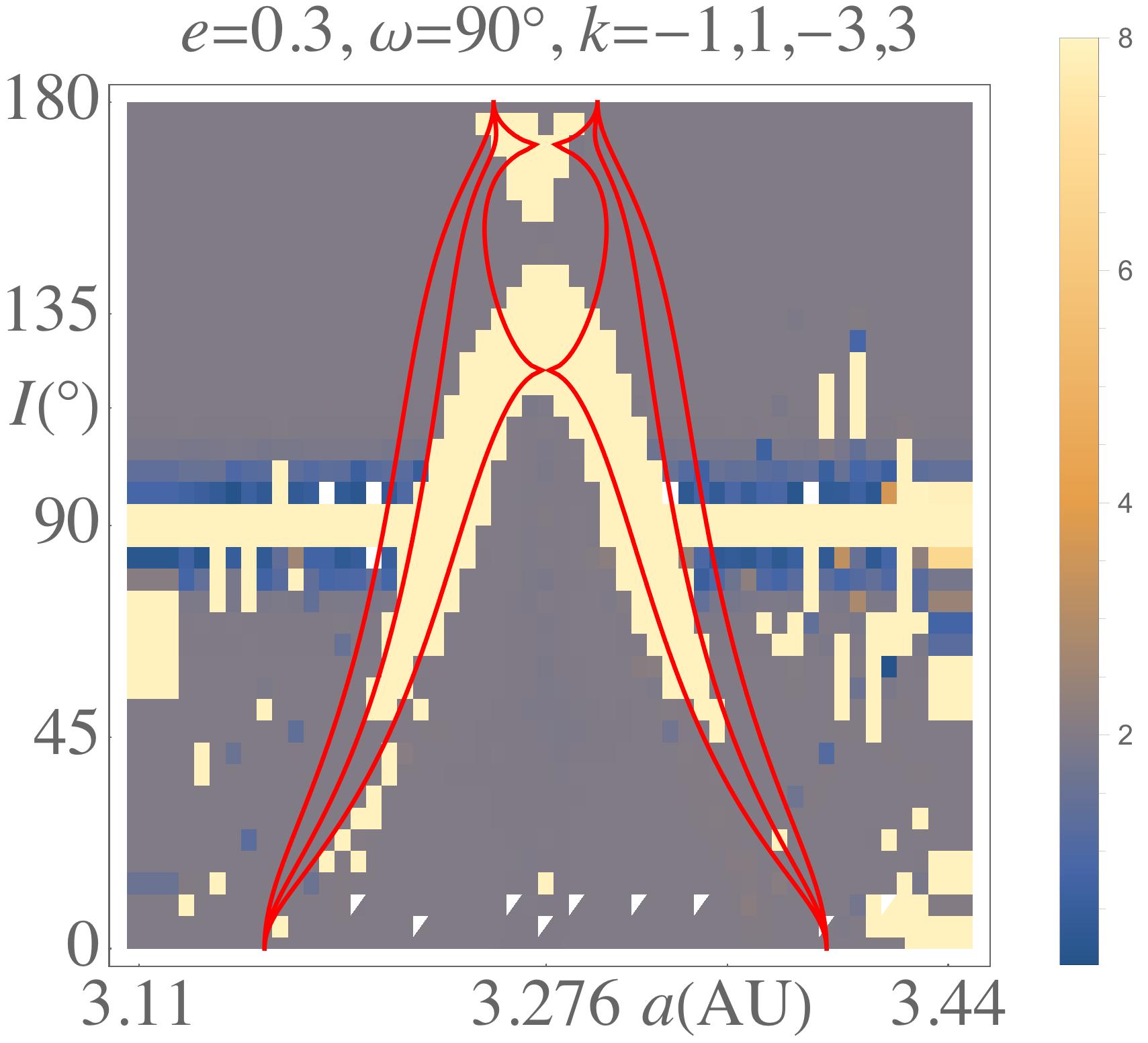}\hspace{3mm}\includegraphics[width=37mm]{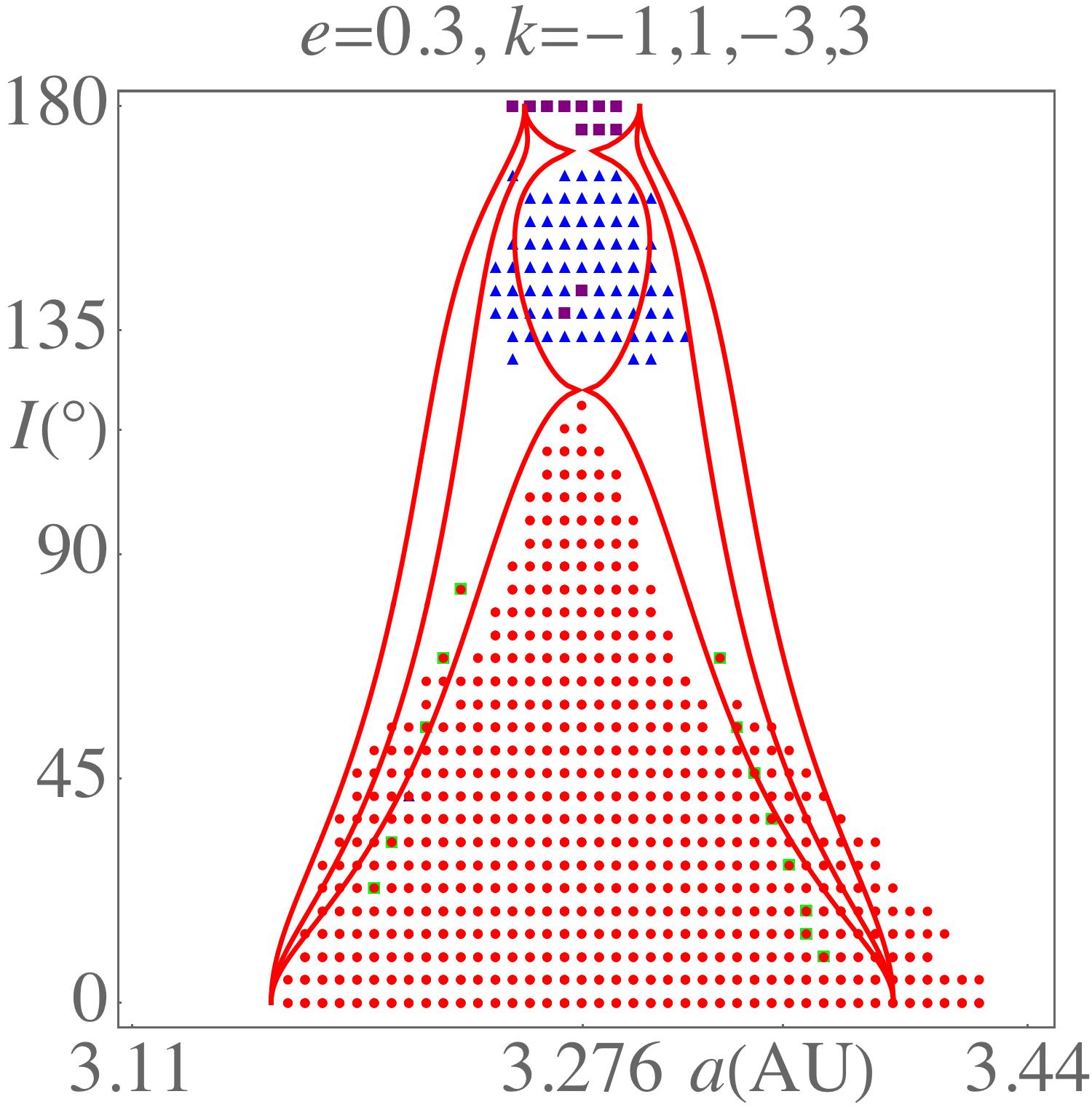}\\
\includegraphics[width=42mm]{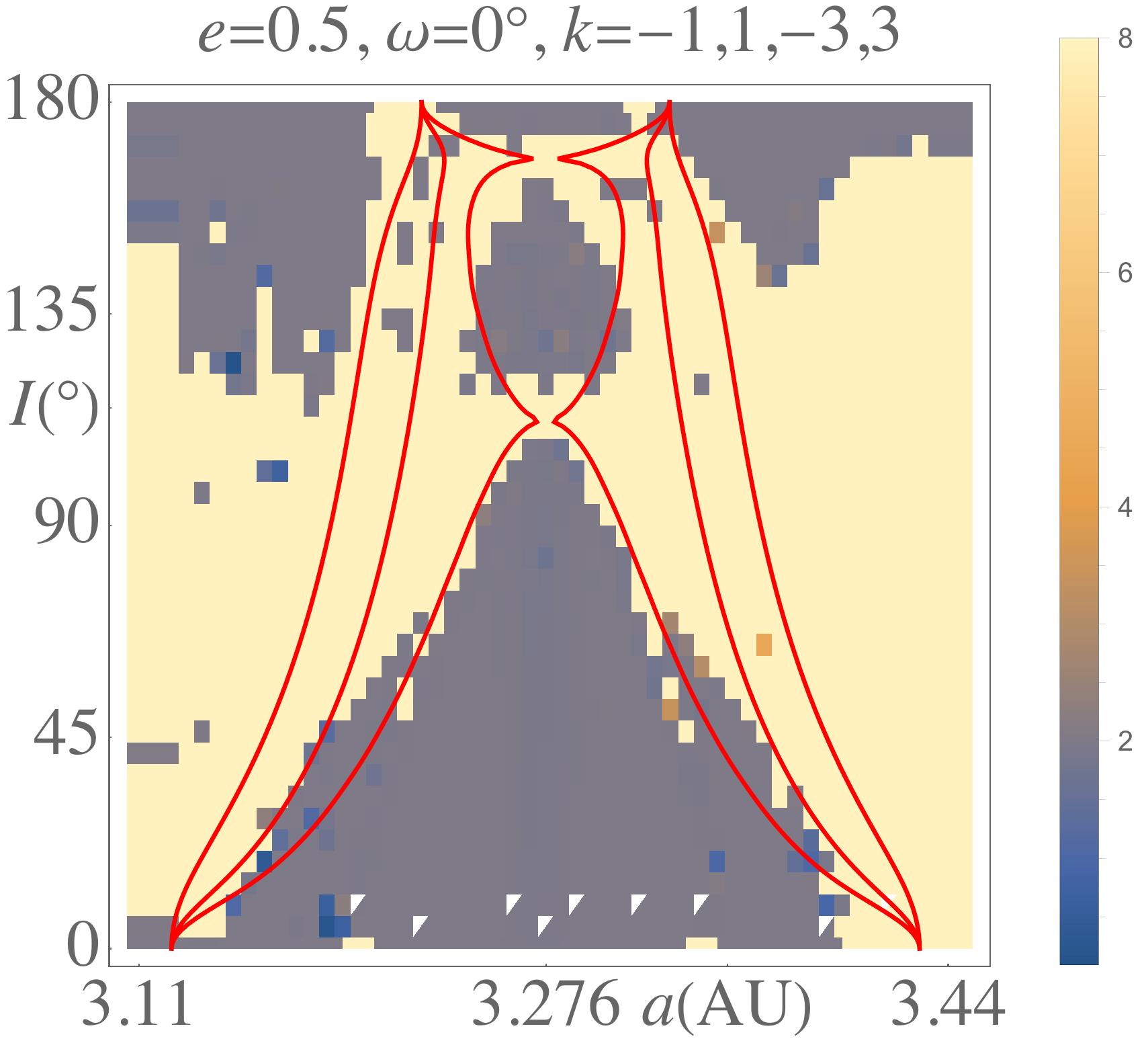}\hspace{3mm}\includegraphics[width=42mm]{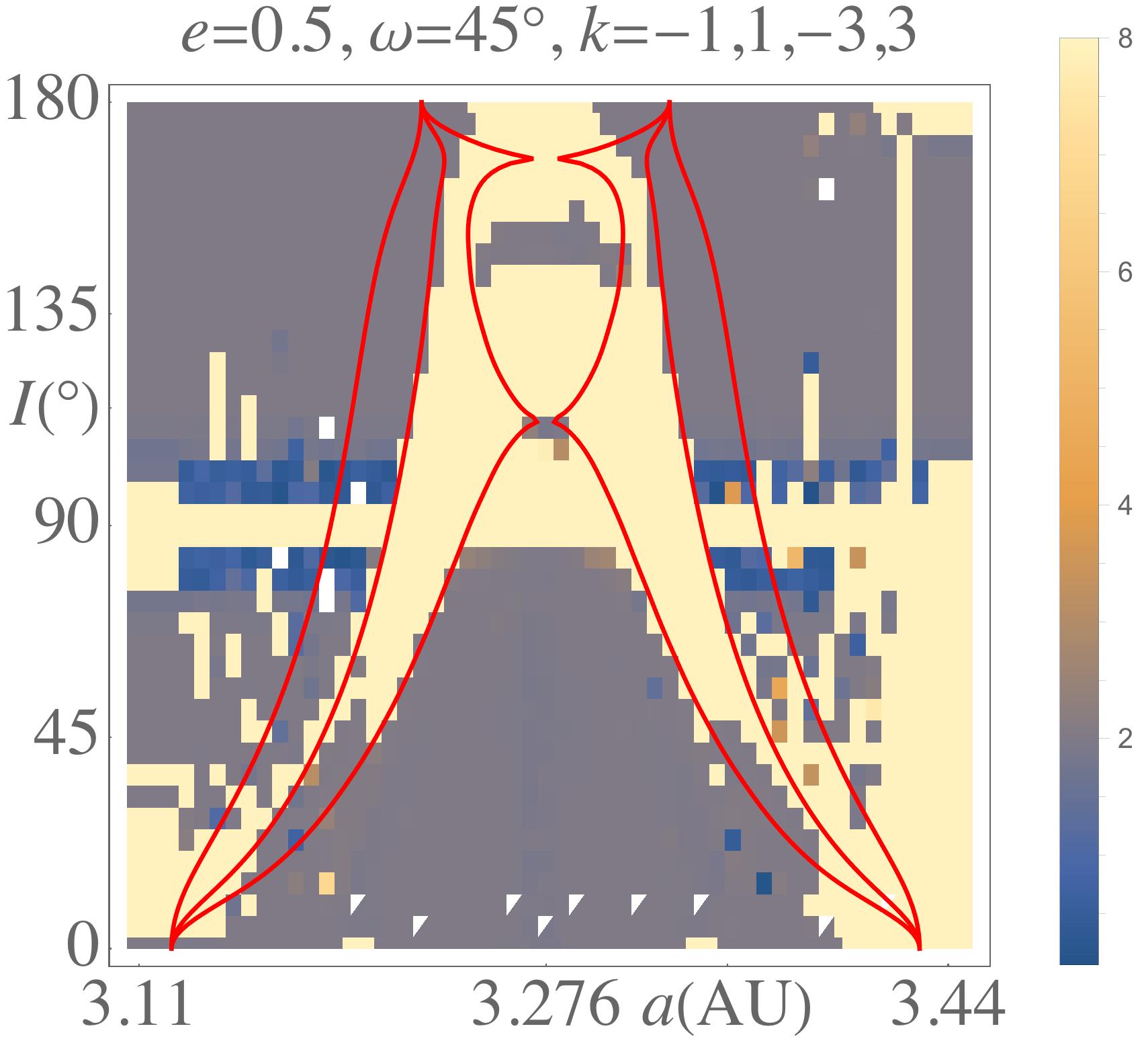}\hspace{3mm}\includegraphics[width=42mm]{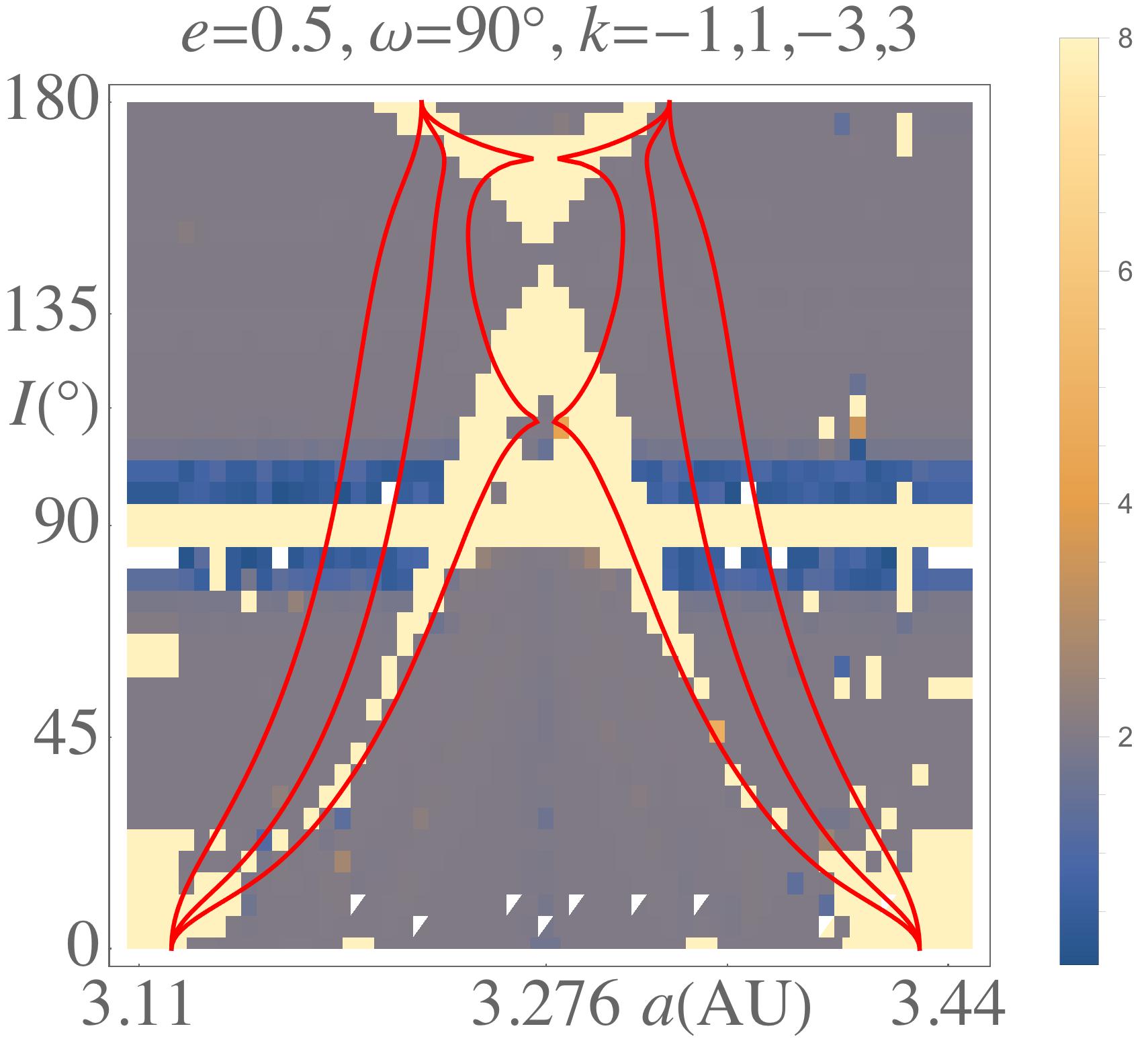}\hspace{3mm}\includegraphics[width=37mm]{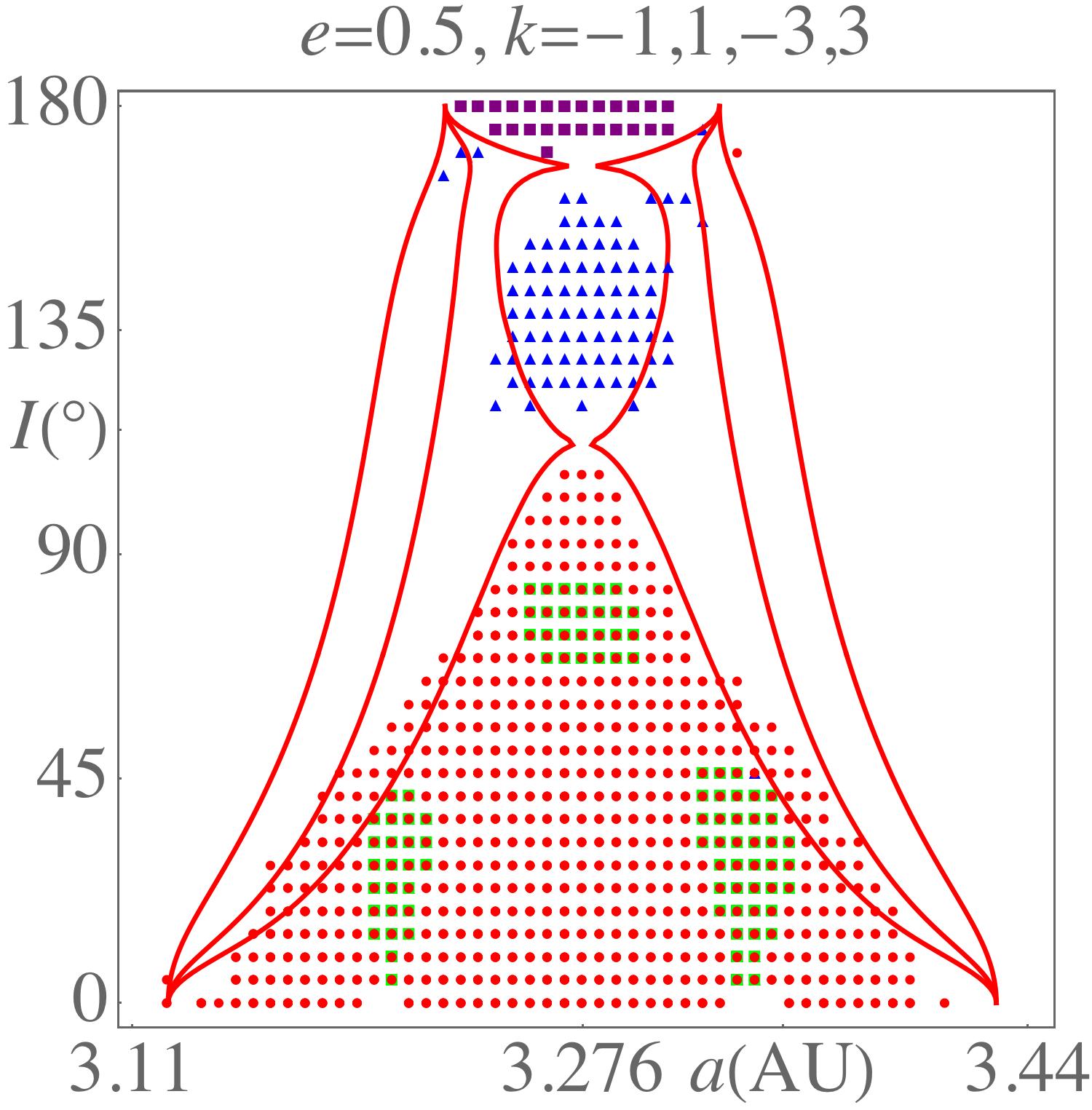}
}
\caption{Jupiter's inner 2:1 resonance. The red curves indicate the resonance width determined from the analytical models for the values of the {modes} $k$ and argument of perihelion $\omega$ indicated above each panel. The bottom two rows show the  {\sc megno} portraits for two eccentricities $e=0.3$ and 0.5 and  different values of $\omega$ as well as  the libration centers for all values of $\omega$ in the rightmost panel.  Superimposed on the {\sc megno} and libration portraits are the three resonance widths from the simultaneous libration of $\{ k: -1,1,-3,3\}$ obtained for each value of $\omega$. In the libration center panels, $k=-1$ is denoted with a red filled circle, $k=-3$ with and green empty square, $k=1$ with a blue filled triangle and $k=3$ with a purple filled triangle.}\label{fJ2t1}
\end{figure*}

\begin{figure*}
{ 
\includegraphics[width=37mm]{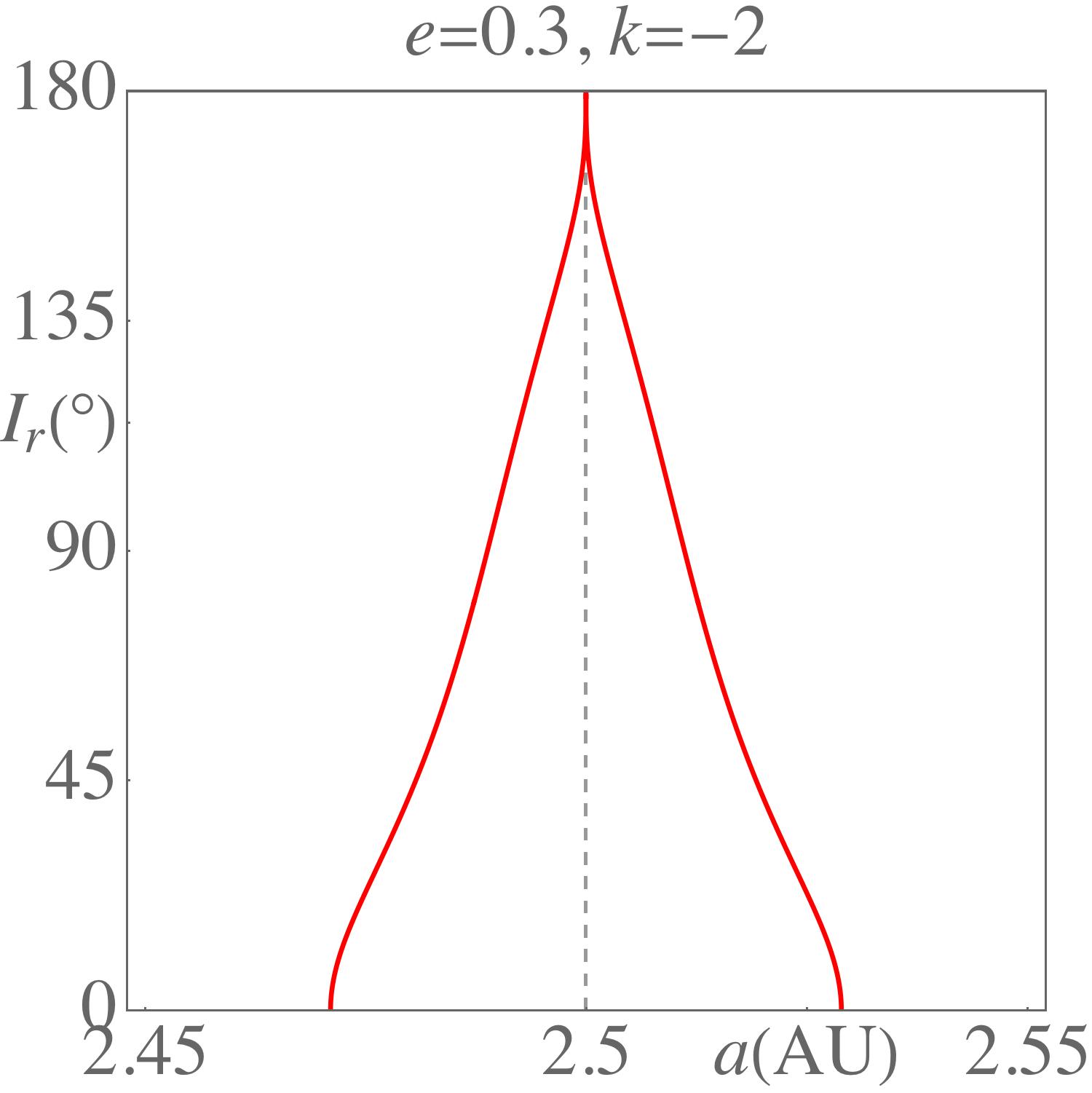}\hspace{9mm}\includegraphics[width=37mm]{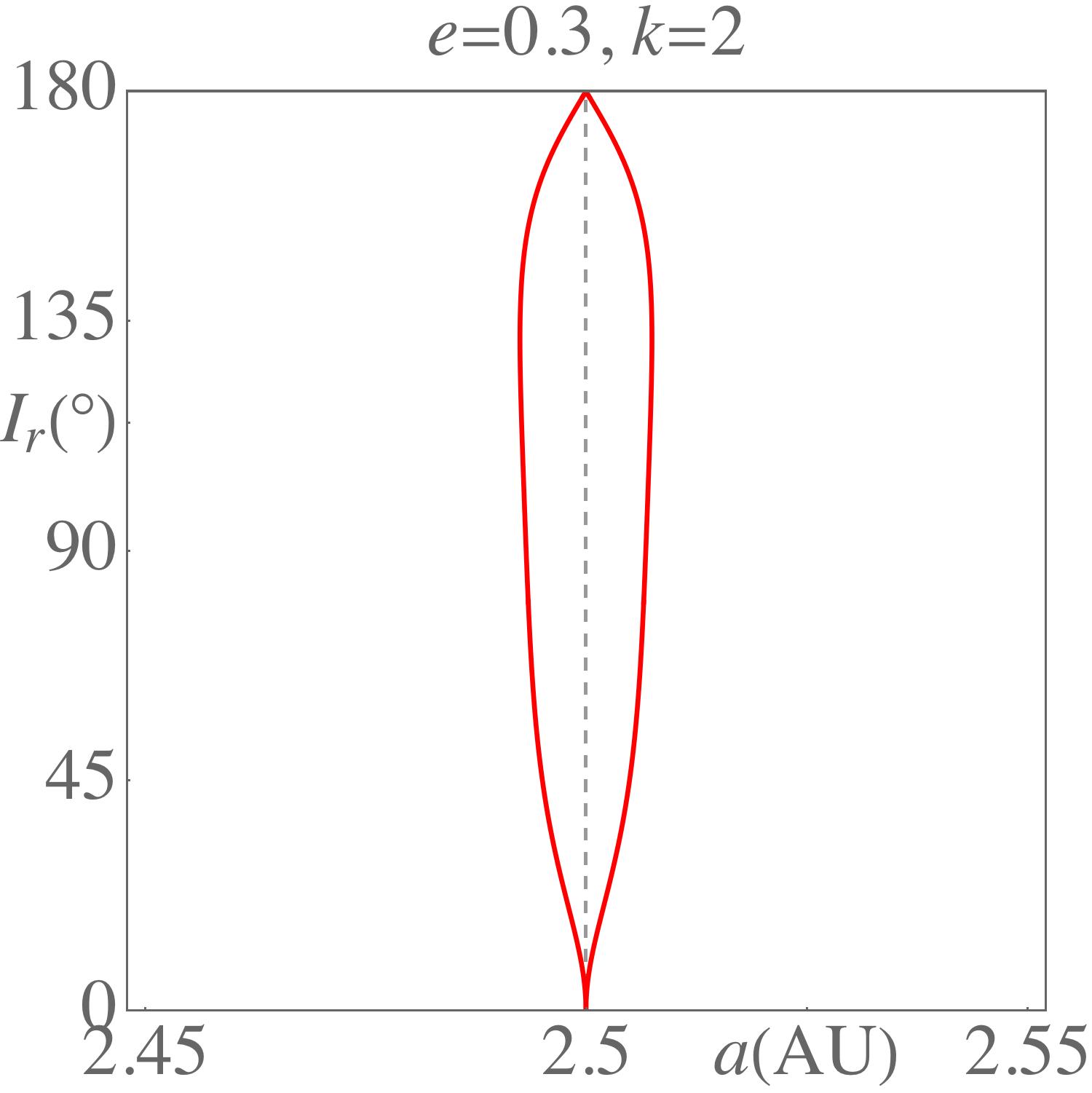}\hspace{9mm}\includegraphics[width=37mm]{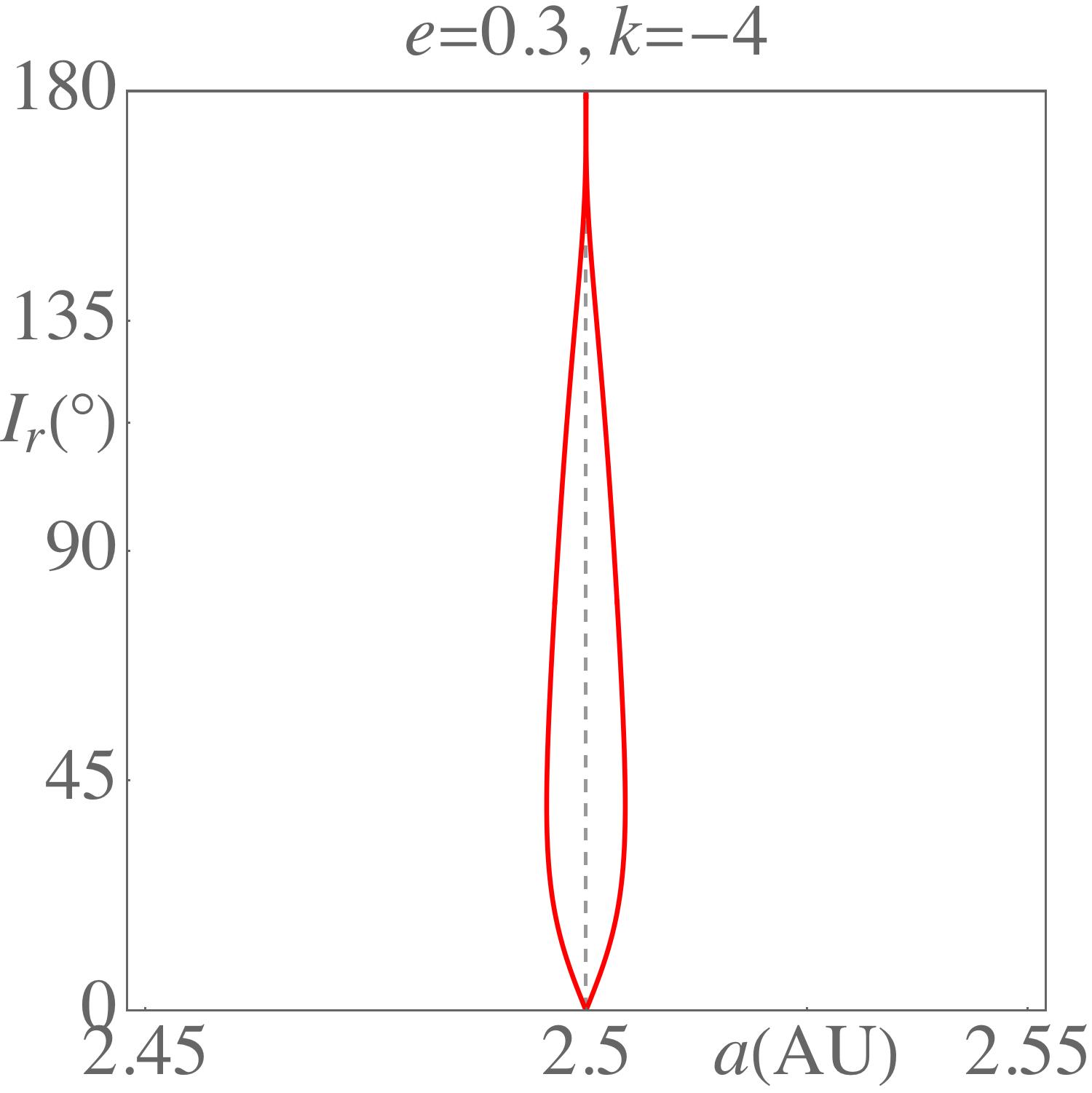}\hspace{9mm}\includegraphics[width=37mm]{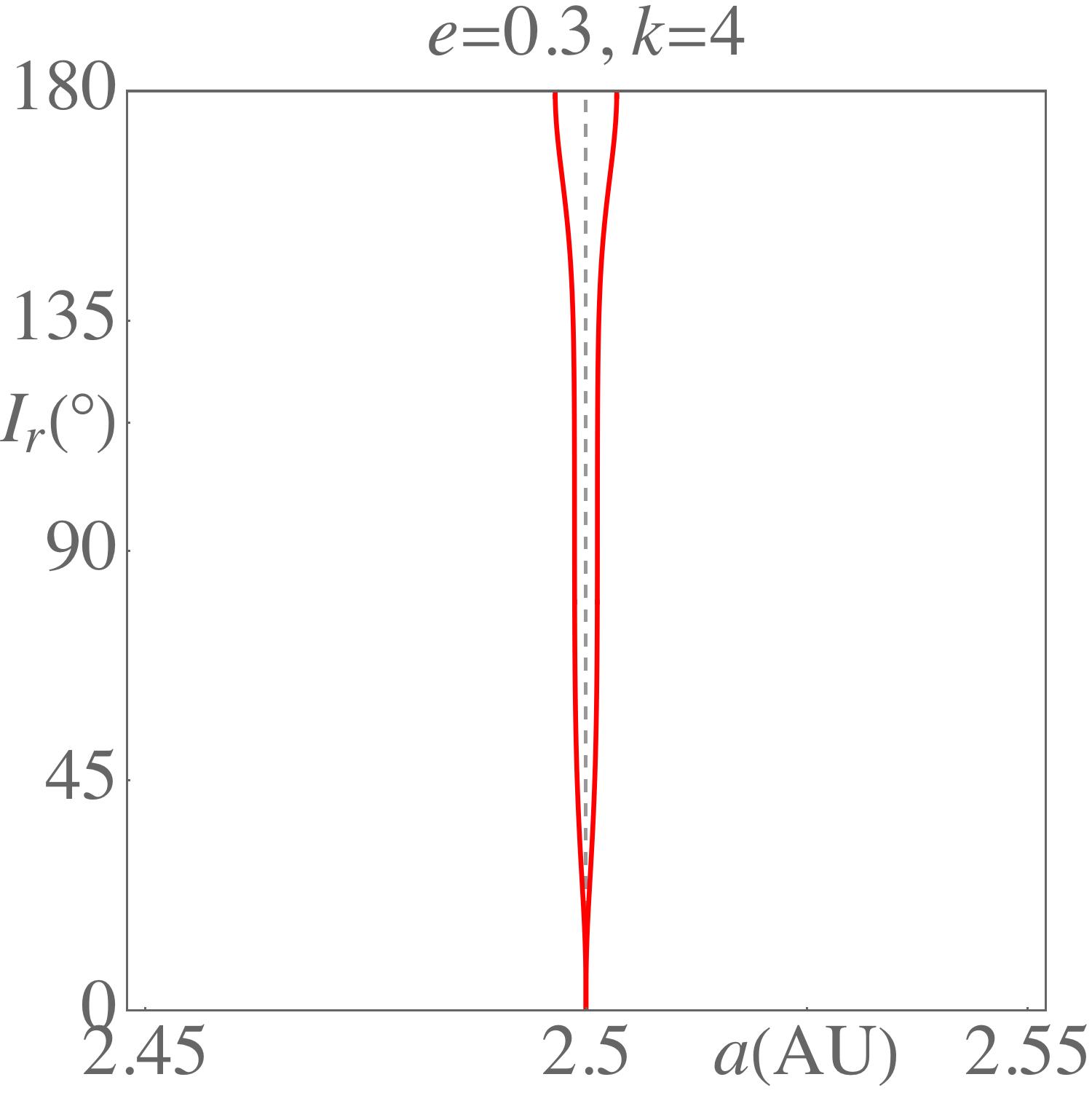}\\
\includegraphics[width=37mm]{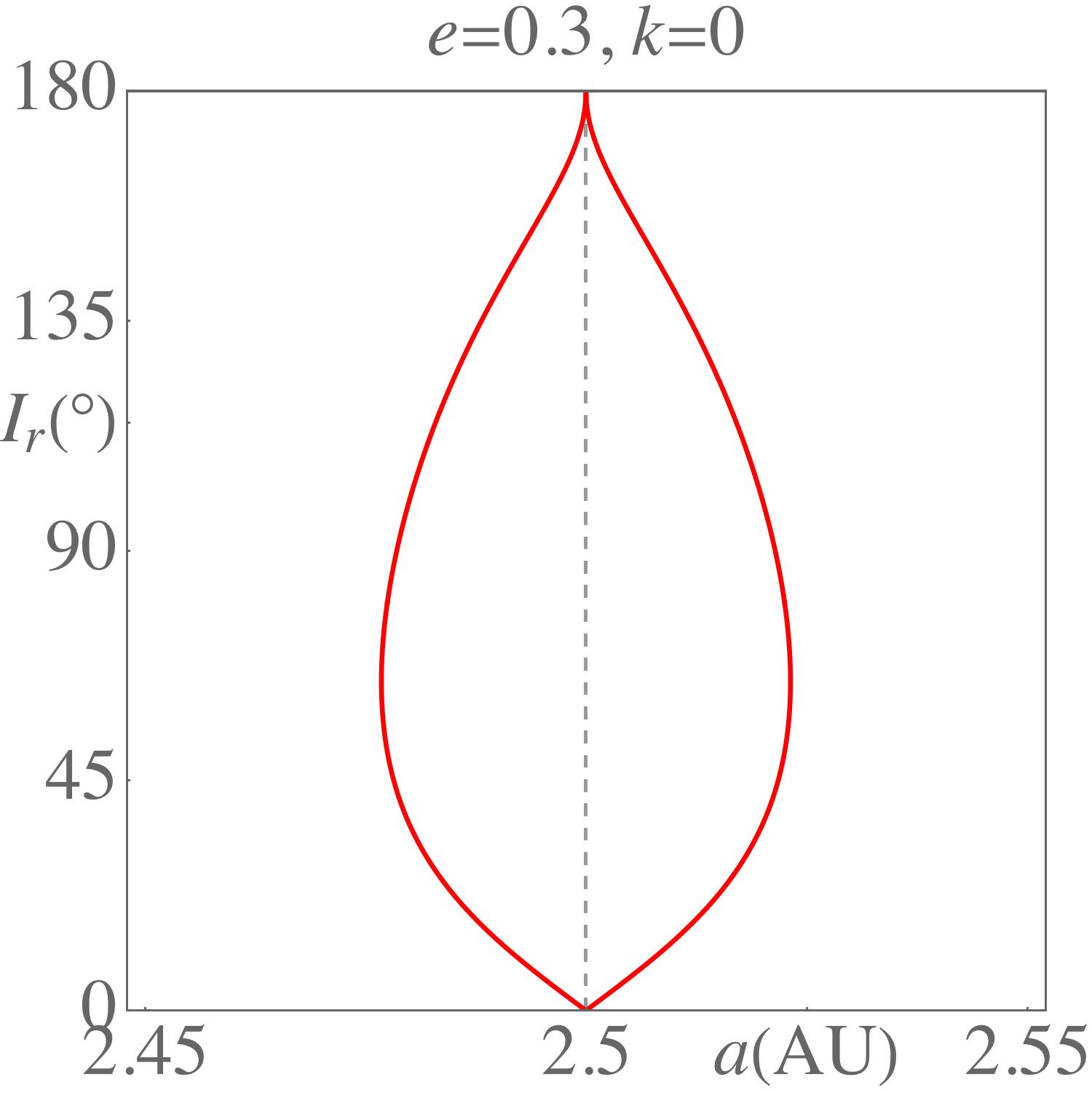}\hspace{8mm}\includegraphics[width=37mm]{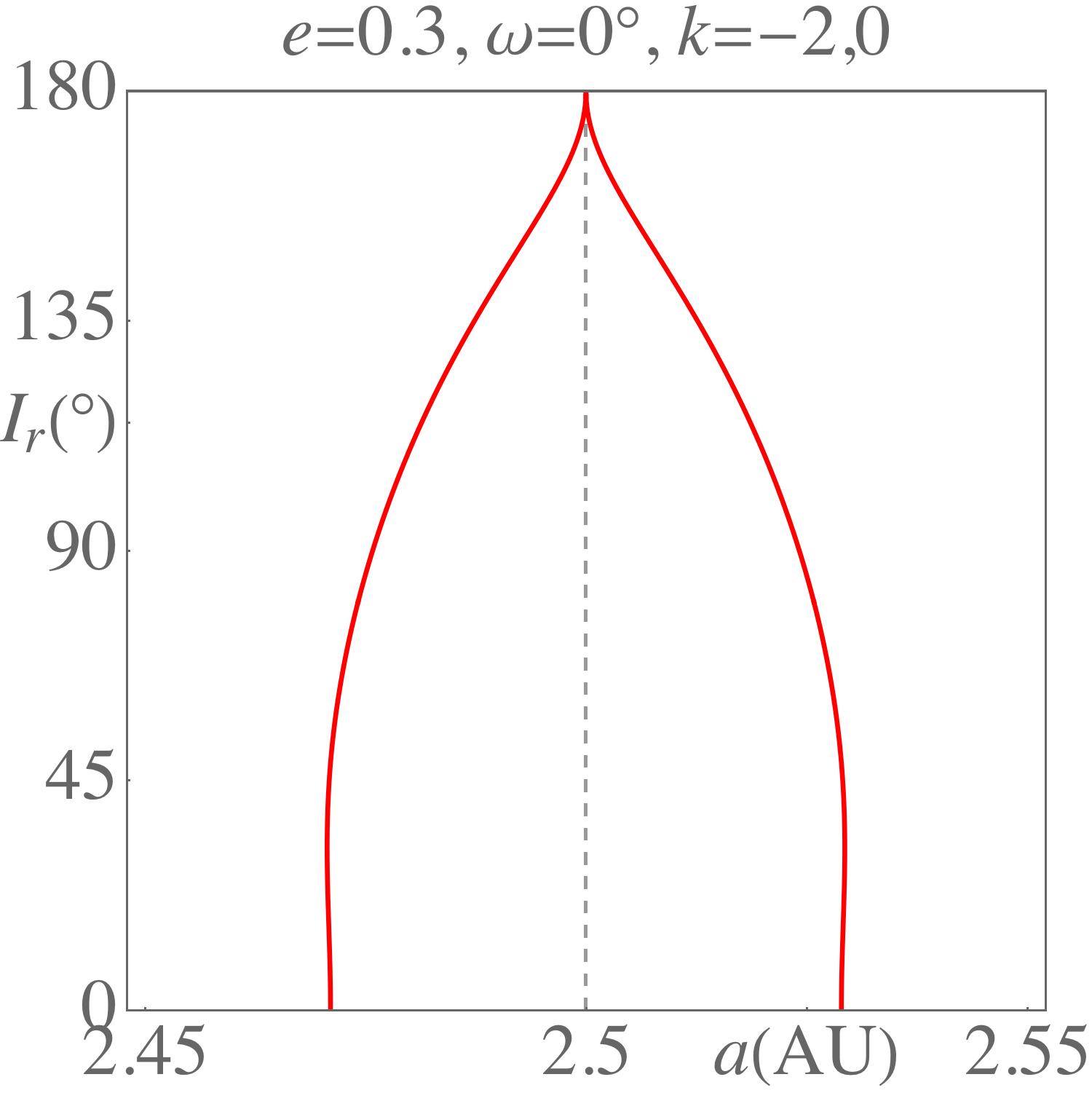}\hspace{8mm}\includegraphics[width=37mm]{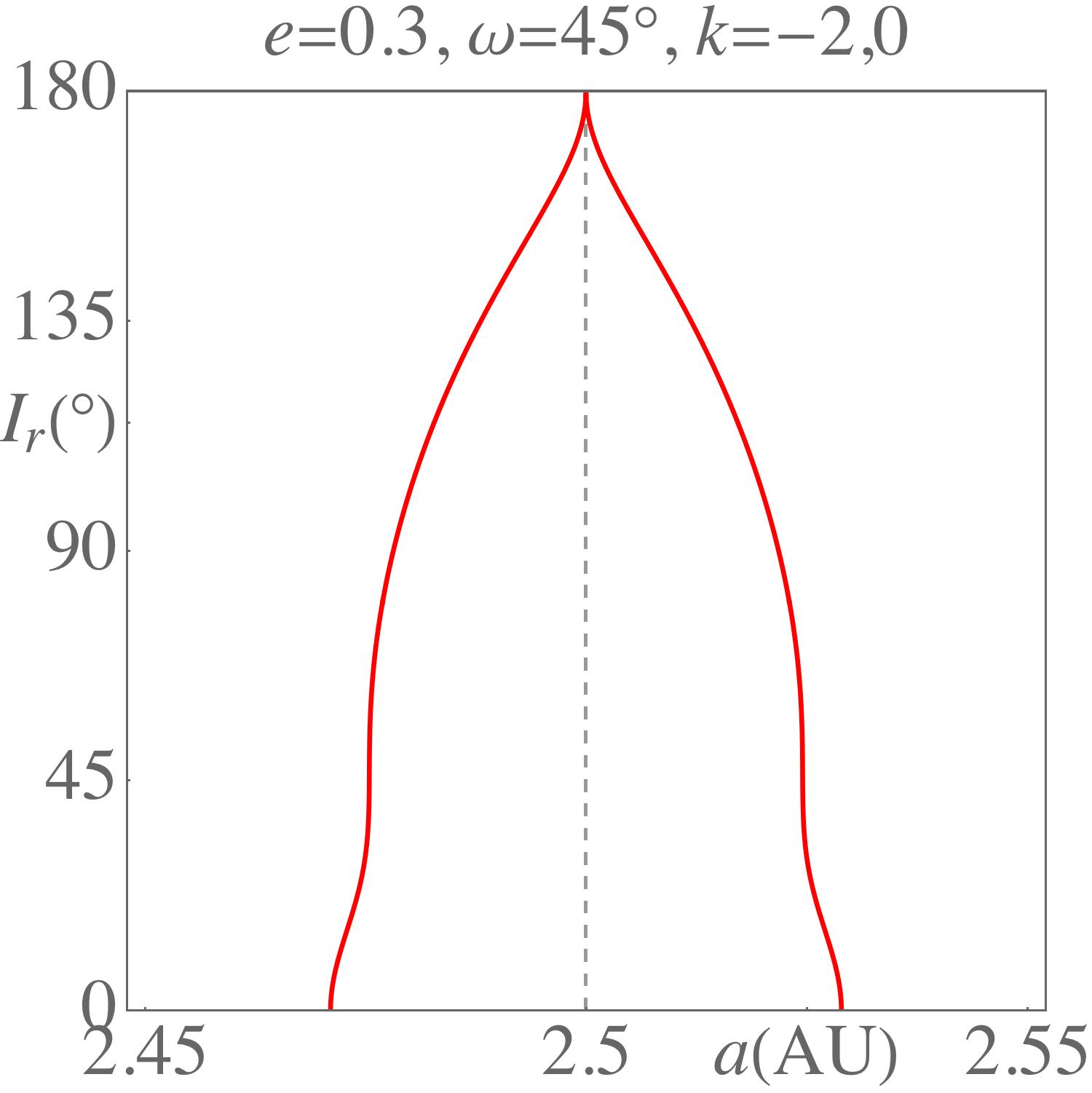}\hspace{8mm}\includegraphics[width=37mm]{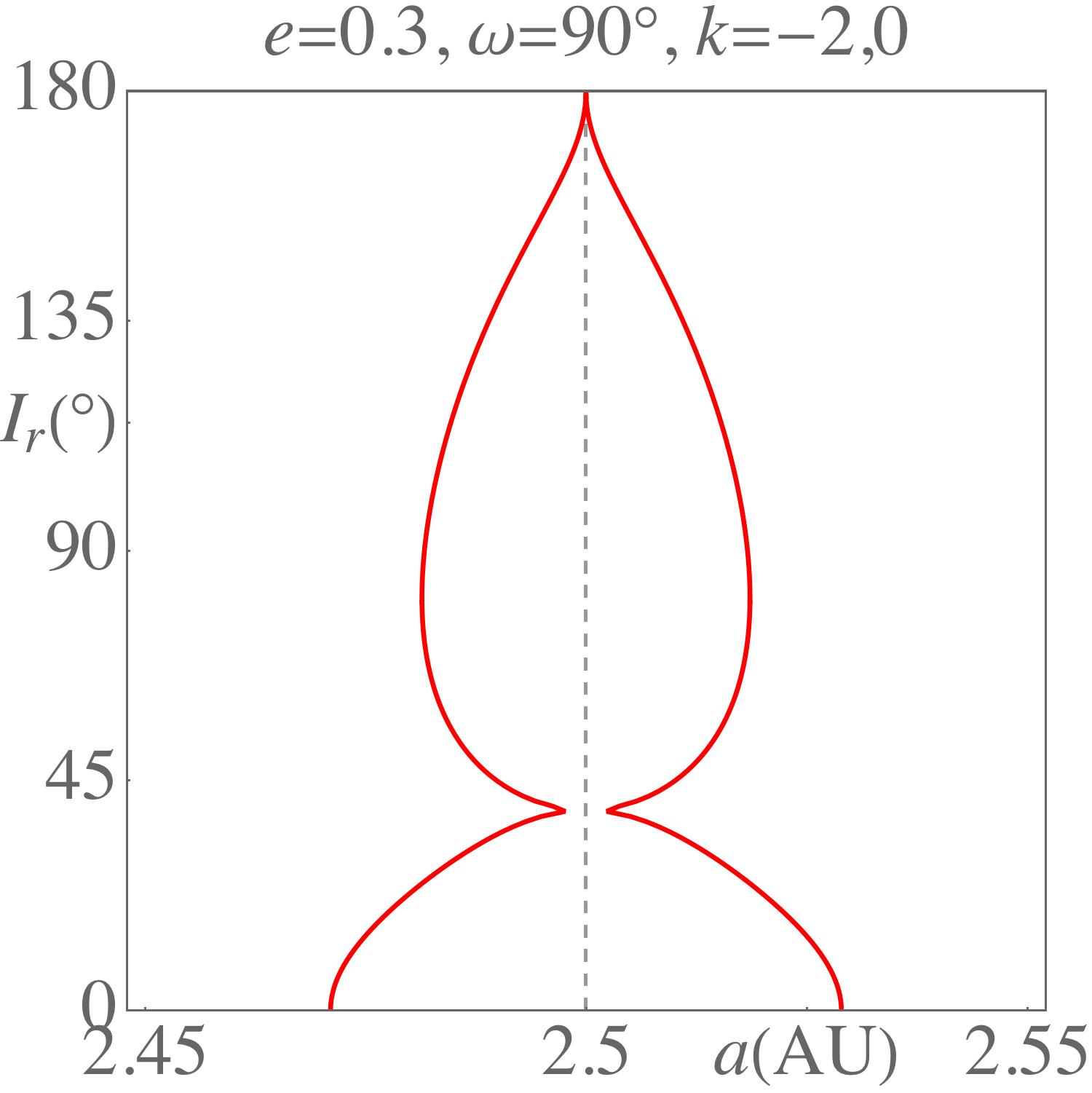}\hspace{8mm}\\
\includegraphics[width=42mm]{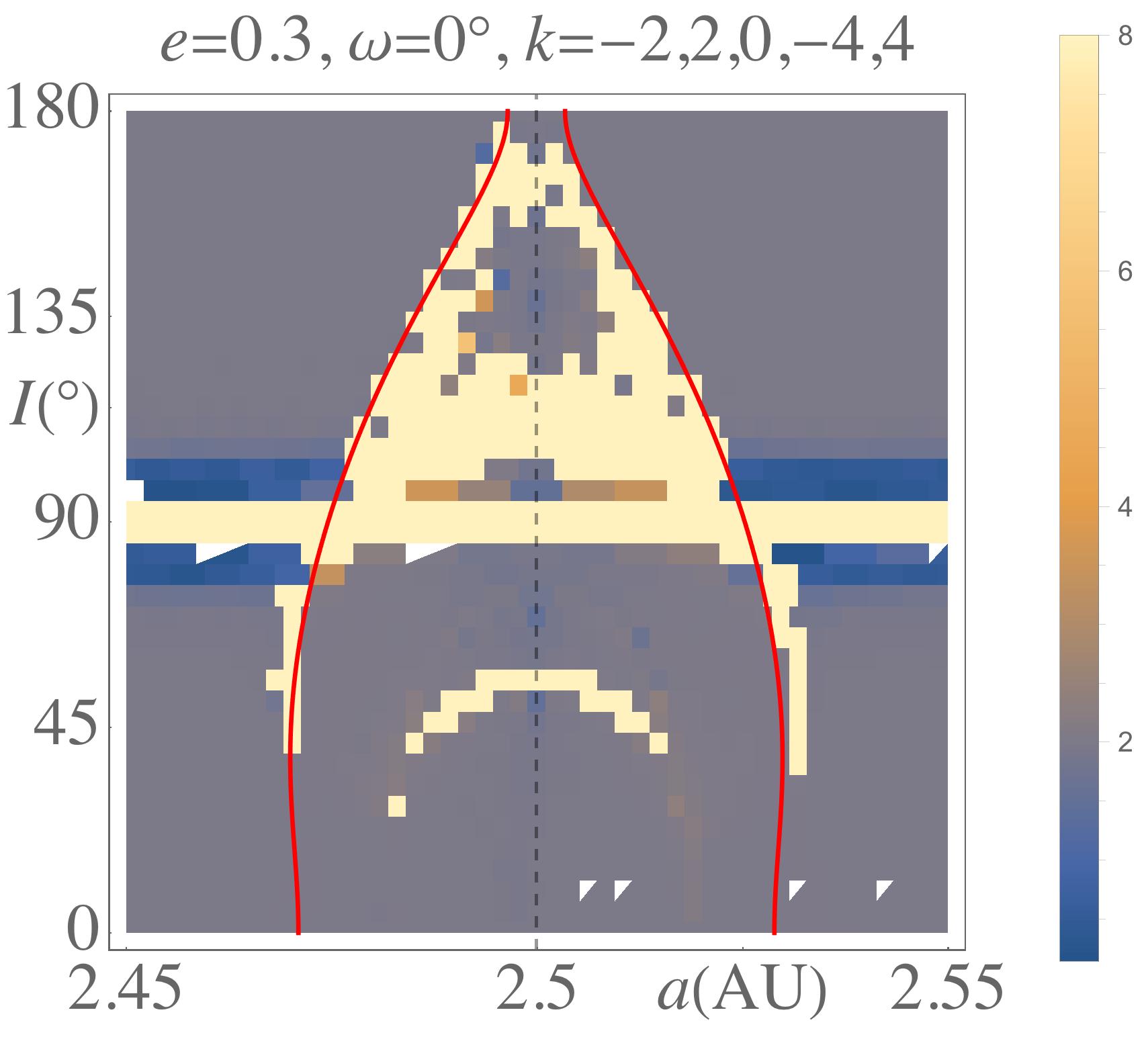}\hspace{3mm}\includegraphics[width=42mm]{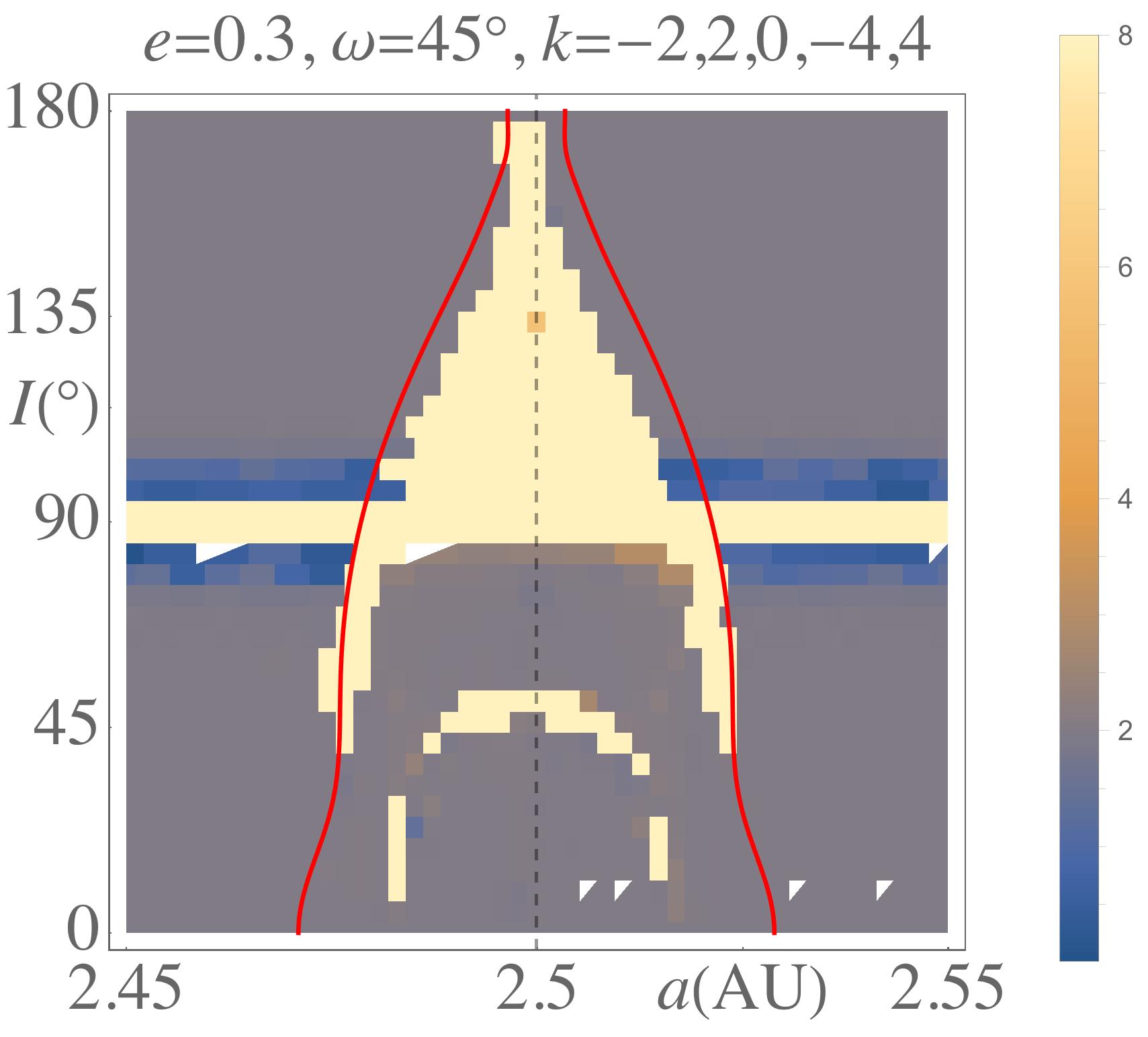}\hspace{3mm}\includegraphics[width=42mm]{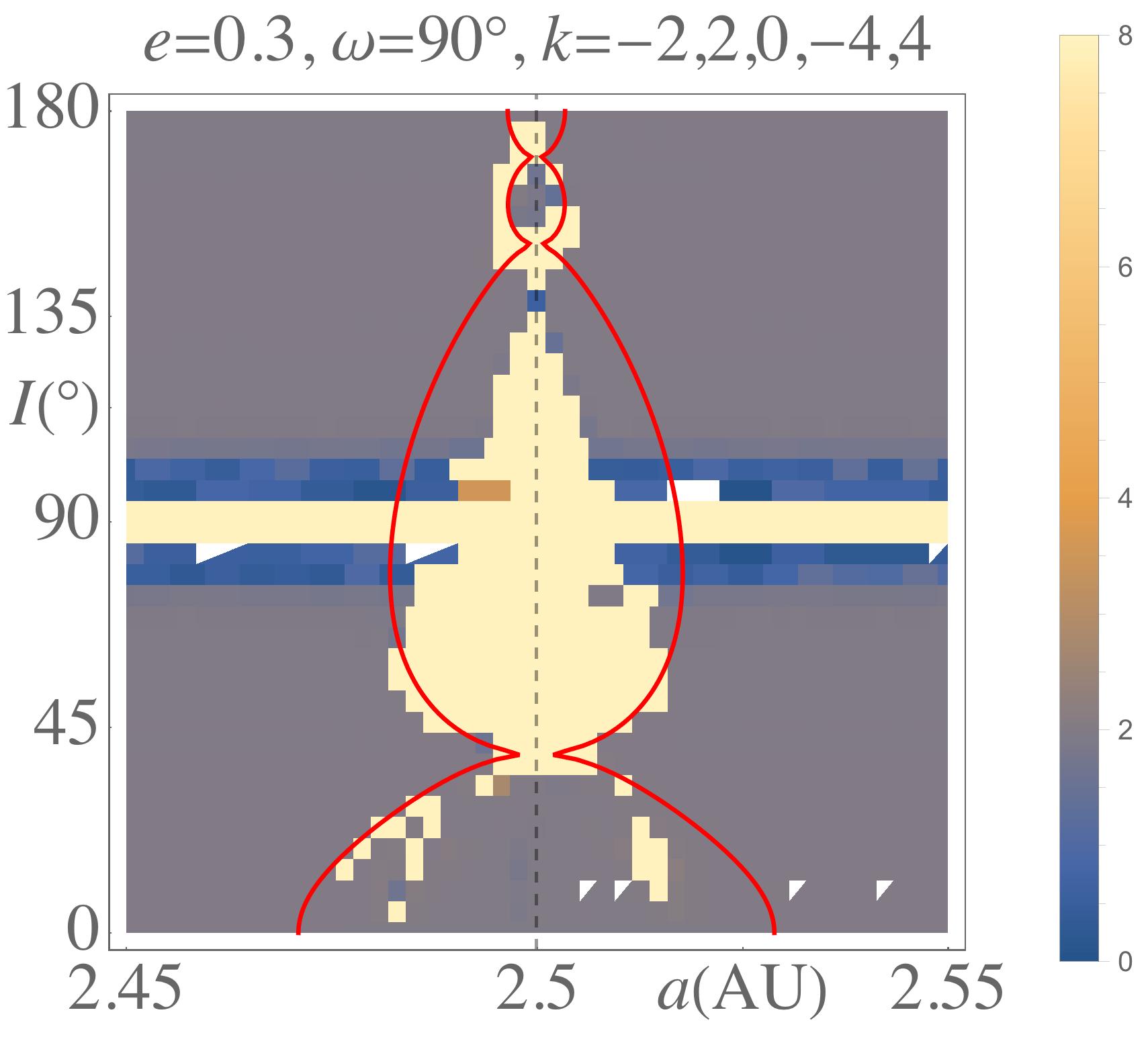}\hspace{3mm}\includegraphics[width=42mm]{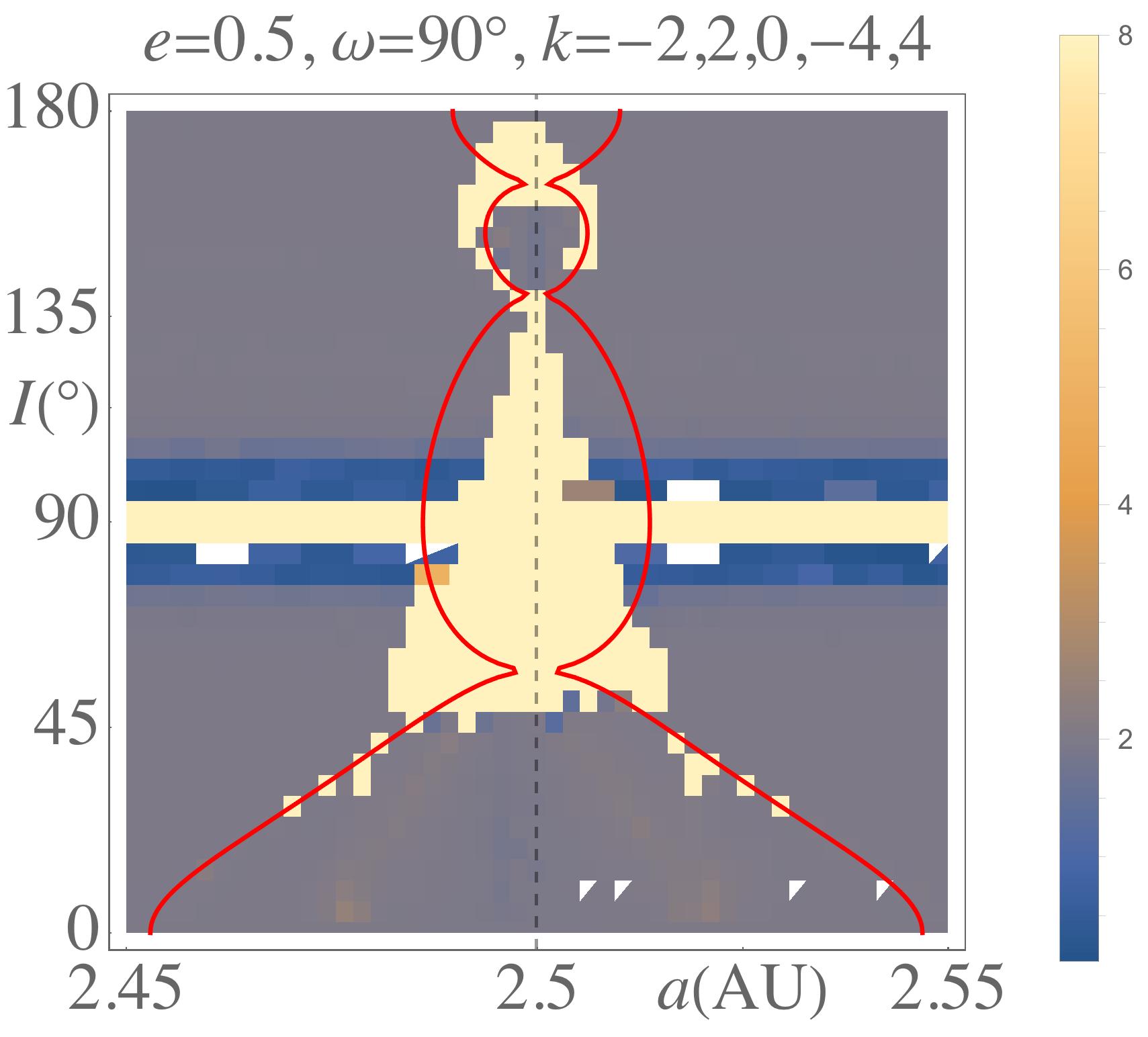}\\
\includegraphics[width=37mm]{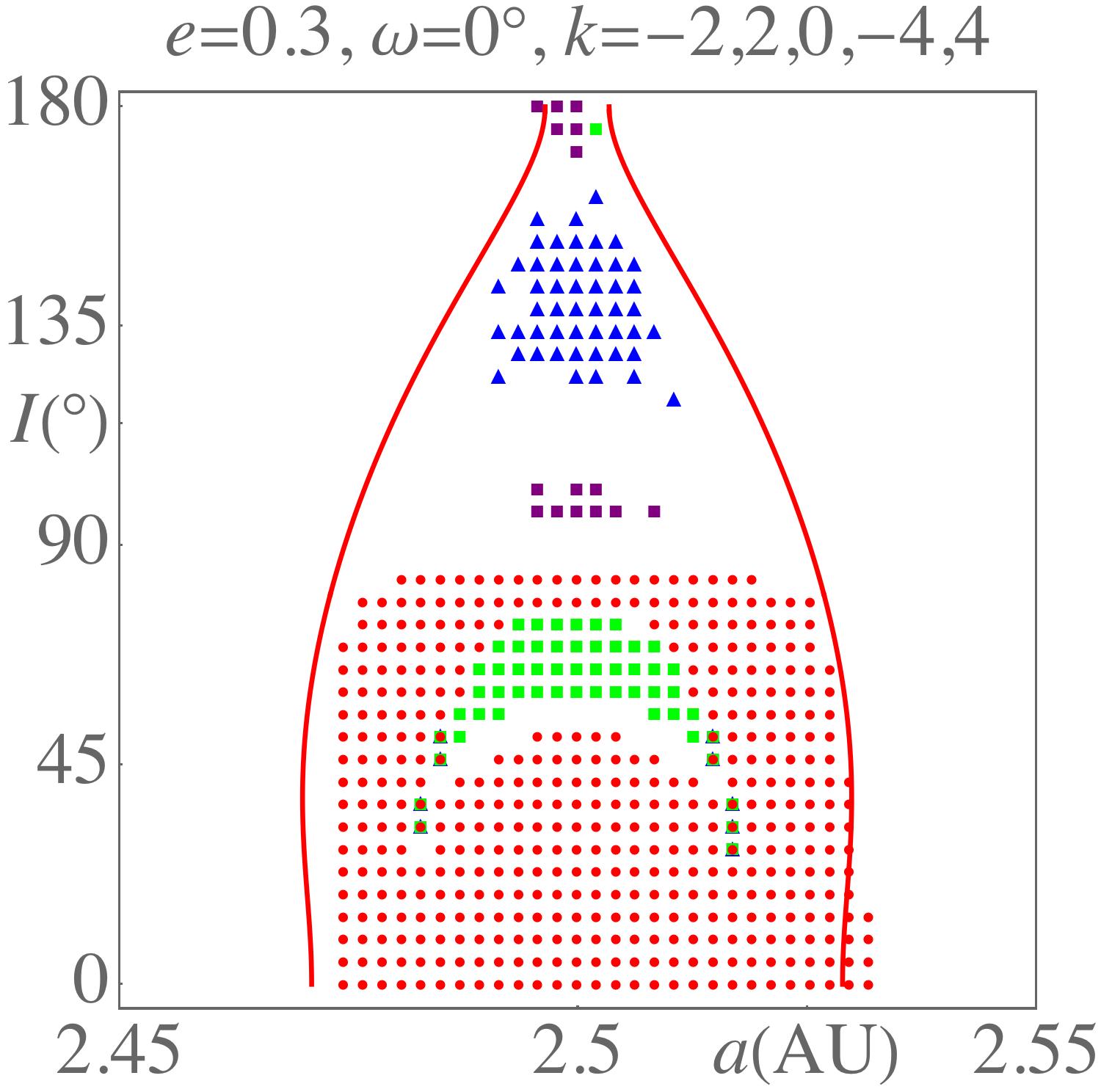}\hspace{8mm}\includegraphics[width=37mm]{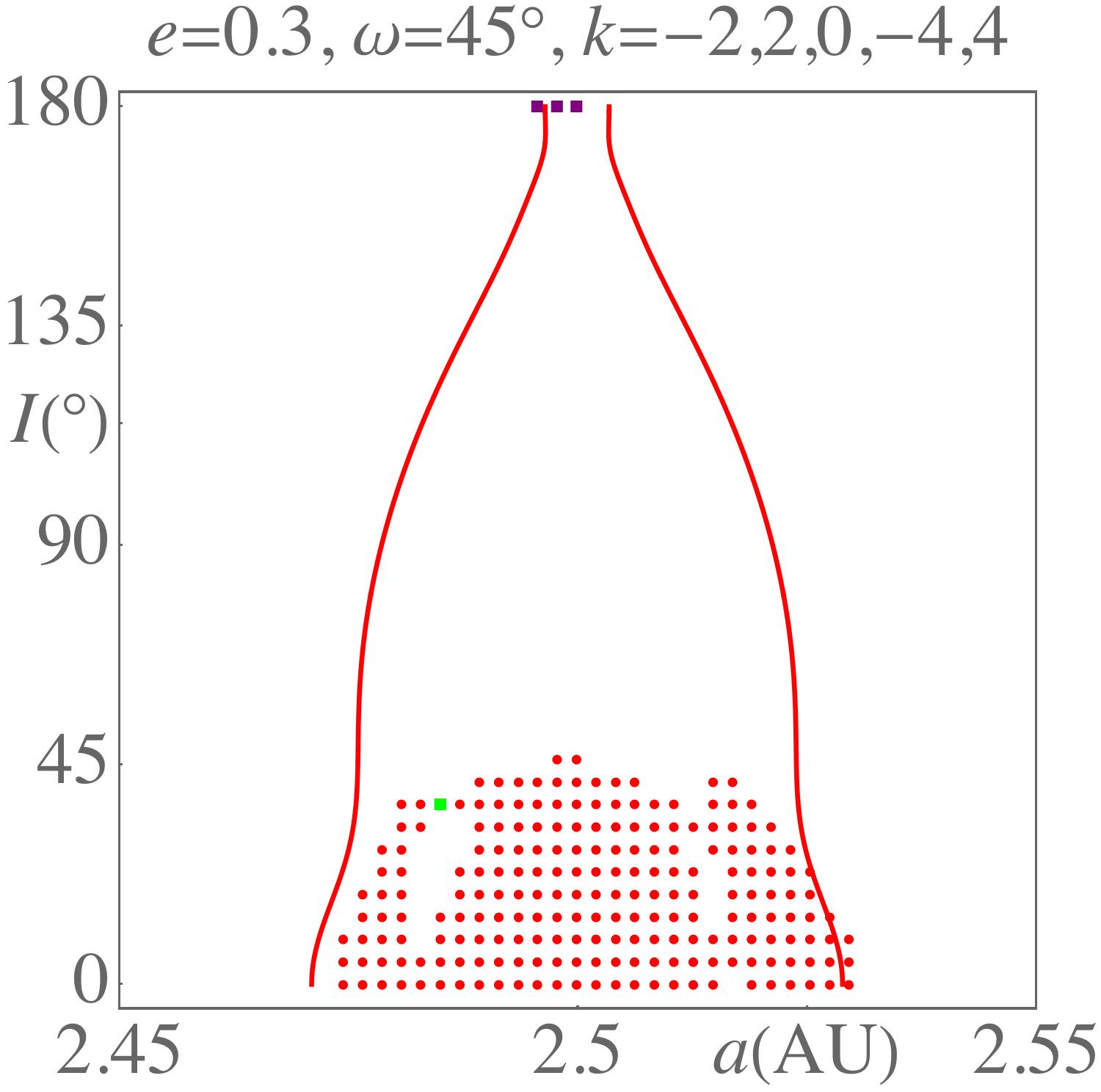}\hspace{8mm}\includegraphics[width=37mm]{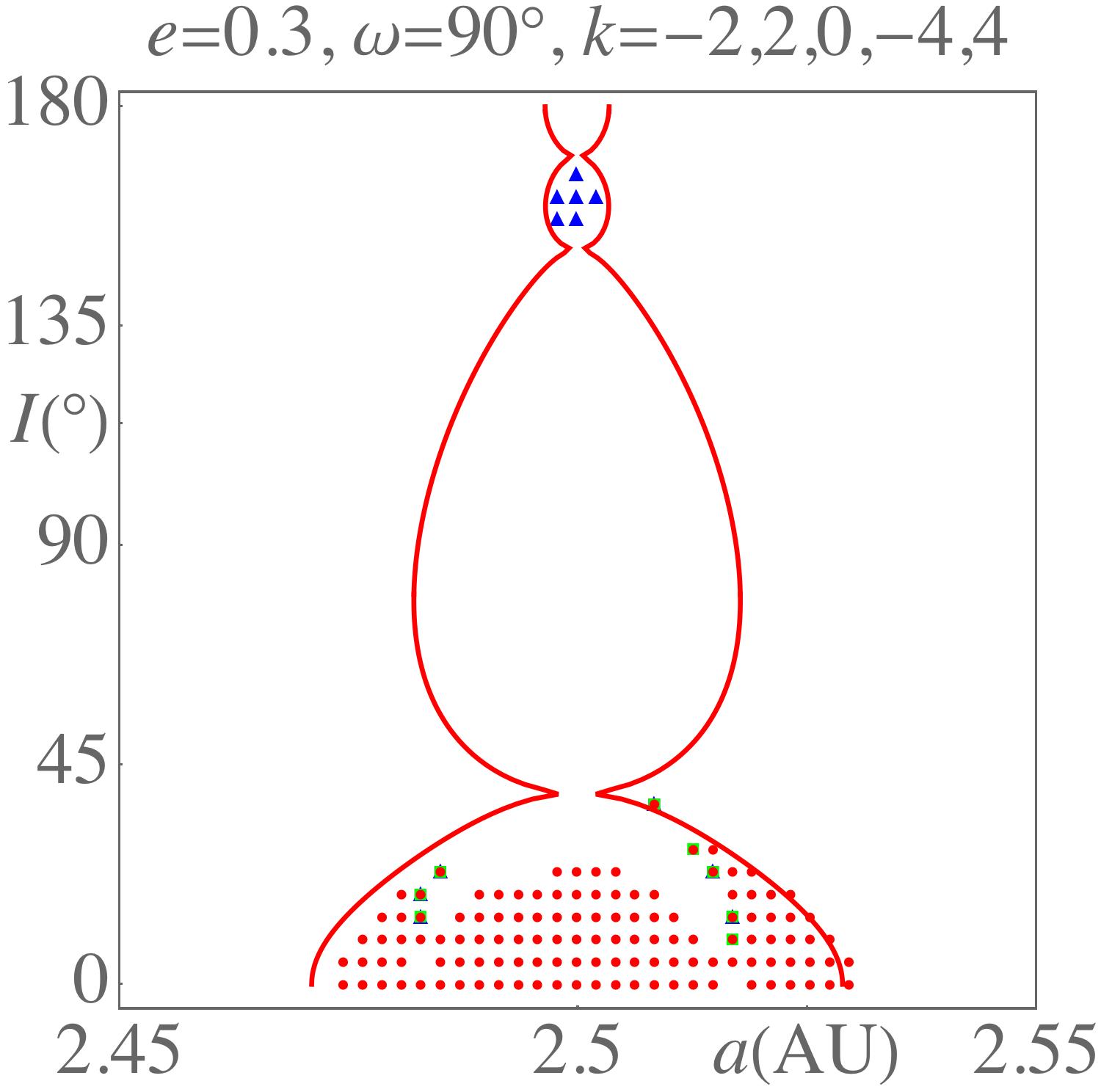}\hspace{8mm}\includegraphics[width=37mm]{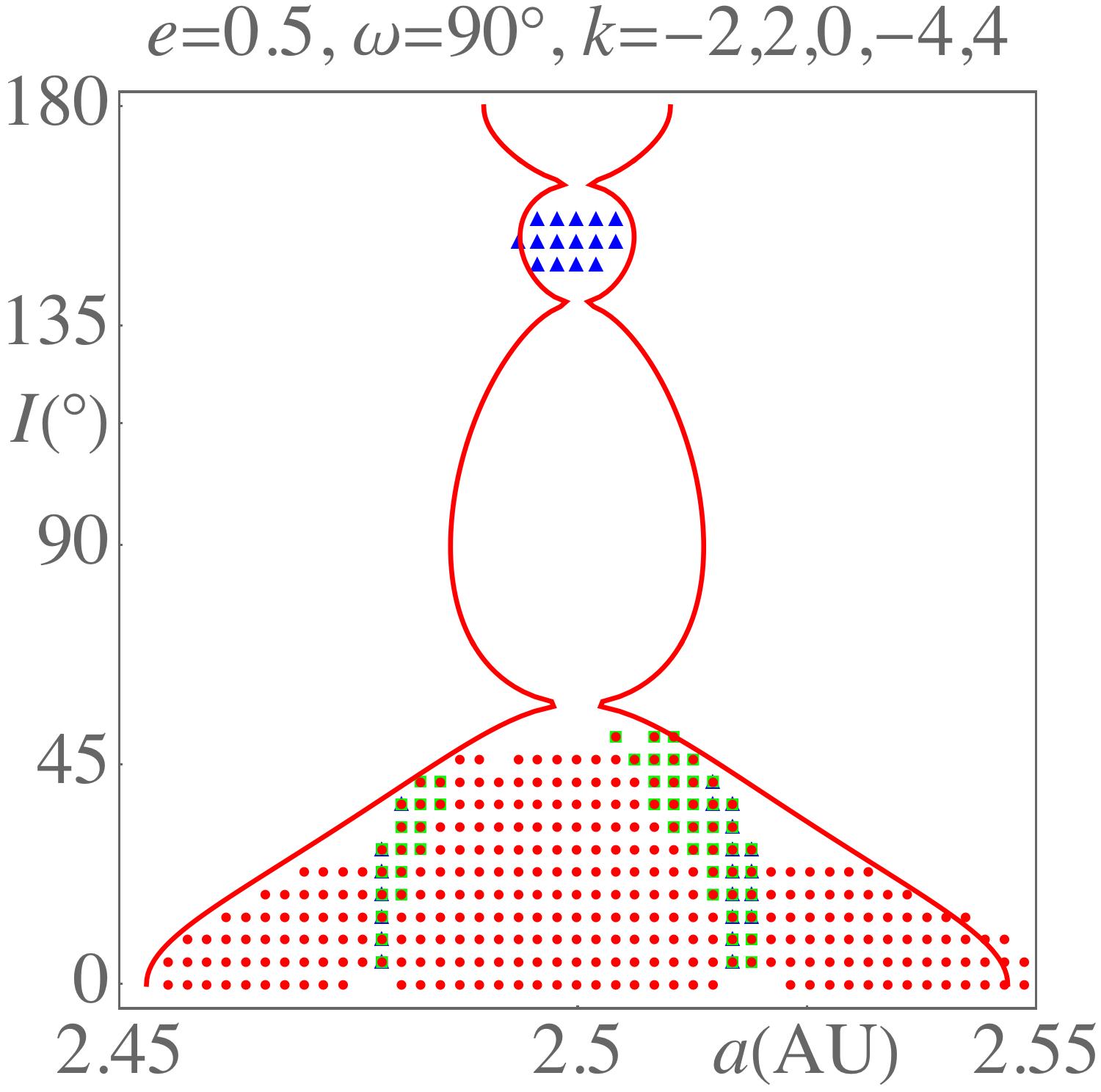}
}
\caption{Jupiter's inner 3:1 resonance. The red curves indicate the resonance width determined from the analytical models for the values of the {modes} $k$ and argument of perihelion $\omega$ indicated above each panel. The third row shows the  {\sc megno} portraits for two eccentricities $e=0.3$ and 0.5 and  different values of $\omega$ whereas the corresponding libration centers are shown in the bottom row.  Superimposed on the  {\sc megno} and libration portraits is the resonance width from the simultaneous libration of $\{ k: -2,2,-4,4,0\}$ obtained for each value of $\omega$. In the libration center panels, $k=-2$ is denoted with a red filled circle, $k=0$ with and green empty square, $k=2$ with a blue filled triangle and $k=4$ with a purple filled triangle.}\label{fJ3t1}
\end{figure*}

\subsection{Jupiter's inner 2:1 resonance}
In the first row of Figure \ref{fJ2t1}, the resonance widths of the first four fundamental modes  $\{k: -1, 1, -3, 3\}$ are shown for an eccentricity $e=0.3$. They were obtained using the width (\ref{reswidth0}) of the one-argument one-harmonic classical  pendulum model and the force amplitudes given in Table A1. Since the resonance order is odd, even values of $k$ are inadmissible. In particular, the  pure inclination $k=0$ mode is irrelevant for the inner 2:1 resonance {as} it appears only as a second harmonic $\phi^{4:2}_0$.  It is not required in order to model precisely an inner resonance like it is for an outer resonance and its asymmetric librations. The $k=-1$ mode defines the prograde pure eccentricity resonance $\phi^{2:1}_{-1}=\lambda-2 \lambda^\prime+\varpi$. Its width was first derived in Paper I.  The $k=3$ mode defines the retrograde pure eccentricity resonance $\phi^{2:1}_{3}=\lambda^\star-2\lambda^\prime-3\varpi^\star$. Its width was first derived in Paper I. The mode is recognizable through its finite width for planar retrograde motion at $I=180^\circ$ and zero width at planar prograde motion $I=0^\circ$. Only pure eccentricity modes have finite widths at $0^\circ$ or $180^\circ$. All inclination modes' widths vanish at exact coplanarity. 

The largest overall width corresponds to $k=-1$. The second largest mode is the inclination mode $k=1$. Depending on the value of $\omega$ (see Section 4.1.1) $k=1$ can counter or reinforce resonance. The leftmost panel of the second row shows the resonance width of the simultaneous librations of $k=-1$ and $k=1$ for $\omega=90^\circ$ obtained using the expressions (\ref{reswidth0},\ref{ampgk1k2}). Resonance is cancelled near $I=130^\circ$ and enlarged near $I=160^\circ$. 

In order not to overcrowd the paper with figures, we do not show the possible binary combinations of all modes. Instead we combine {the first four fundamental} modes using the model of Section 4.1.2 (Equations \ref{reswidth0}, \ref{gmultiple}) and show them in the last three-panels of the second row for the three values of $\omega$ that are used in the {\sc megno}  and libration portraits. For $\omega=0^\circ$, all mode amplitudes add up and the resulting resonance width is largest. The opposite occurs for $\omega=90^\circ$ where resonance width is the smallest and vanishes not only near $I=120^\circ$ but also near $I=170^\circ$ because of the addition of  the retrograde mode $k=3$  that  introduces a finite width at $I=180^\circ$ for all values of $\omega$. The width associated with  $\omega=45^\circ$ is intermediate between those of $\omega=0^\circ$ and $90^\circ$. We chose these three typical values for the perihelion because the secular potential for arbitrary inclination that {describes} the long term evolution of perihelion is the Kozai-Lidov potential whose characteristic period for the argument of perihelion is $90^\circ$ (\cite{Kozai62}, \cite{Lidov62} see also Paper II).

The bottom two rows of Figure \ref{fJ2t1} show the  portraits superimposed on the three resonance widths of the three perihelion values obtained with the four fundamental modes. For an eccentricity $e=0.3$, the chaotic domain shown in yellow is contained within the two resonance width extrema of $\omega=0^\circ$ and $90^\circ$. The agreement with the analytical estimates is remarkable. The horizontal polar  instability domain visible across all semi-major axis values is caused by the Kozai-Lidov resonance (see Paper II for a discussion of that instability). Increasing the eccentricity to $e=0.5$ shows that the chaotic domains away from the Kozai-Lidov instability zone is  located between the two extremal resonance width curves. The reason for this is the effect of  the secular potential that forces the circulation of the argument of perihelion.  As $\omega$ circulates  over $5\times 10^5$ Jupiter periods, the {resonance} separatrix will sweep over the space between the two extremal widths of $\omega=0^\circ$ and $90^\circ$ making any asteroid that falls into that region  unstable.  It is therefore not surprising that the stable motion domain corresponds mainly to the separatrix of $\omega=90^\circ$ with the smallest resonance width seen also in the rightmost panels showing the librating modes over $5 \times 10^5$ Jupiter periods.

All librations shown in Figure \ref{fJ2t1} occur on the secular timescale because of the large integration timespan. All four fundamental modes $k=-1,1,-3$ and 3 do shape the full resonance width as we have confirmed with the {\sc megno} portraits but they still do so on the resonant timescale as discussed in Section 4. What we report in the bottom rightmost panels are librations on the integration timescale of  $5\times 10^5$ Jupiter periods. In this way, we access information on mode selection over secular timescales that cannot be given by the pendulum model as it ignores secular interactions. 
We find that libration in the 2:1 resonance for  $e=0.3$ is dominated mainly by the pure eccentricity mode $k=-1$ up to $120^\circ$-inclination. The island formed by $k=-1,1,$ and 3 at large retrograde inclination is occupied by $k=1$ librations whereas the retrograde pure eccentricity mode is confined to near co-planar retrograde motion (relative inclination $\sim10^\circ$). Double argument libration in the modes $k=-1$ and $k=-3$ occurs in smaller islands within the prograde motion domain. As eccentricity is increased, the resonance width is larger for prograde inclination but extends to smaller retrograde inclinations at the expense of a larger $k=1$-libration island.  The extent of double argument libration with $k=-1,-3$ is also increased. The argument shifts (\ref{ampdelta}) and (\ref{deltamultiple}) are found in all cases not to exceed $0.1^\circ$. Lastly, we have examined whether {modes} $|k|\geq 5$ have a significant effect on the resonance width and concluded that they did not.

\subsection{Jupiter's inner 3:1 resonance}
In the first two rows of Figure \ref{fJ3t1}, the resonance widths of the first five fundamental modes  $\{k: -2, 2, -4, 4, 0\}$ are shown for an eccentricity $e=0.3$. As for the 2:1 resonance, the widths were obtained using  the classical  pendulum model and the force amplitudes given in Table A2. Since the resonance order is even, odd values of $k$ are inadmissible. 

The $k=-2$ mode defines the prograde pure eccentricity resonance $\phi^{3:1}_{-1}=\lambda-3 \lambda^\prime+2\varpi$.  The $k=4$ mode defines the retrograde pure eccentricity resonance $\phi^{3:1}_{4}=\lambda^\star-3\lambda^\prime-4\varpi^\star$.  The widths of $k=-2$ and $k=4$ were first derived in Paper I. The $k=0$ mode defines the pure inclination resonance $\phi^{3:1}_{0}=\lambda-3\lambda^\prime+2\Omega$. 

The largest overall widths corresponds to $k=-2$ and $k=0$. Modes $k=2$ and $k=4$ dominate resonance at large retrograde inclination and nearly coplanar retrograde motion respectively.  The width resulting from the simultaneous librations of modes $k=-2$ and $k=0$ {(Figure  \ref{fJ3t1}, second row)} {depends} crucially on the value of the argument of perihelion $\omega$ like the case of the 2:1 resonance. The overall width is largest for $\omega=0^\circ$ and smallest for $\omega=90^\circ$. In particular, resonance width vanishes at $I=40^\circ$.  {By} including the  remaining three modes, shown directly on the  {\sc megno} and libration portraits separately for each value of $\omega$, a new island in the inclination range $150^\circ\leq I\leq 170^\circ$ appears because of the $k=2$ mode in a similar way to the 2:1 resonance. 

However, the 3:1  portraits differ from those of the 2:1 resonance in that the chaotic layer that surrounds resonance does not correspond to the extremal widths of $\omega=0^\circ$ and $\omega=90^\circ$ that would result from the secular circulation of the argument of perihelion. Instead, stable motion occurs exactly within the analytical width given by the initial value of $\omega$. Furthermore, excluding the unstable horizontal domain of the Kozai-Lidov resonance for nearly polar orbits, most stable motion is prograde except for small islands dominated by libration in the $k=2$ and $k=4$ modes. It is also interesting to note the presence of a sizable crescent-shaped libration {region} in the pure inclination mode $k=0$ for large prograde inclinations and $\omega=0^\circ$. As explained in Section 4, the pendulum model is not equipped to predict mode selection on secular timescales, {or} measure the extent of chaotic regions. These aspects with be addressed in future work. Increasing the eccentricity to $e=0.5$ (Figure \ref{fJ3t1}, bottom two rows, rightmost panels) shows that the analytical estimates are robust and describe accurately the resonance width of the 3:1 resonance. 

\subsection{Neptune's outer 1:2 resonance}
\begin{figure*}
\begin{center}
{ 
\hspace*{-2mm}\includegraphics[width=36mm]{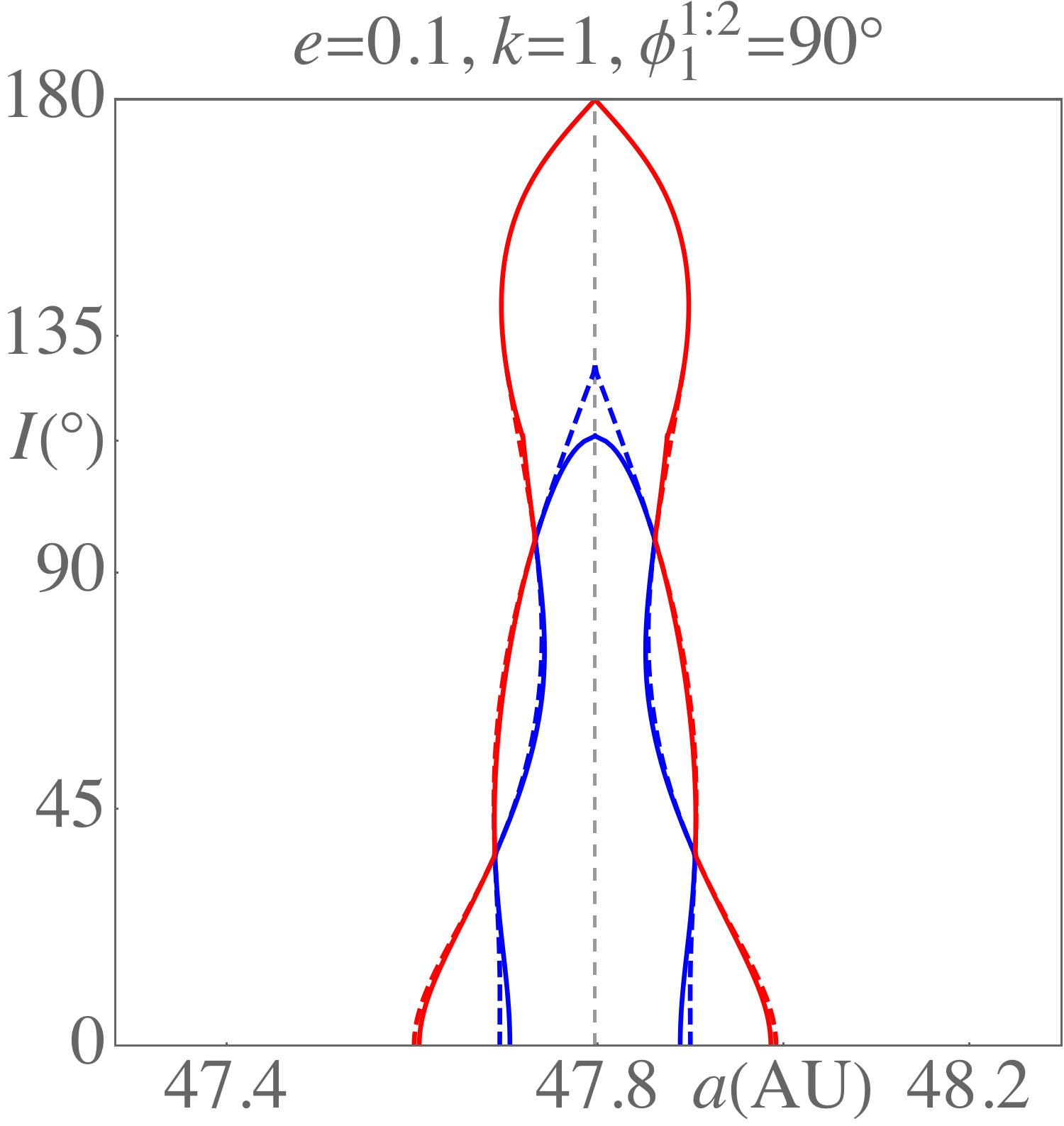}\hspace{7mm}
\includegraphics[width=36mm]{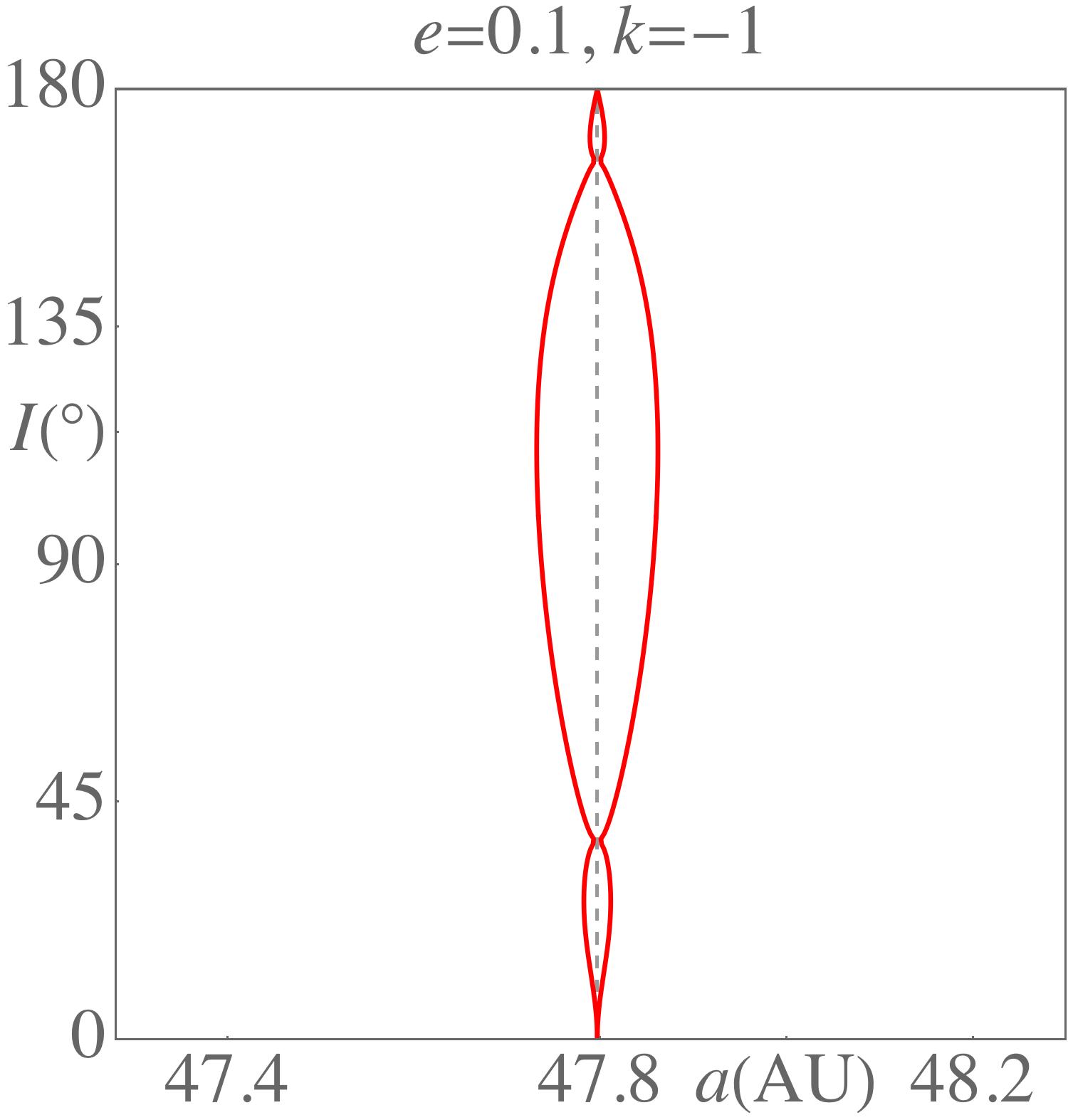}\hspace{6mm}
\includegraphics[width=42mm]{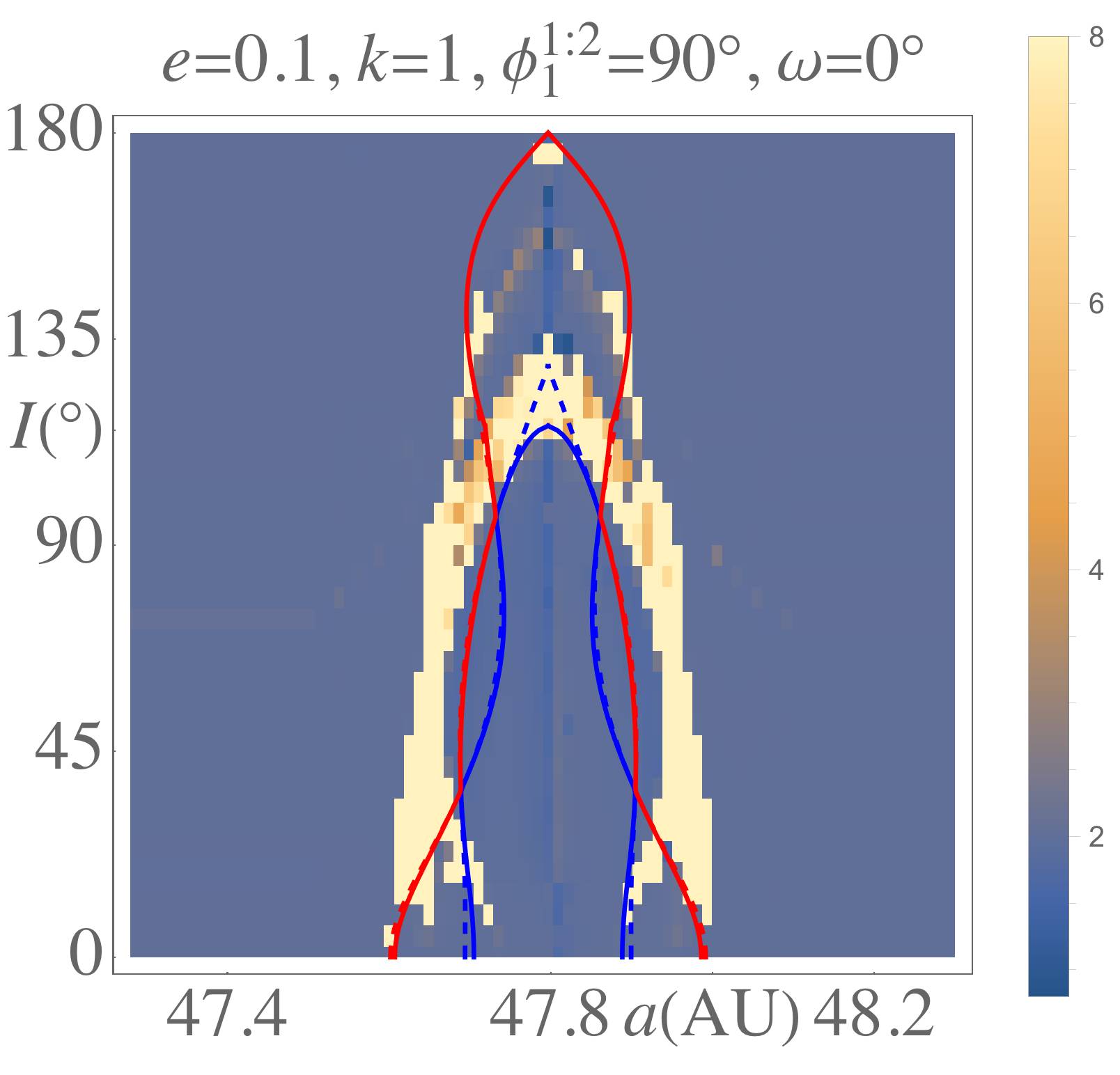}\hspace{4mm}
\includegraphics[width=42mm]{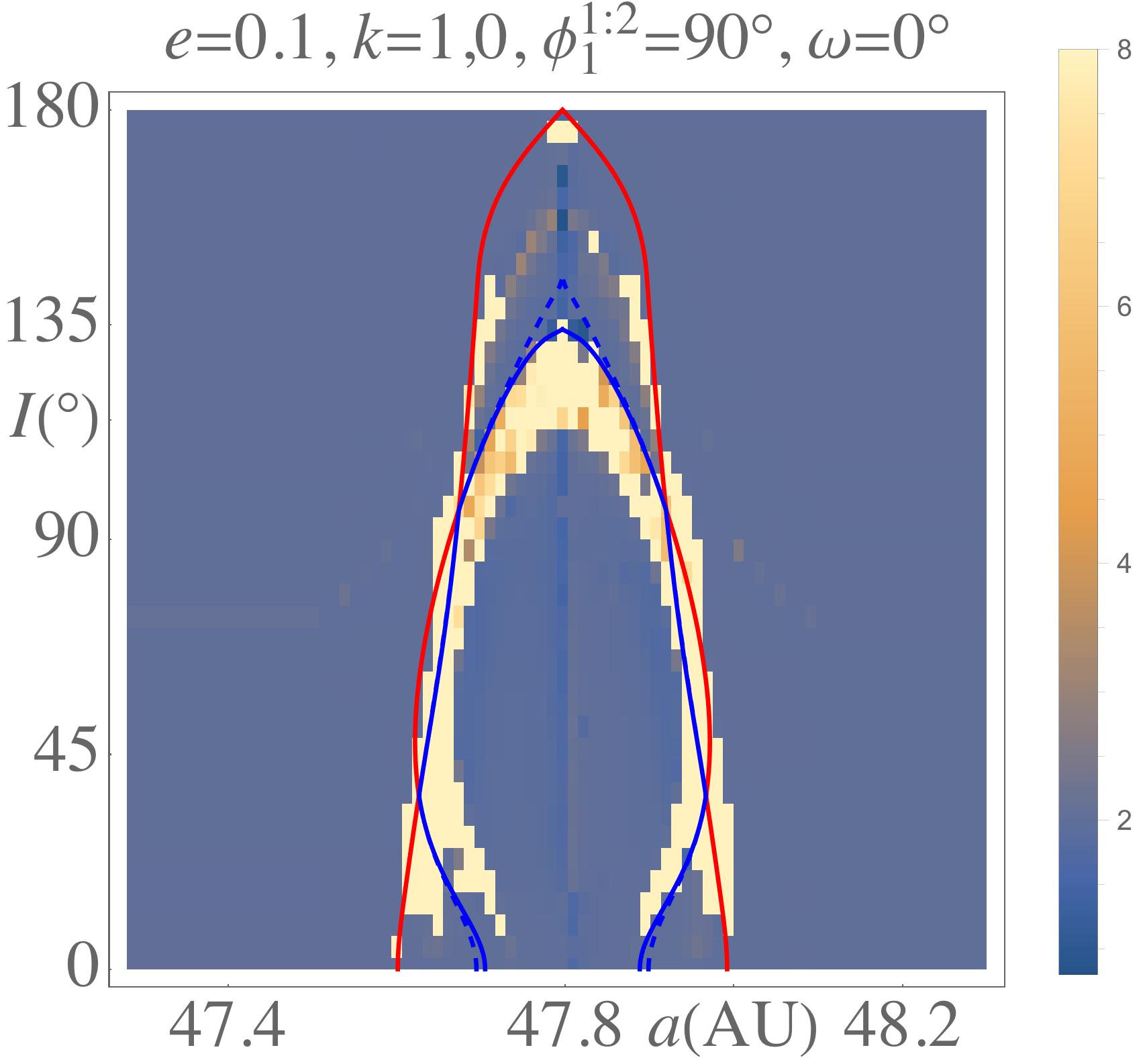}\\
\hspace*{-5mm}\includegraphics[width=36mm]{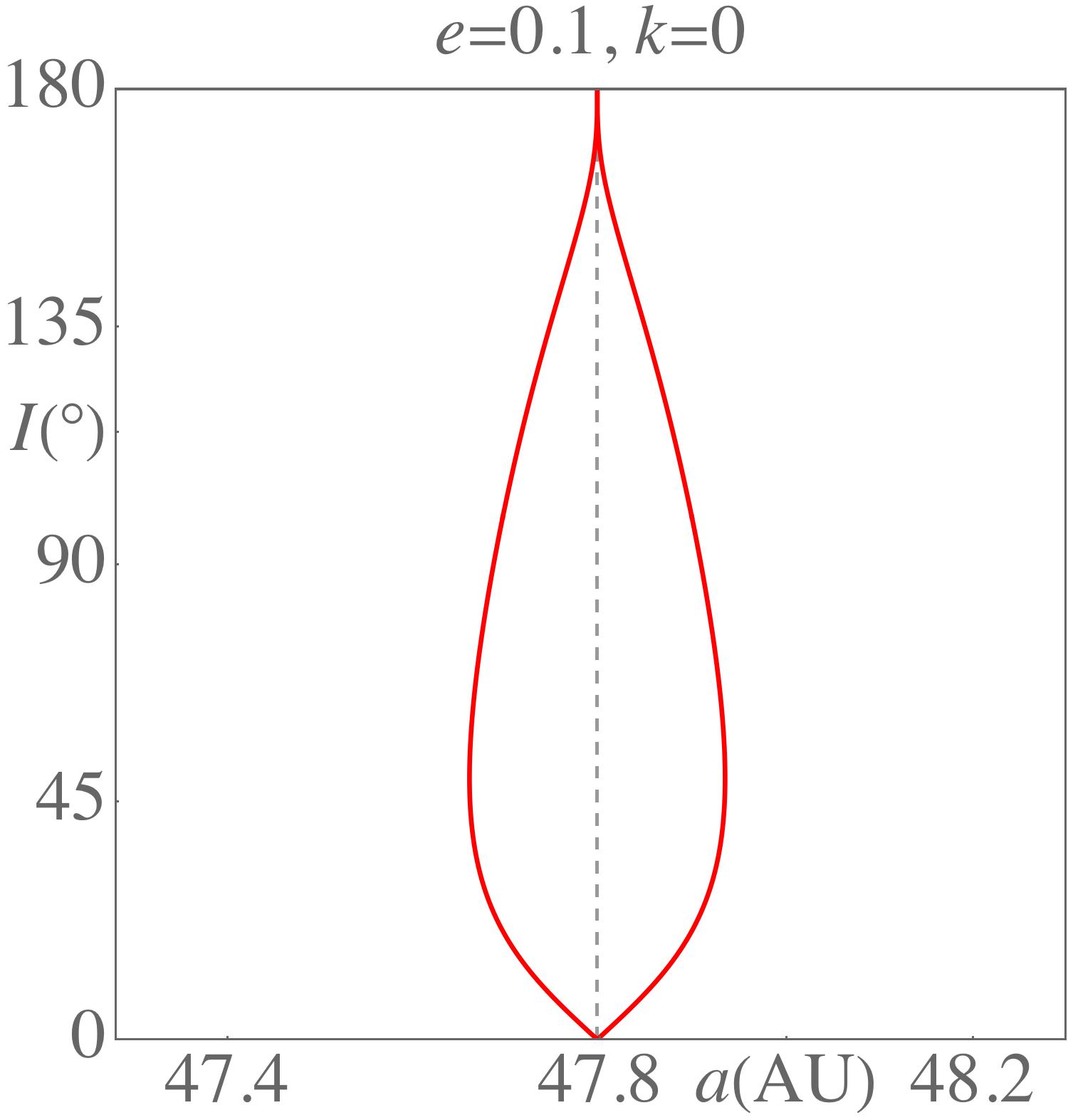}\hspace{9mm}
\includegraphics[width=36mm]{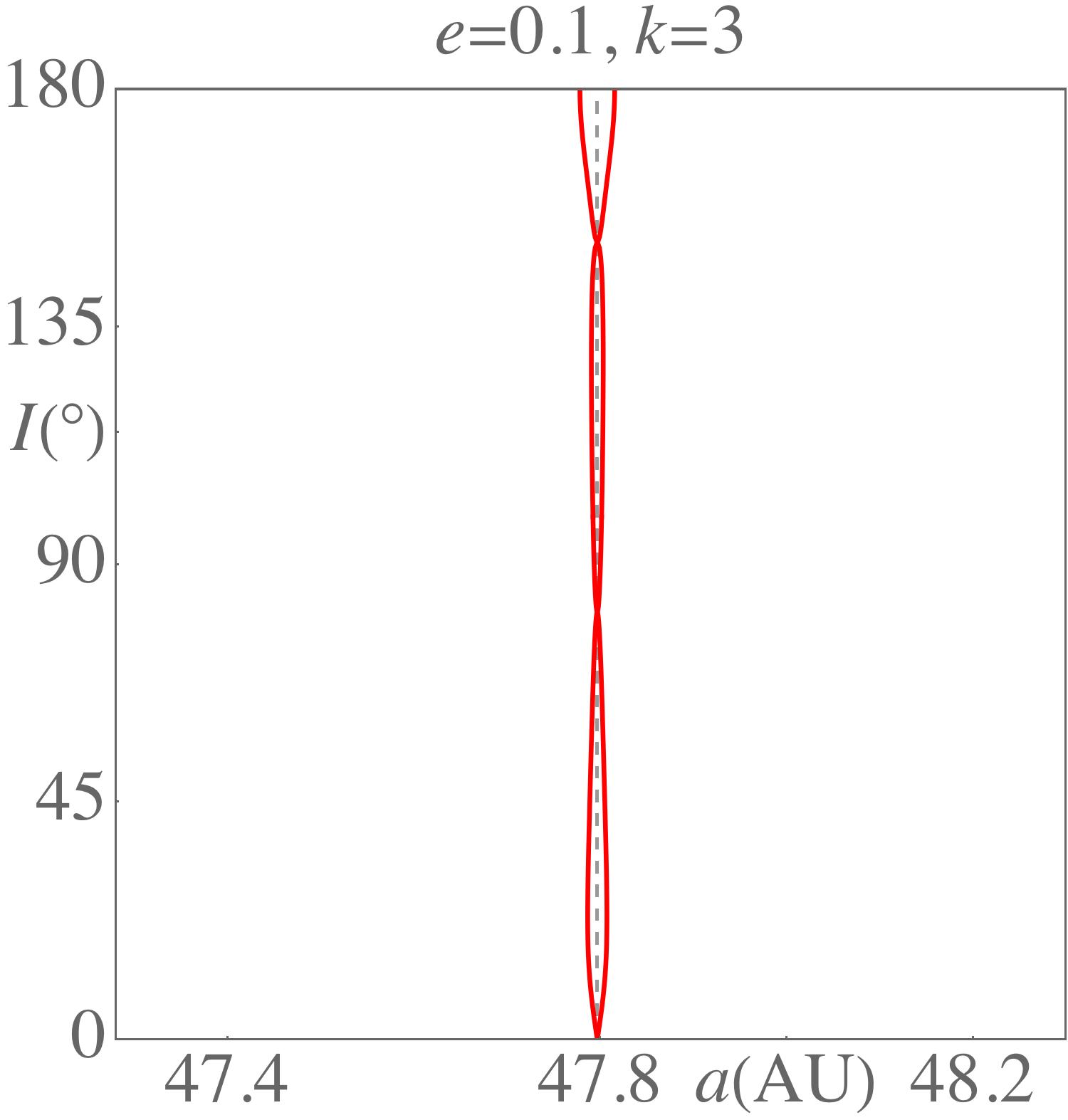}\hspace{6mm}
\includegraphics[width=41mm]{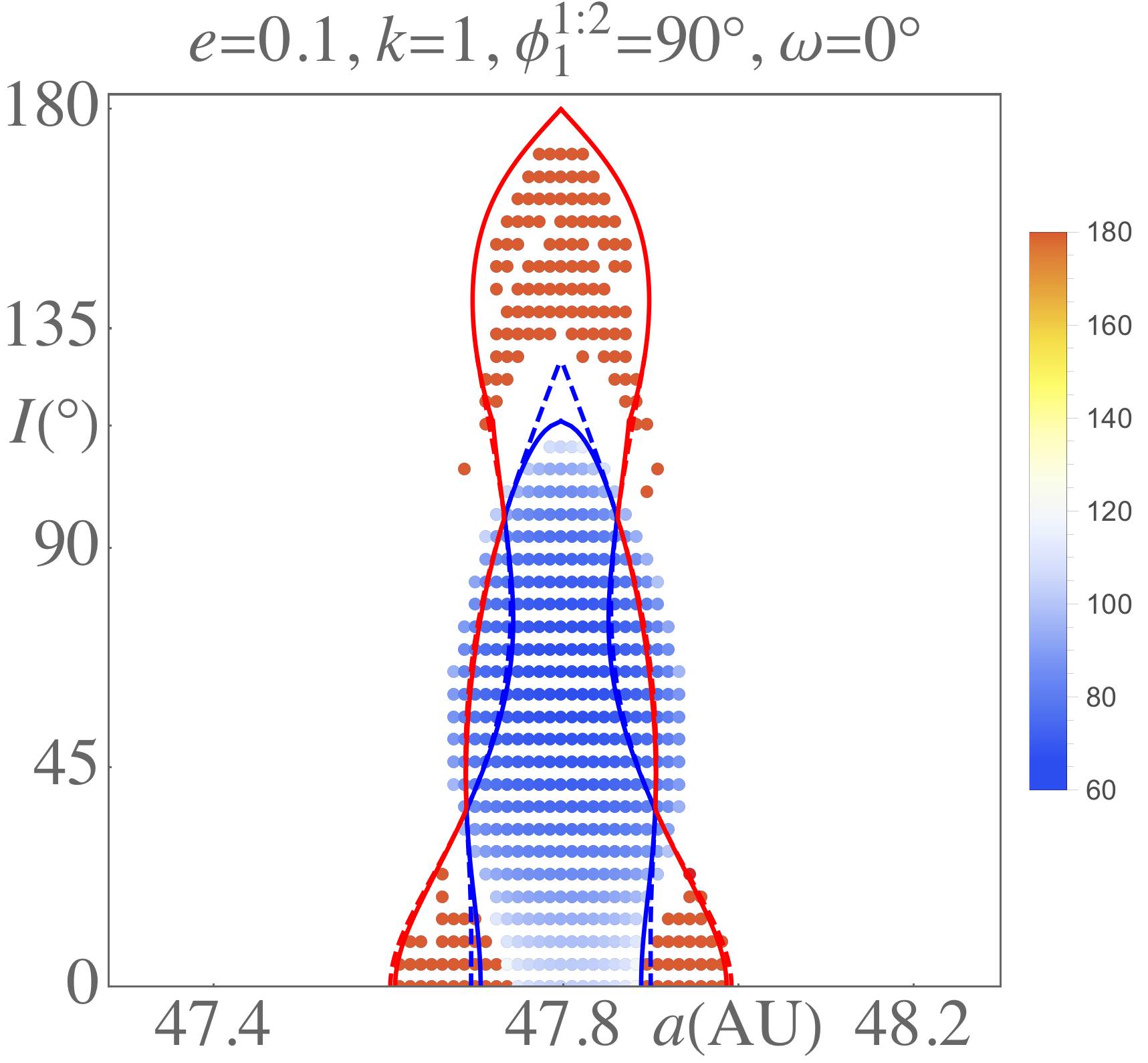}\hspace{4mm}
\includegraphics[width=41mm]{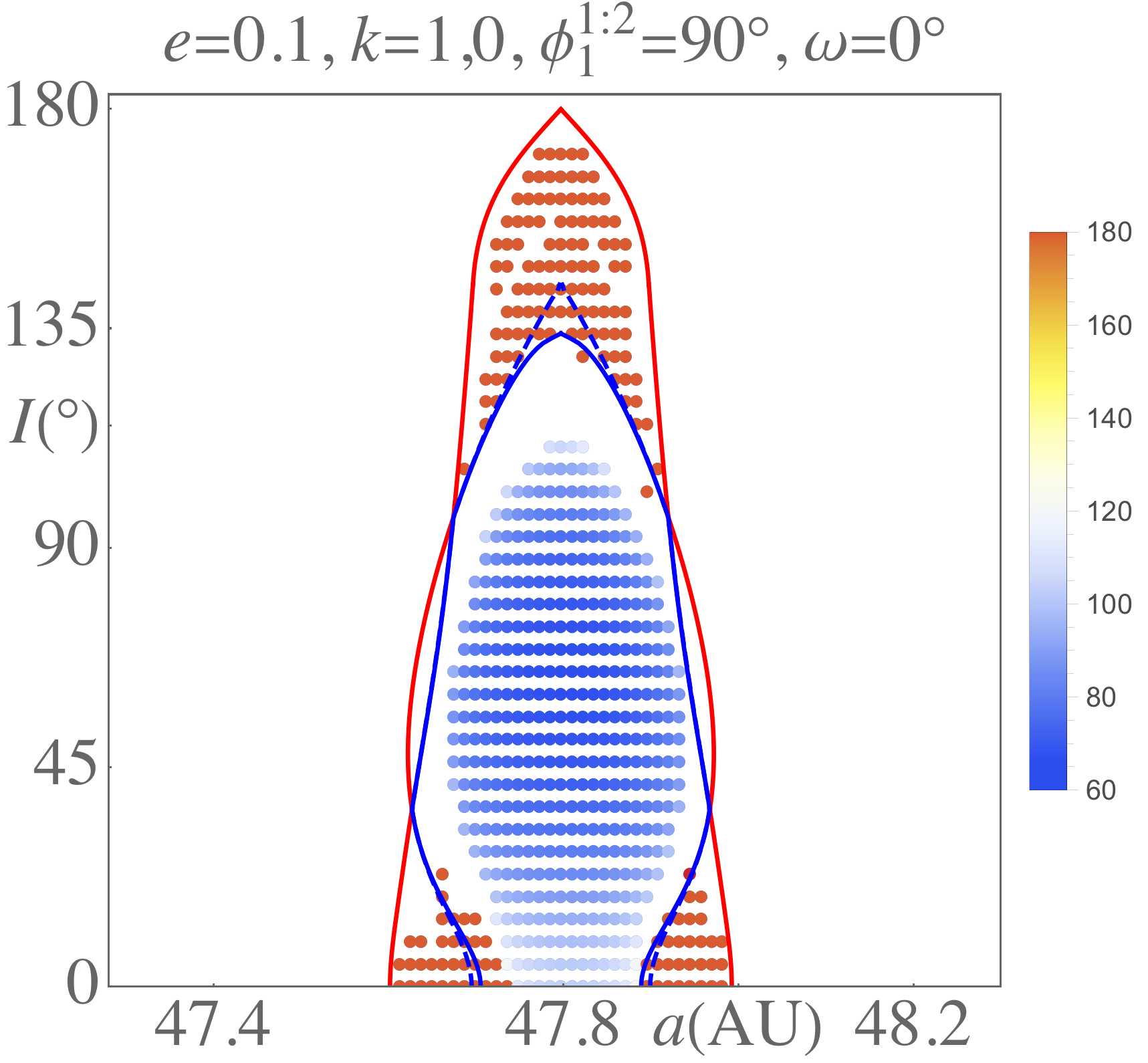}\\
\includegraphics[width=42mm]{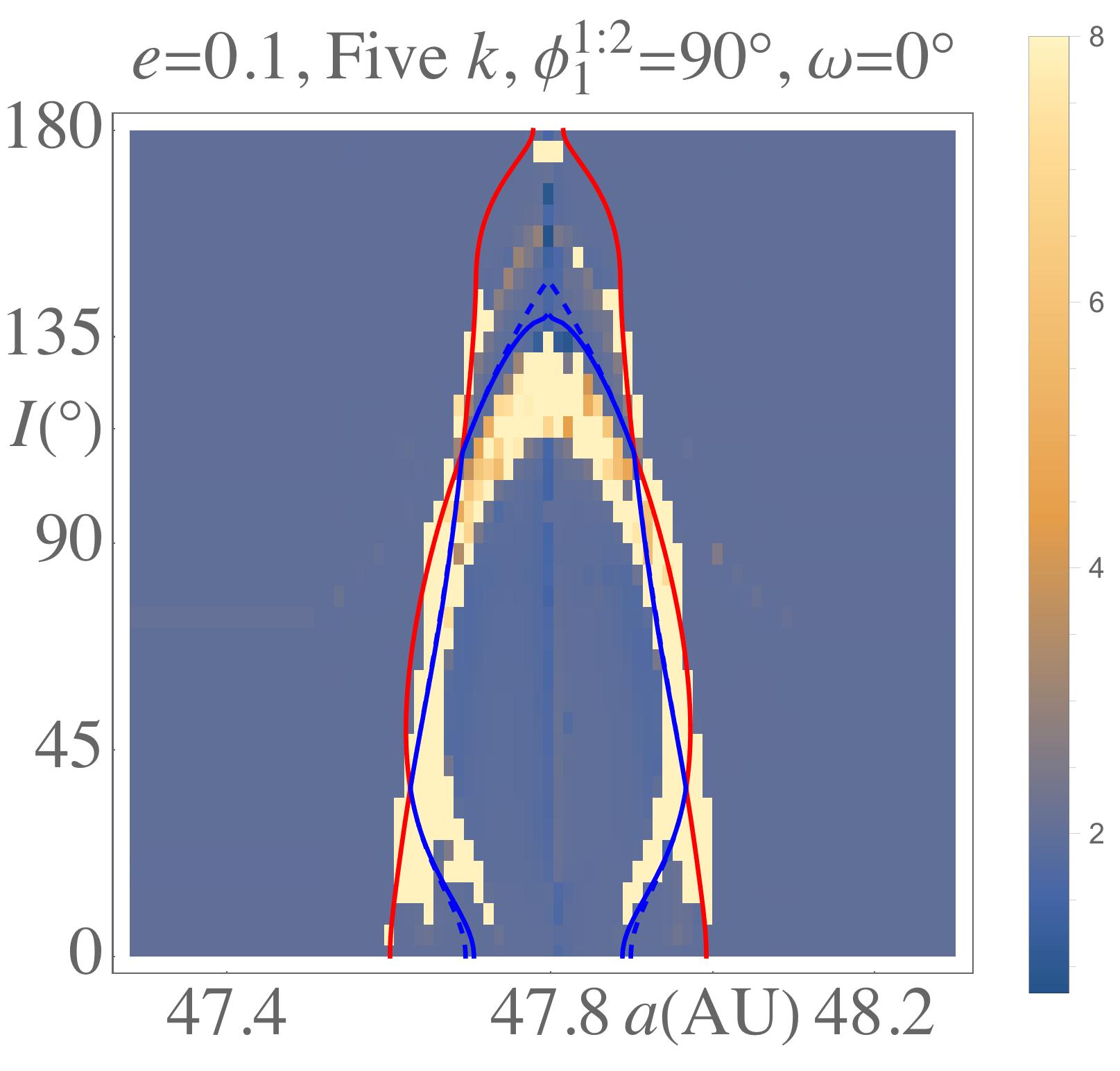}\hspace{2mm}
\includegraphics[width=42mm]{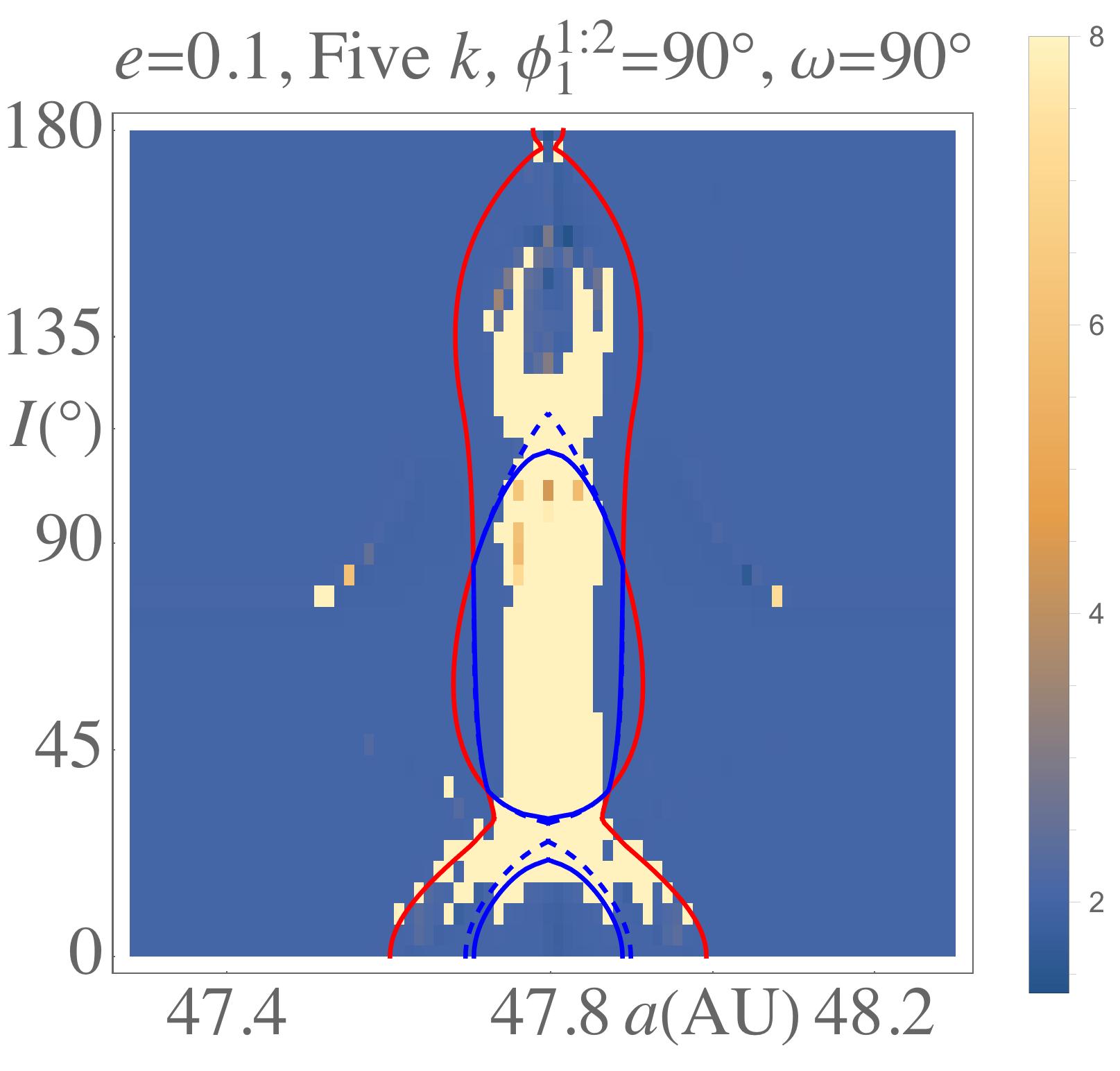}\hspace{2mm}
\includegraphics[width=42mm]{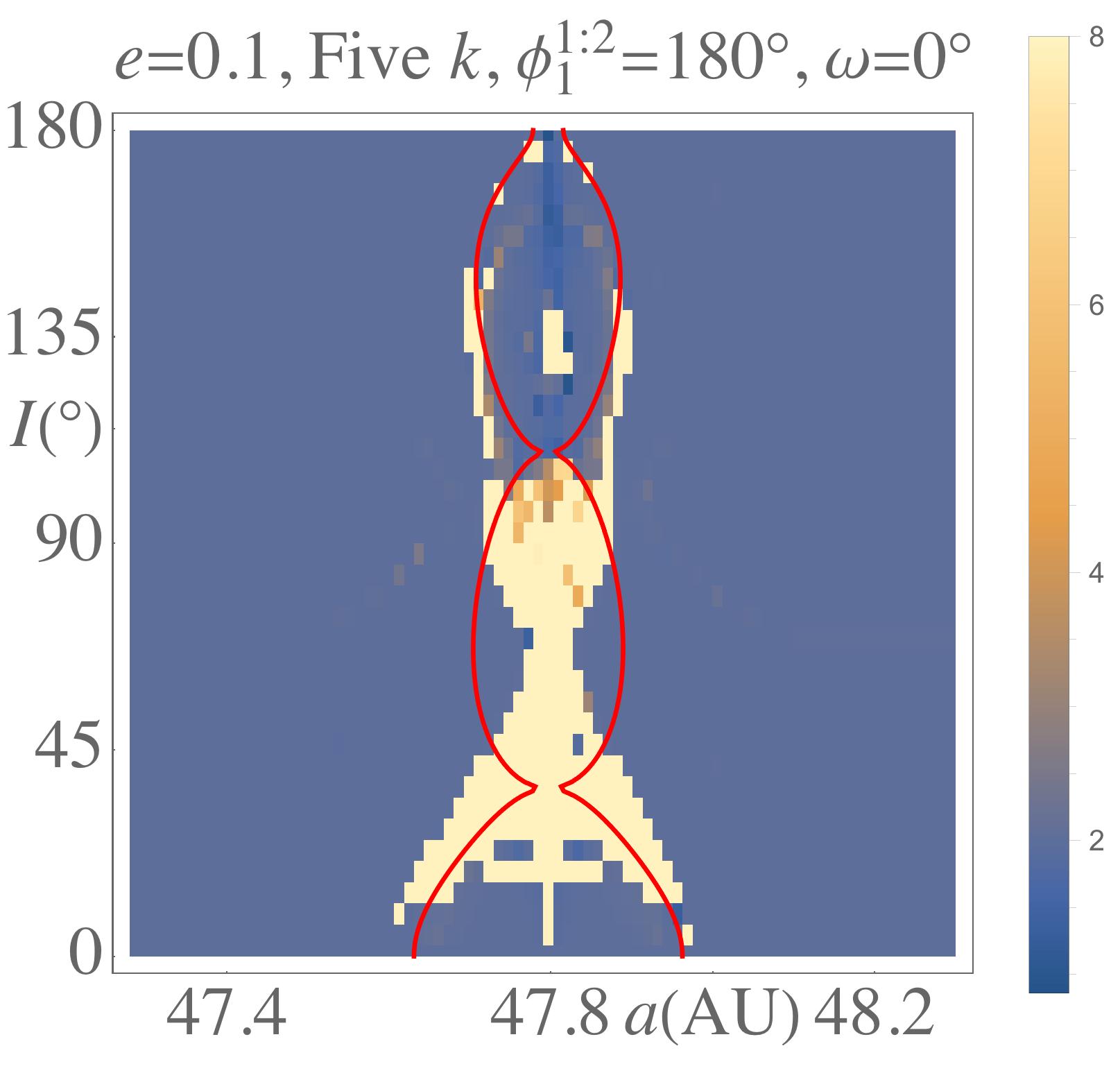}\hspace{2mm}
\includegraphics[width=42mm]{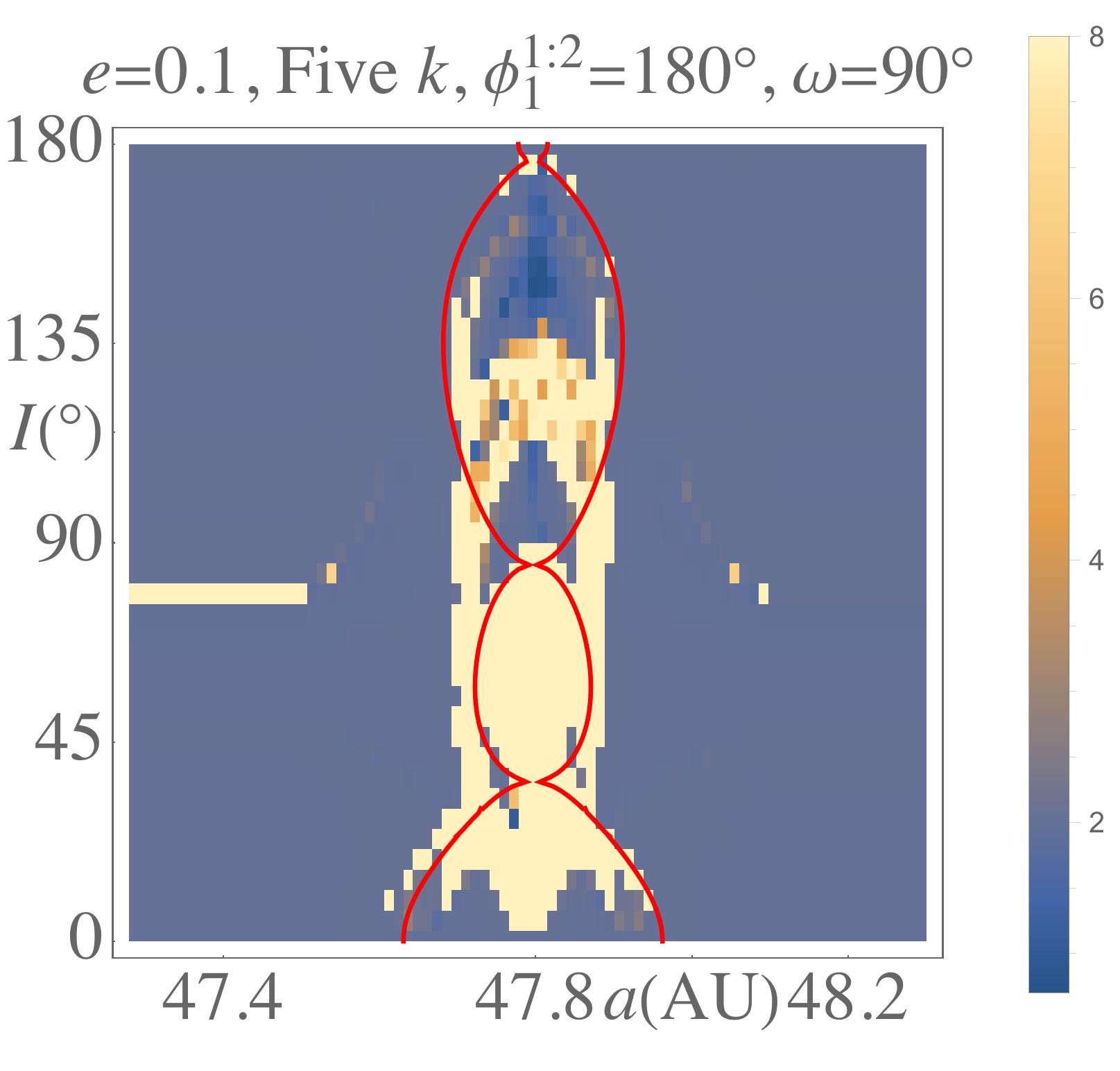}\\
\hspace*{-3mm}\includegraphics[width=41mm]{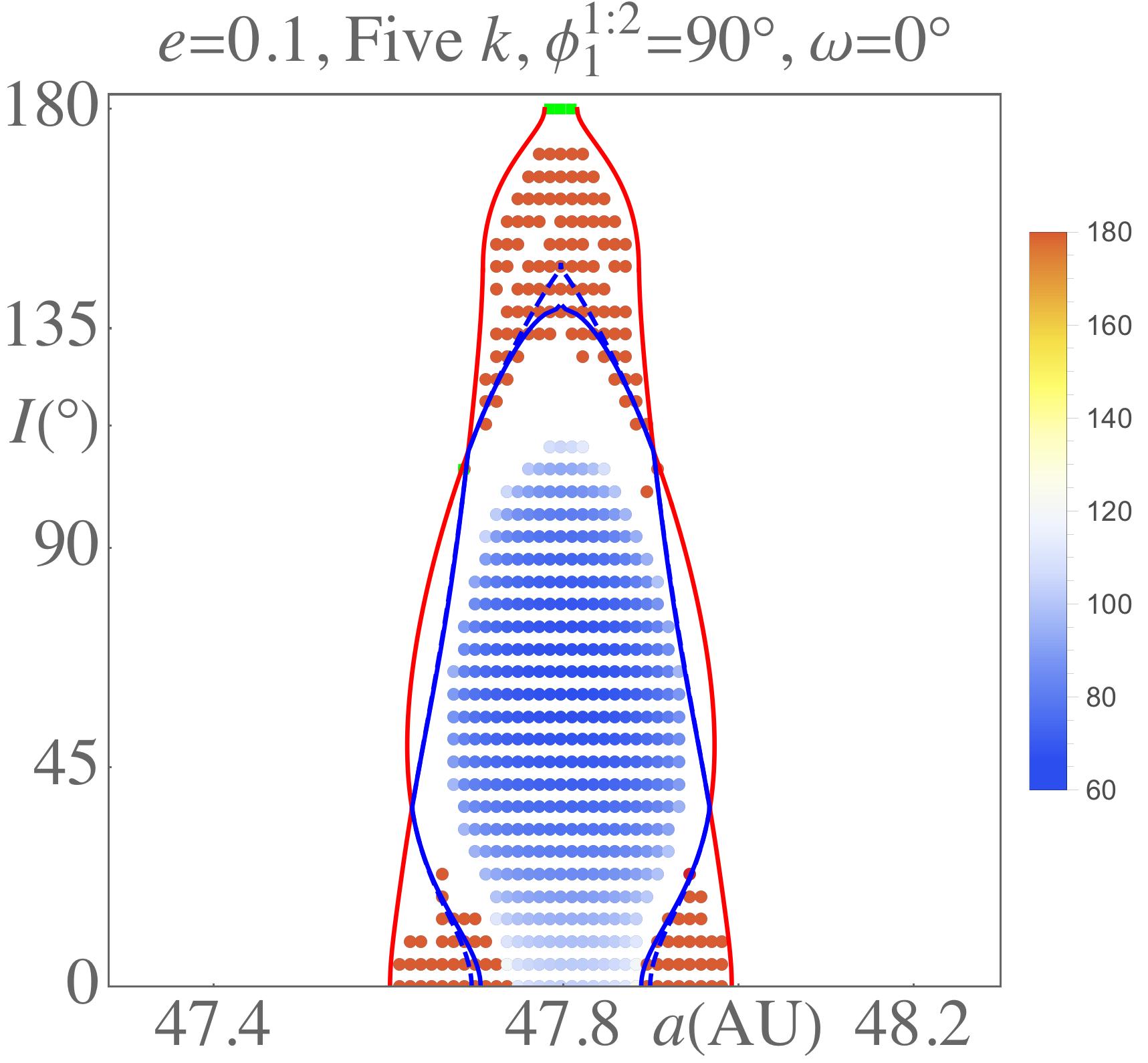}\hspace{3mm}\includegraphics[width=41mm]{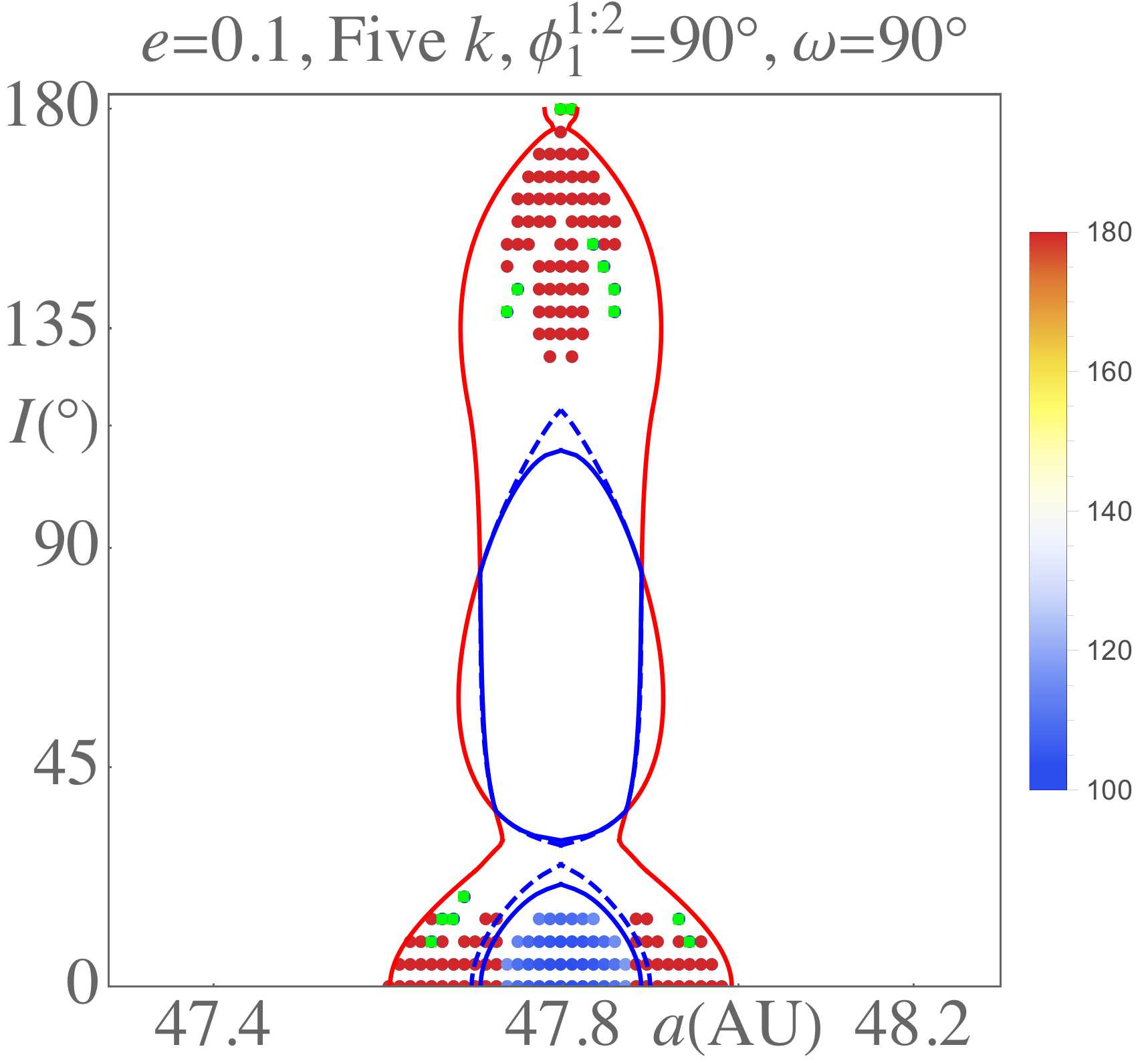}\hspace{3mm}
\includegraphics[width=36mm]{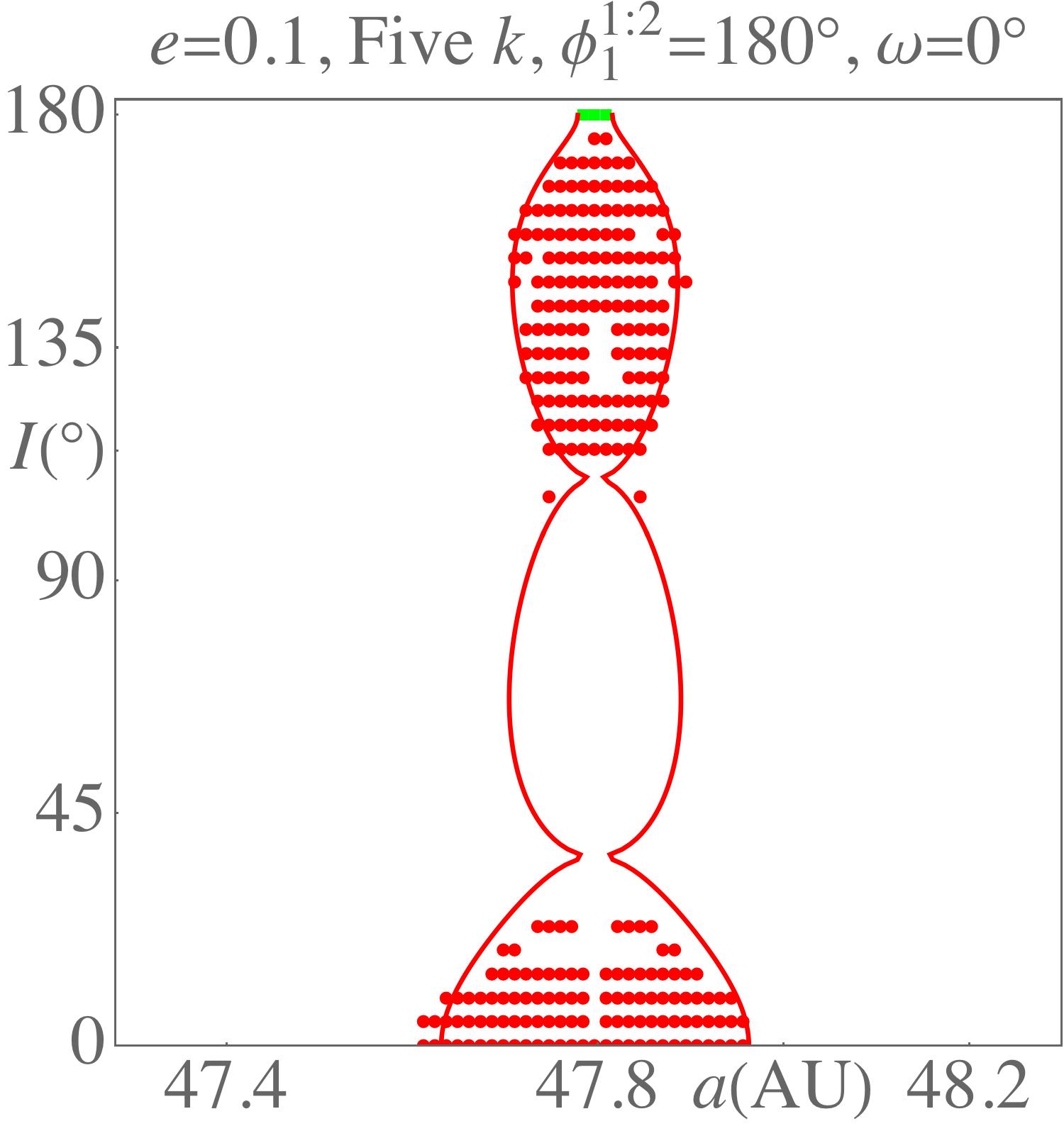}\hspace{8mm}
\includegraphics[width=36mm]{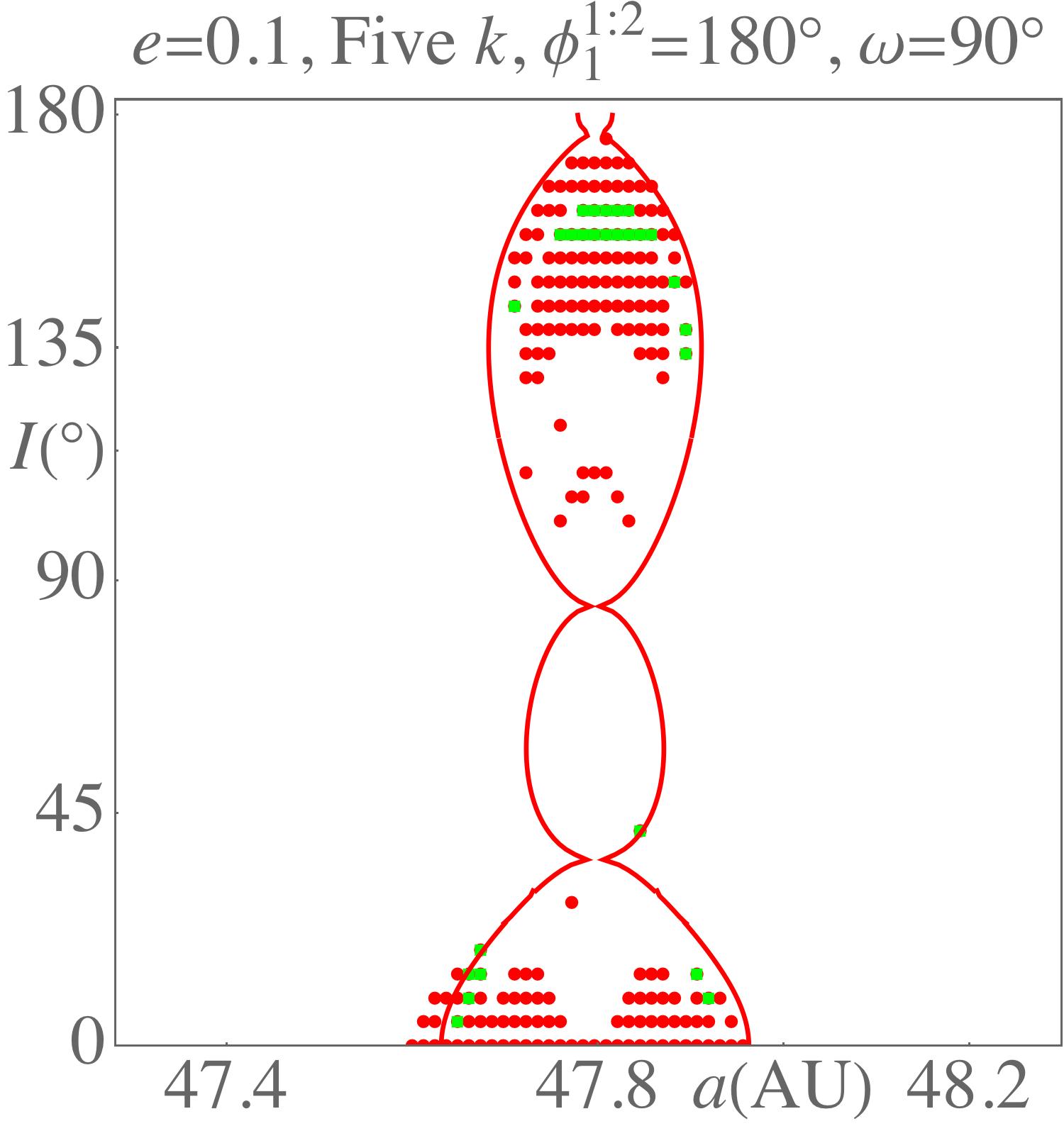}
}
\caption{Outer 1:2 Neptune resonance for $e=0.1$. Initial conditions are shown on top of each panel. Below each {\sc megno}  portrait is the corresponding libration center portrait. Red lines indicate the limit of the libration domain around $180^\circ$ and blue lines indicate the domain of asymmetric librations. Filled circles denote the location of the libration center of $\phi^{1:2}_1$ indicated on the color scale. Green filled  squares denote the librations of the retrograde mode $k=3$. `Five $k$' indicates the set $\{k: 1,-1,3,-3,0\}$.}\label{fN1t2e1}
\end{center}
\end{figure*}

\begin{figure*}
\begin{center}
{ 
\hspace*{-3mm}\includegraphics[width=36mm]{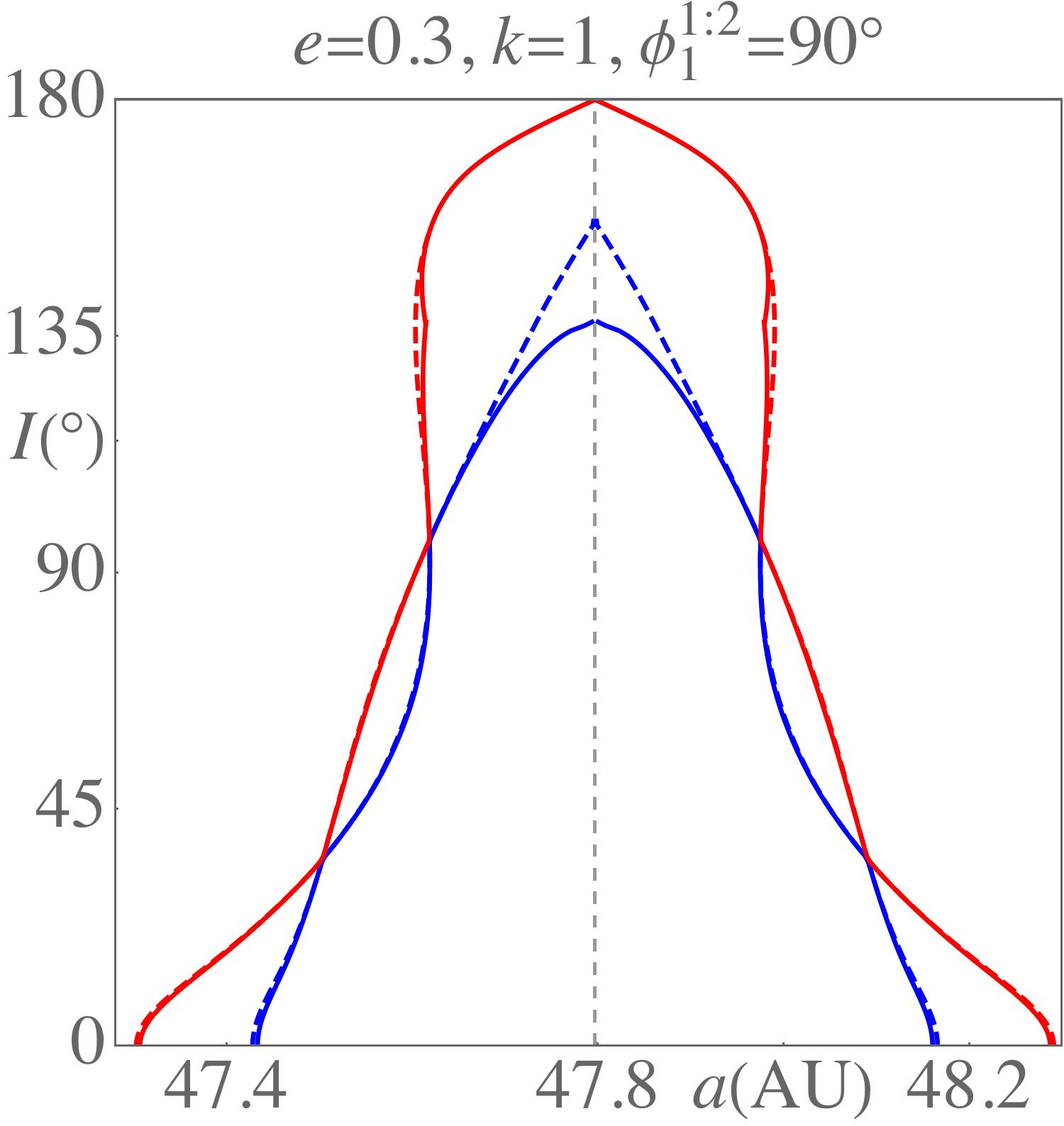}\hspace{8mm}
\includegraphics[width=36mm]{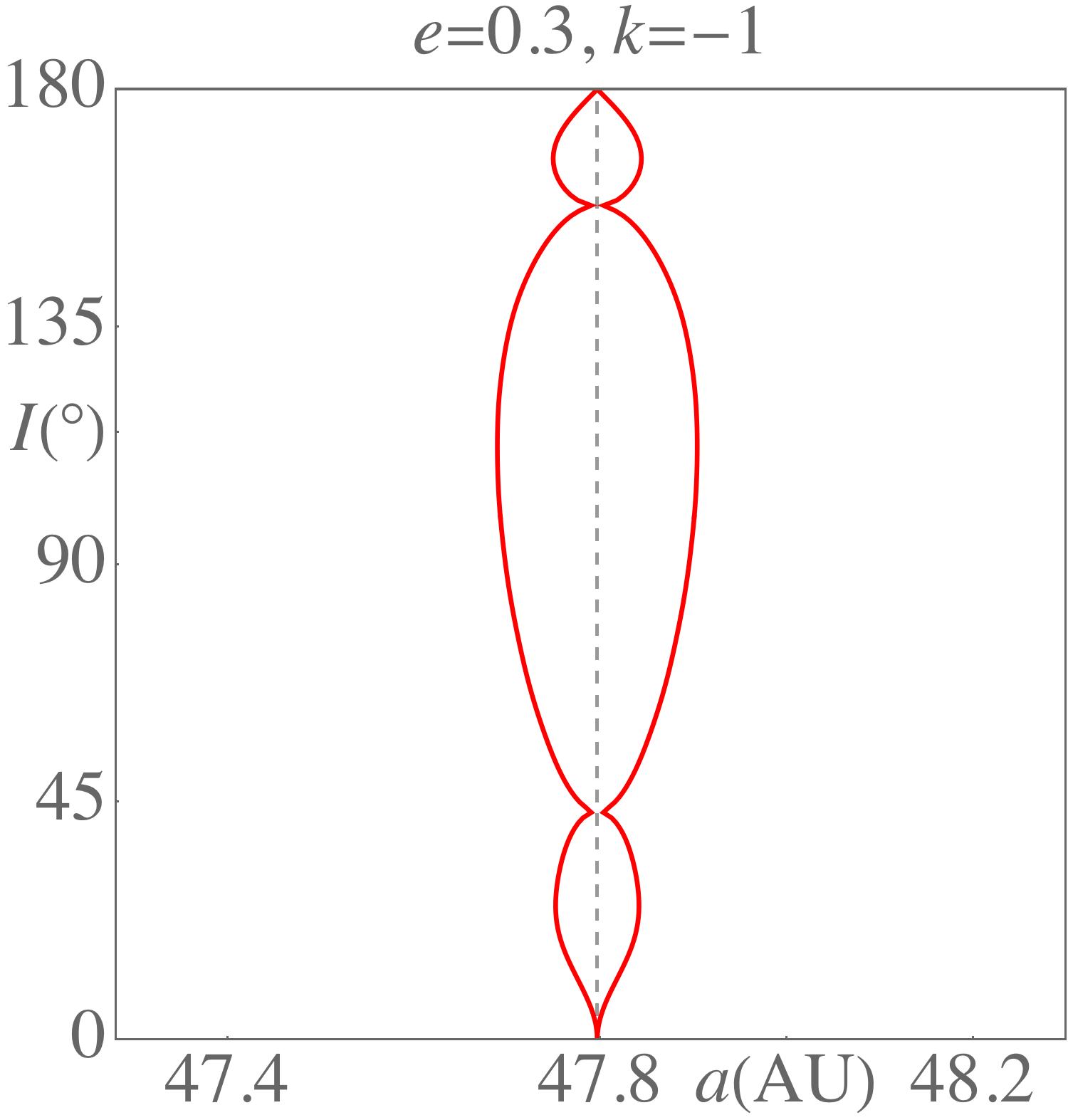}\hspace{8mm}
\includegraphics[width=36mm]{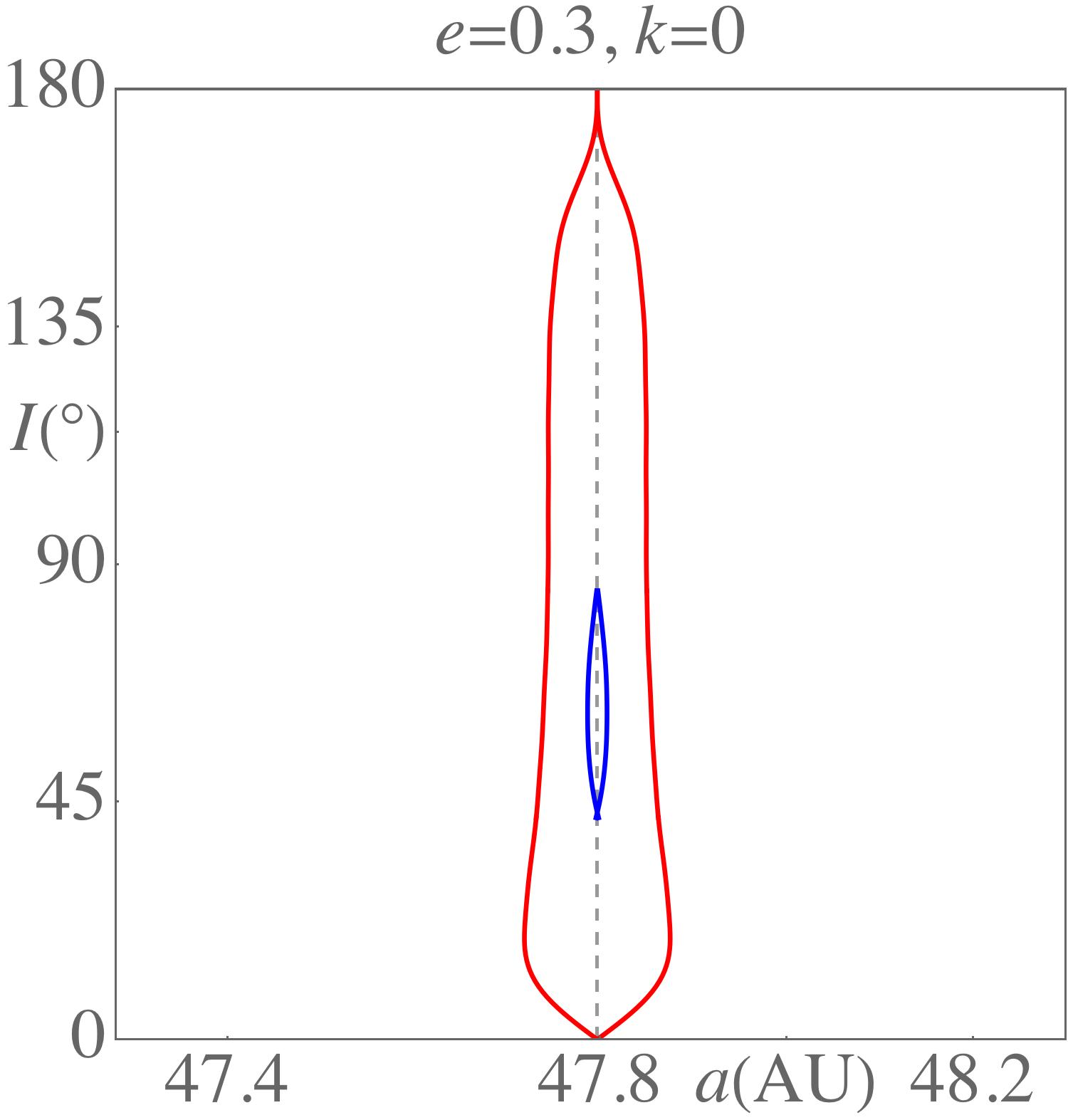}\hspace{8mm}
\includegraphics[width=36mm]{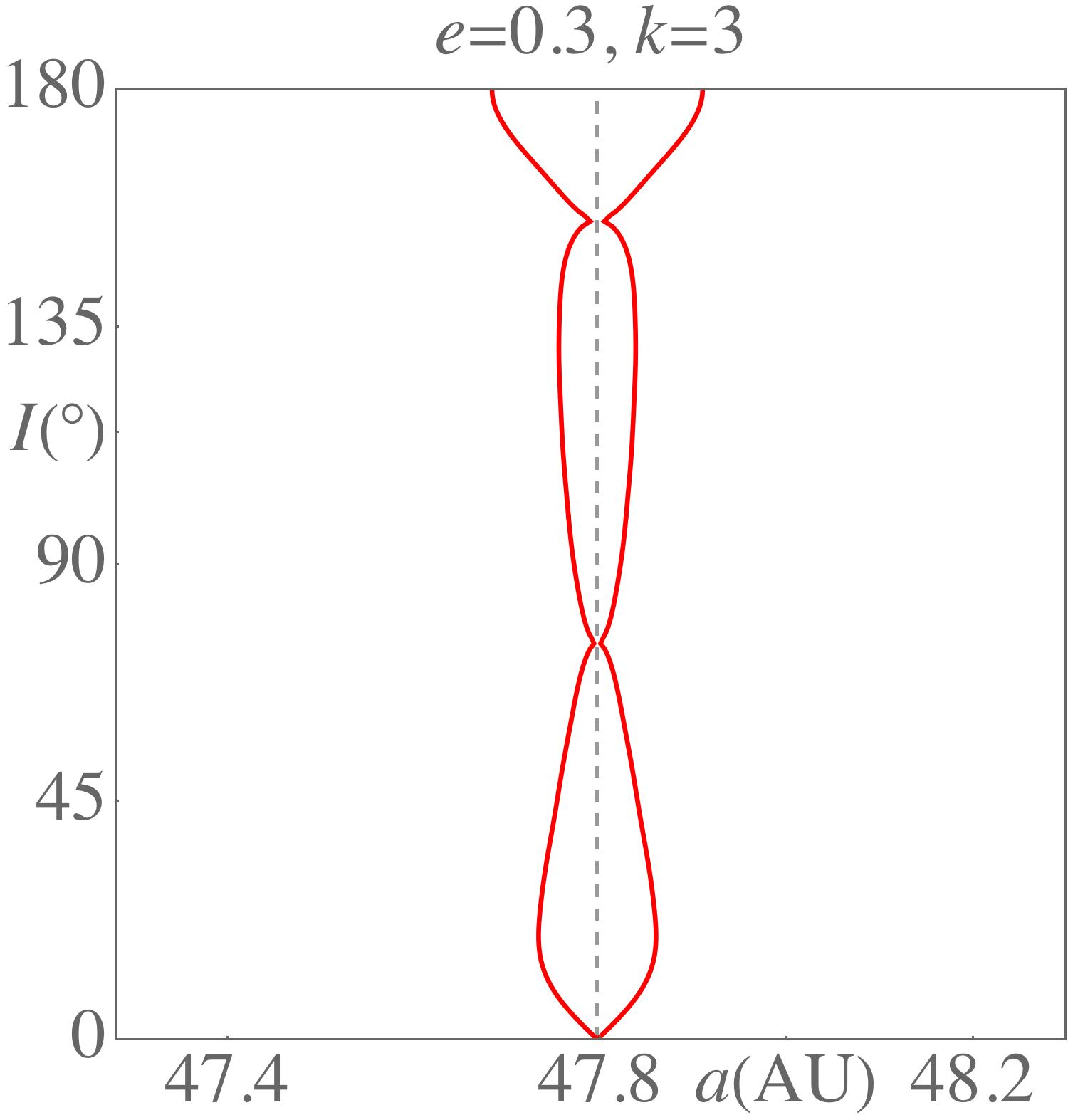}\\
\includegraphics[width=42mm]{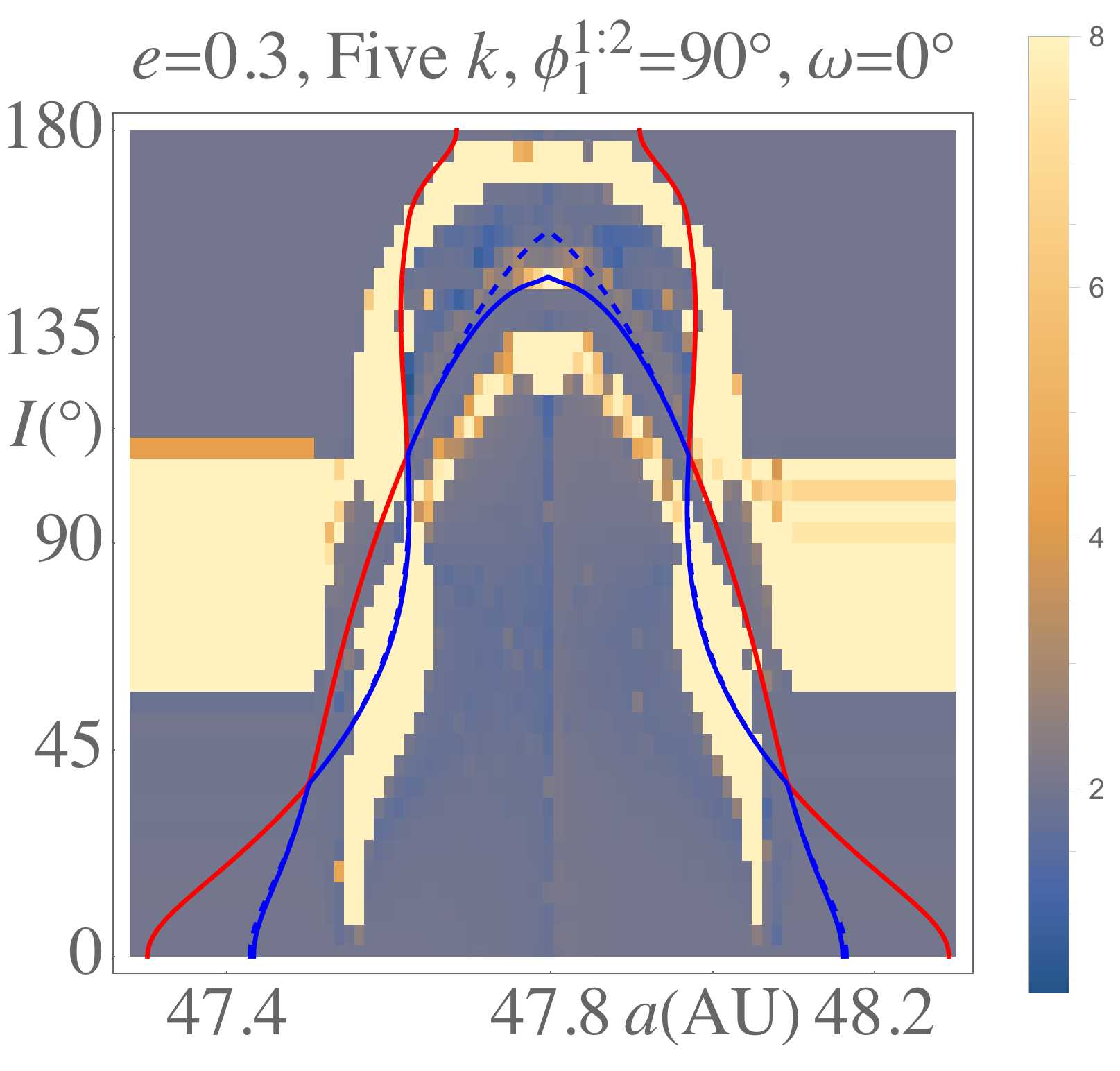}\hspace{2mm}
\includegraphics[width=42mm]{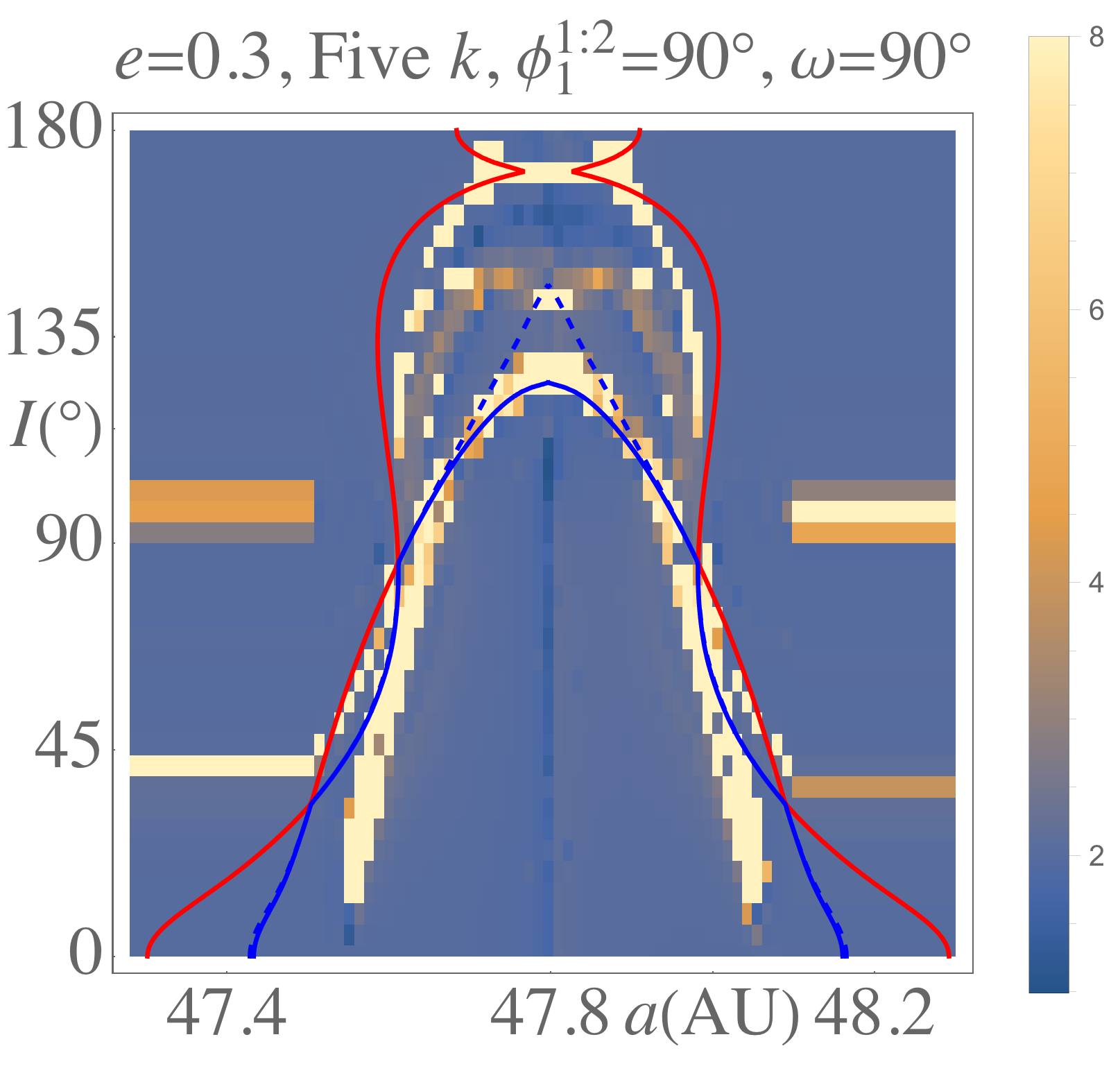}\hspace{2mm}
\includegraphics[width=42mm]{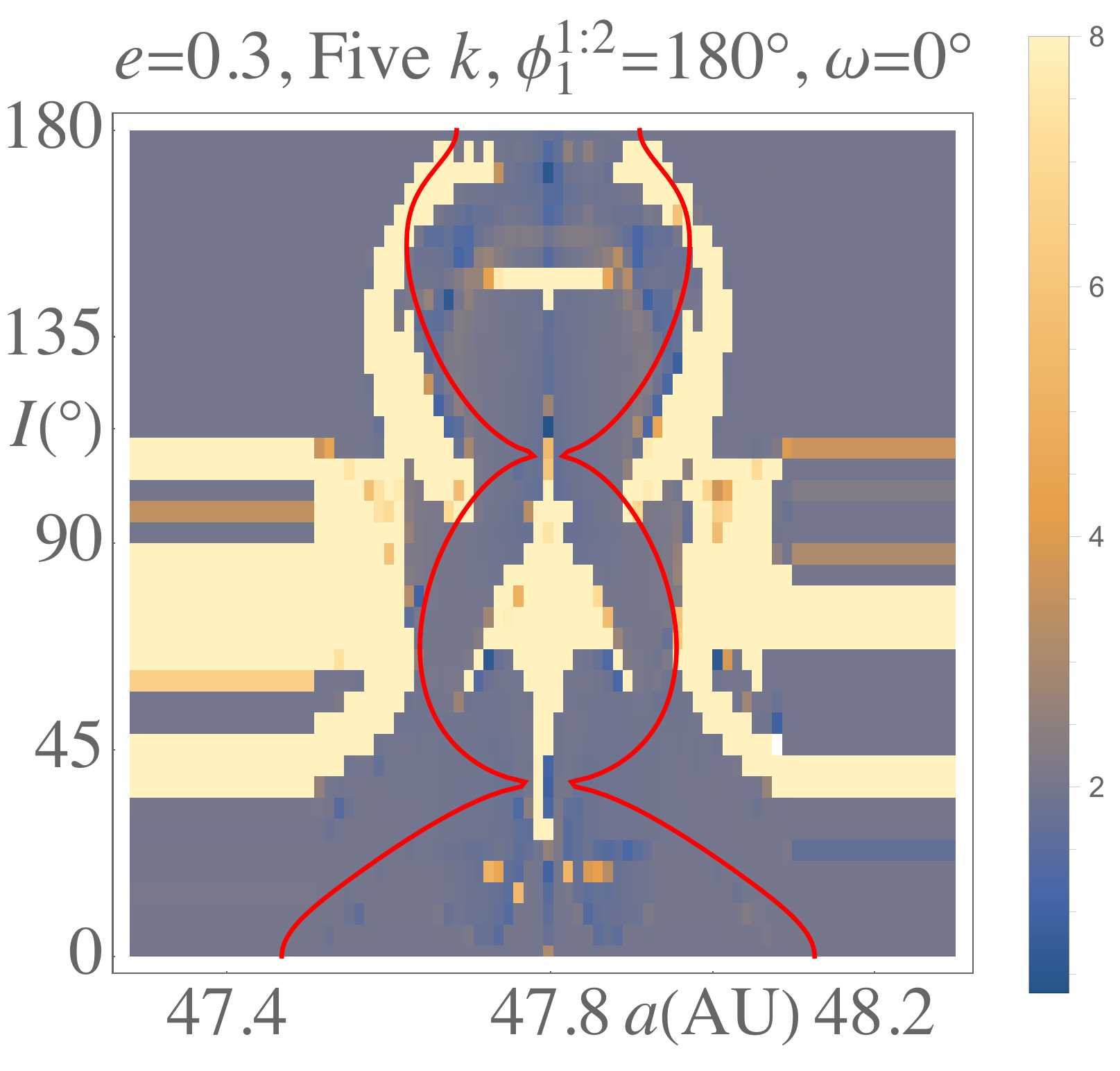}\hspace{2mm}
\includegraphics[width=42mm]{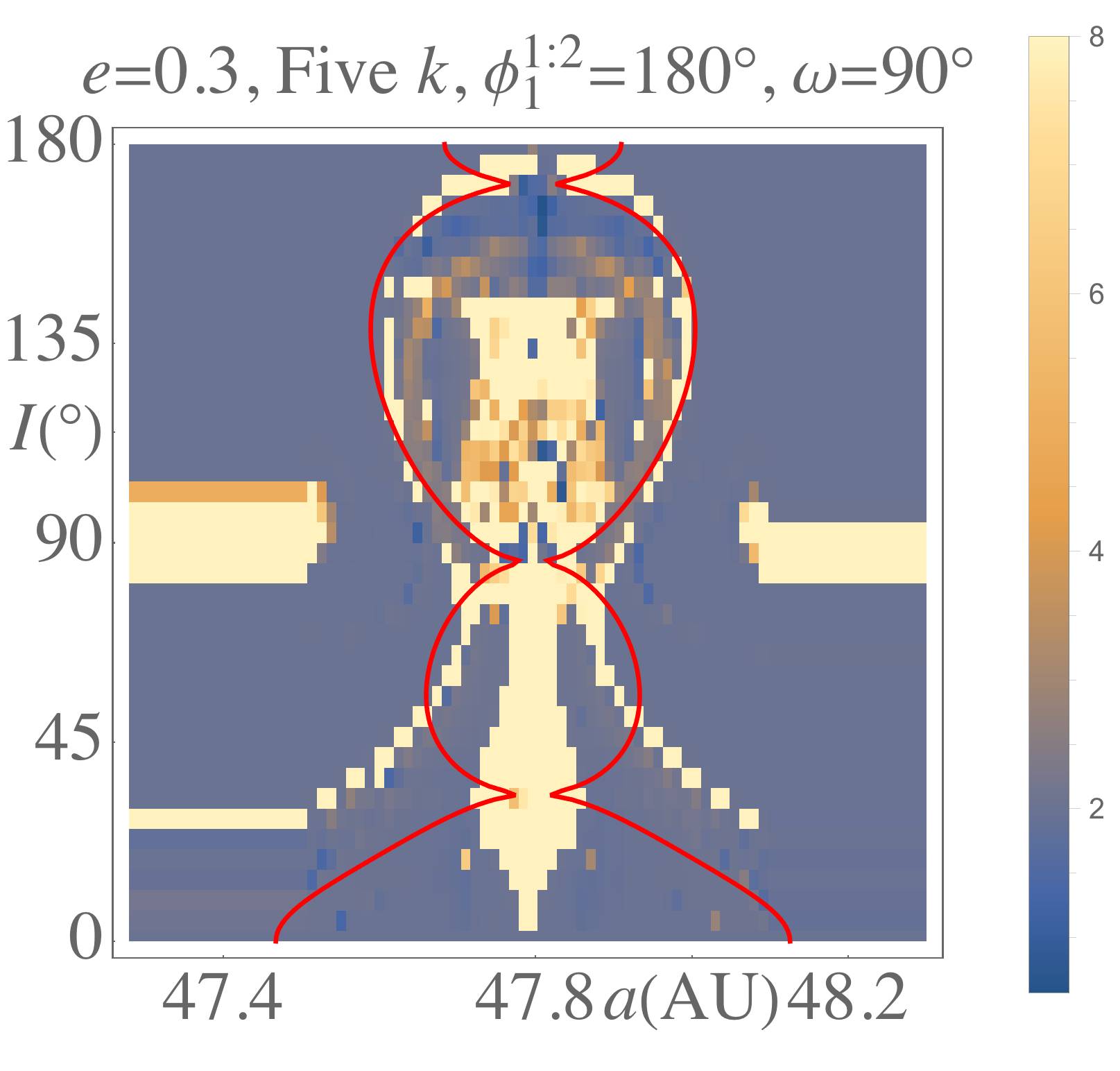}\\
\hspace*{-3mm}\includegraphics[width=41mm]{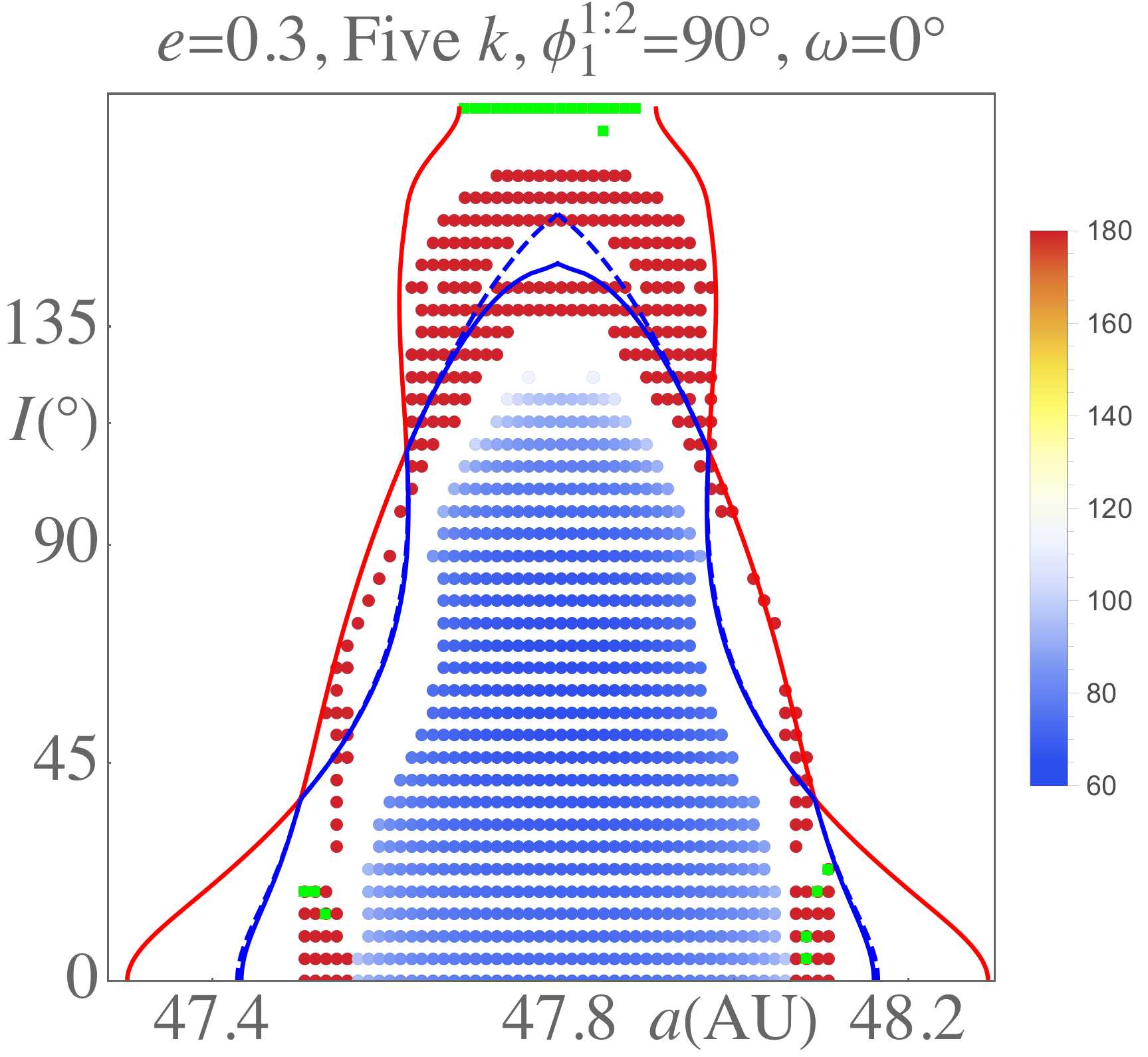}\hspace{3mm}\includegraphics[width=41mm]{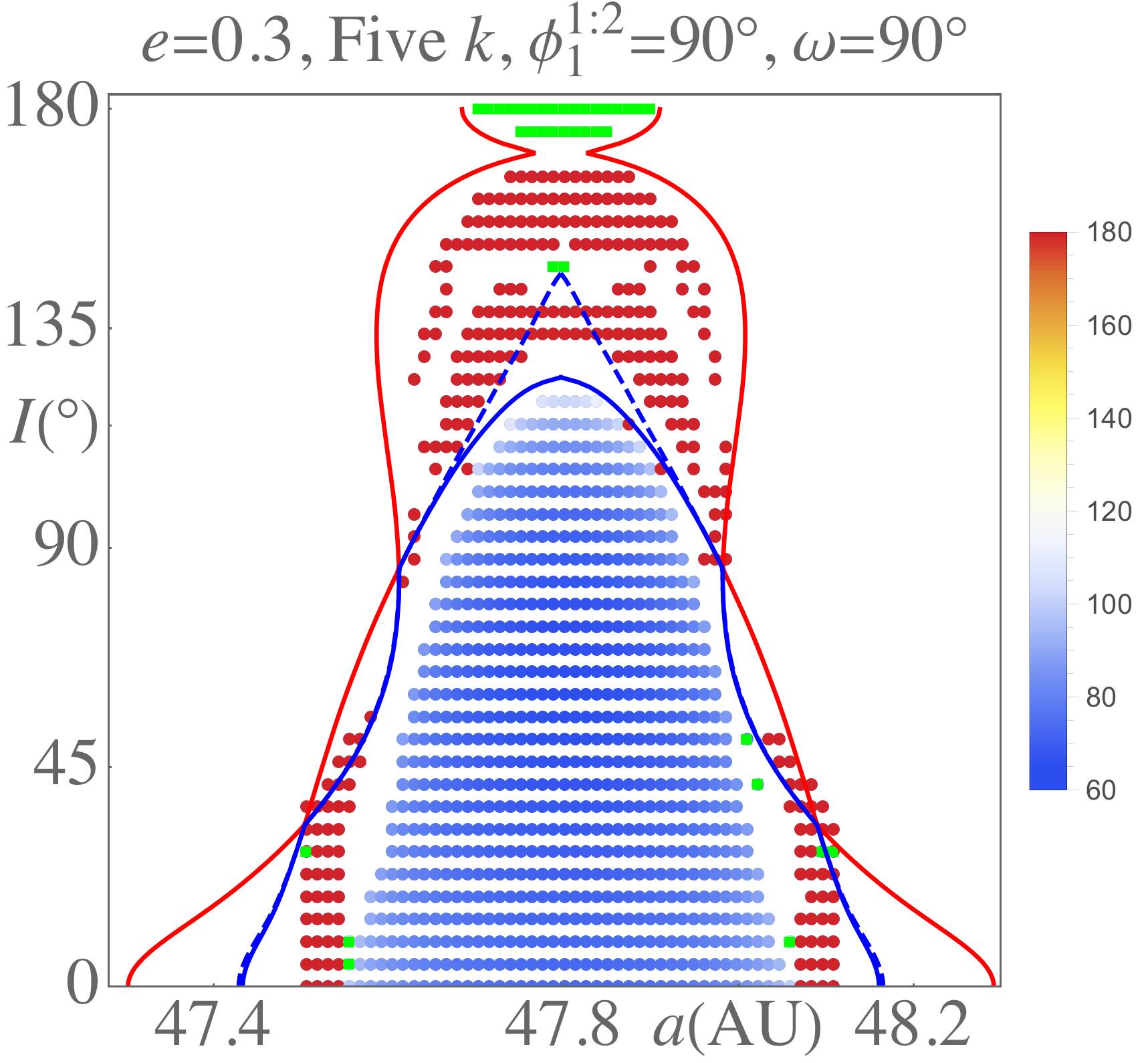}\hspace{3mm}\includegraphics[width=36mm]{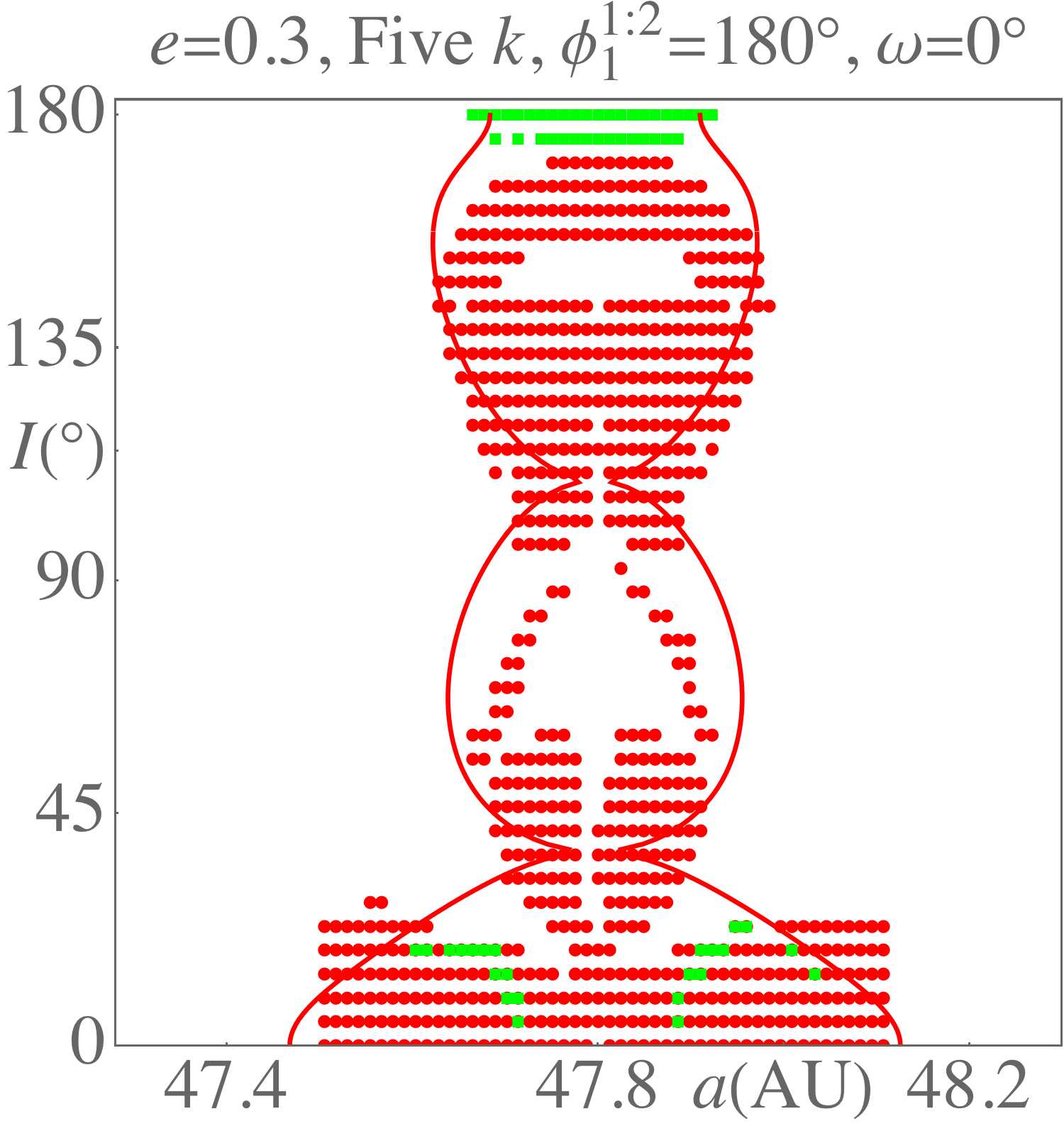}\hspace{8mm}\includegraphics[width=36mm]{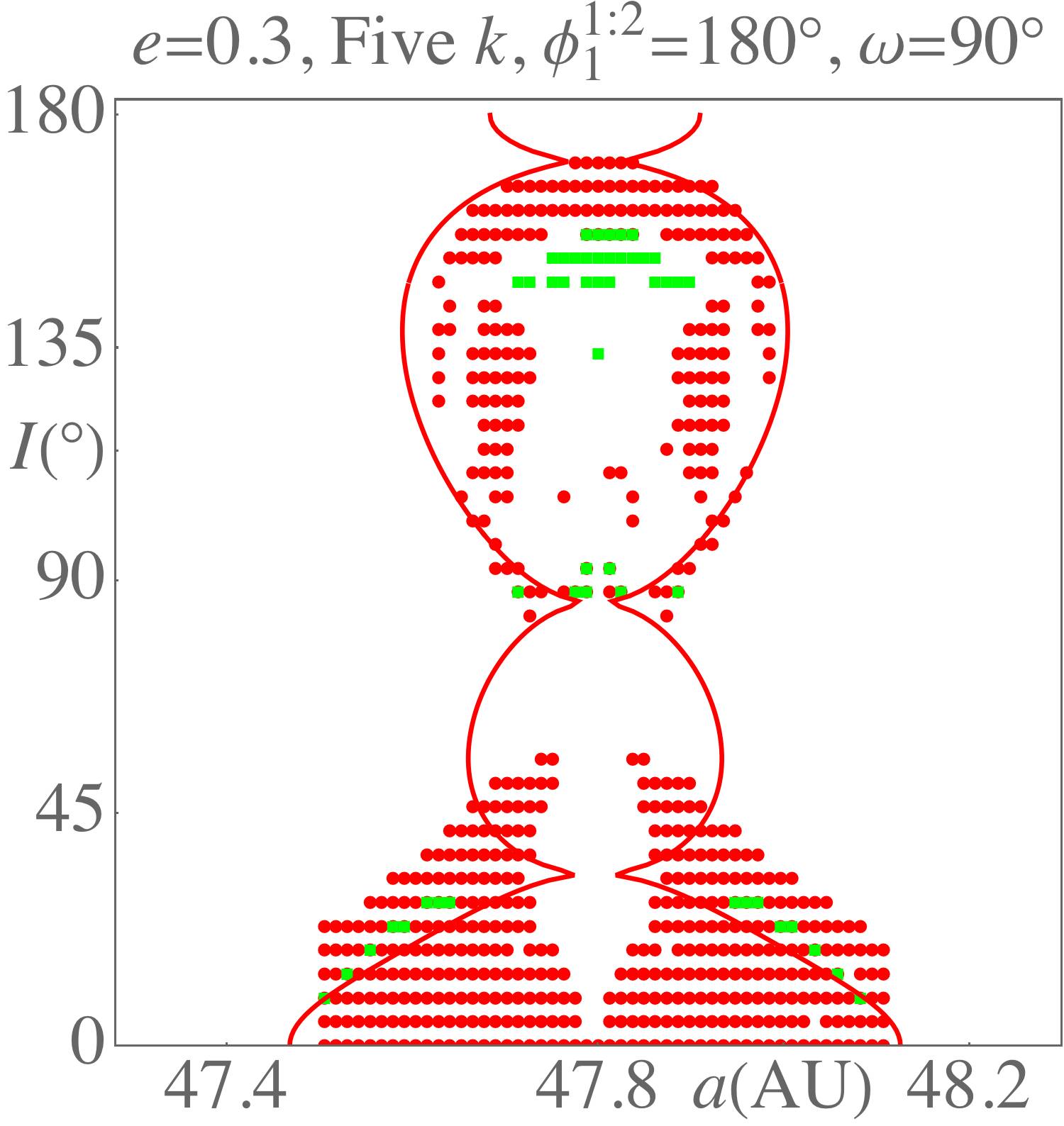}
}
\caption{Outer 1:2 Neptune resonance with $e=0.3$. Symbols and color codes are those of Figure \ref{fN1t2e1}.}\label{fN1t2e3}
\end{center}
\end{figure*}

\begin{figure*}
\begin{center}
{ 
\hspace*{-2mm}\includegraphics[width=36mm]{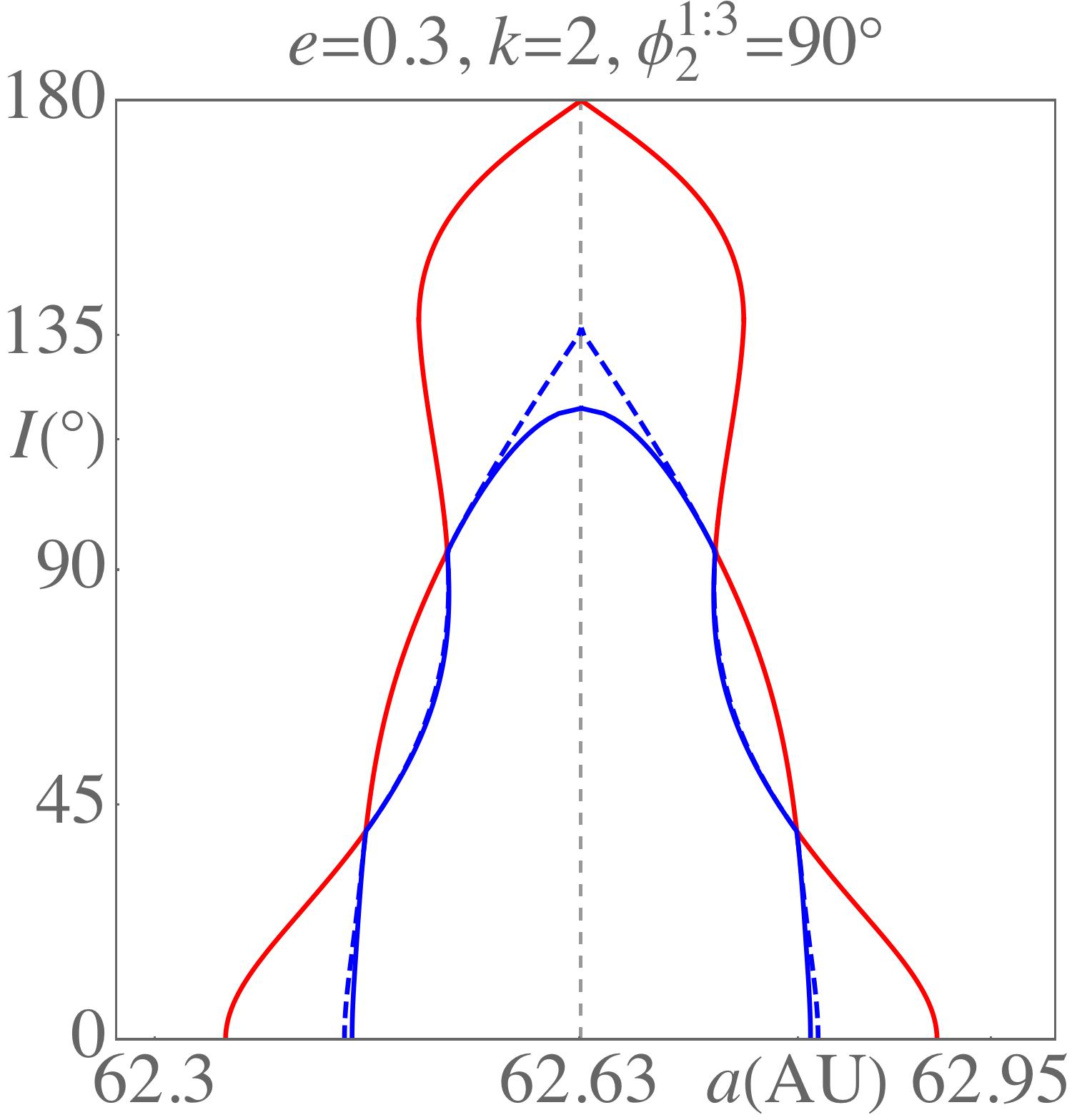}\hspace{8mm}
\includegraphics[width=36mm]{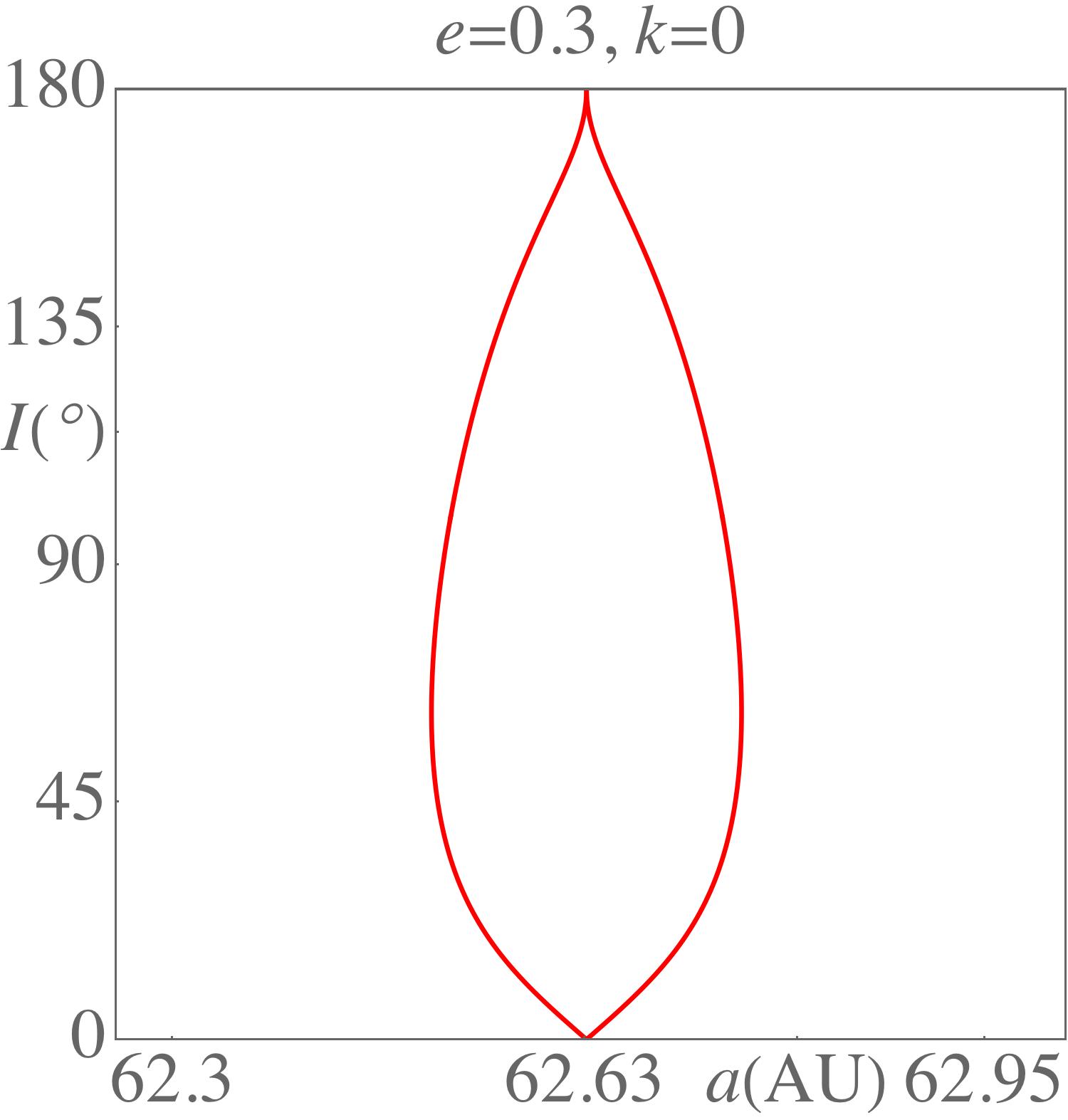}\hspace{8mm}
\includegraphics[width=36mm]{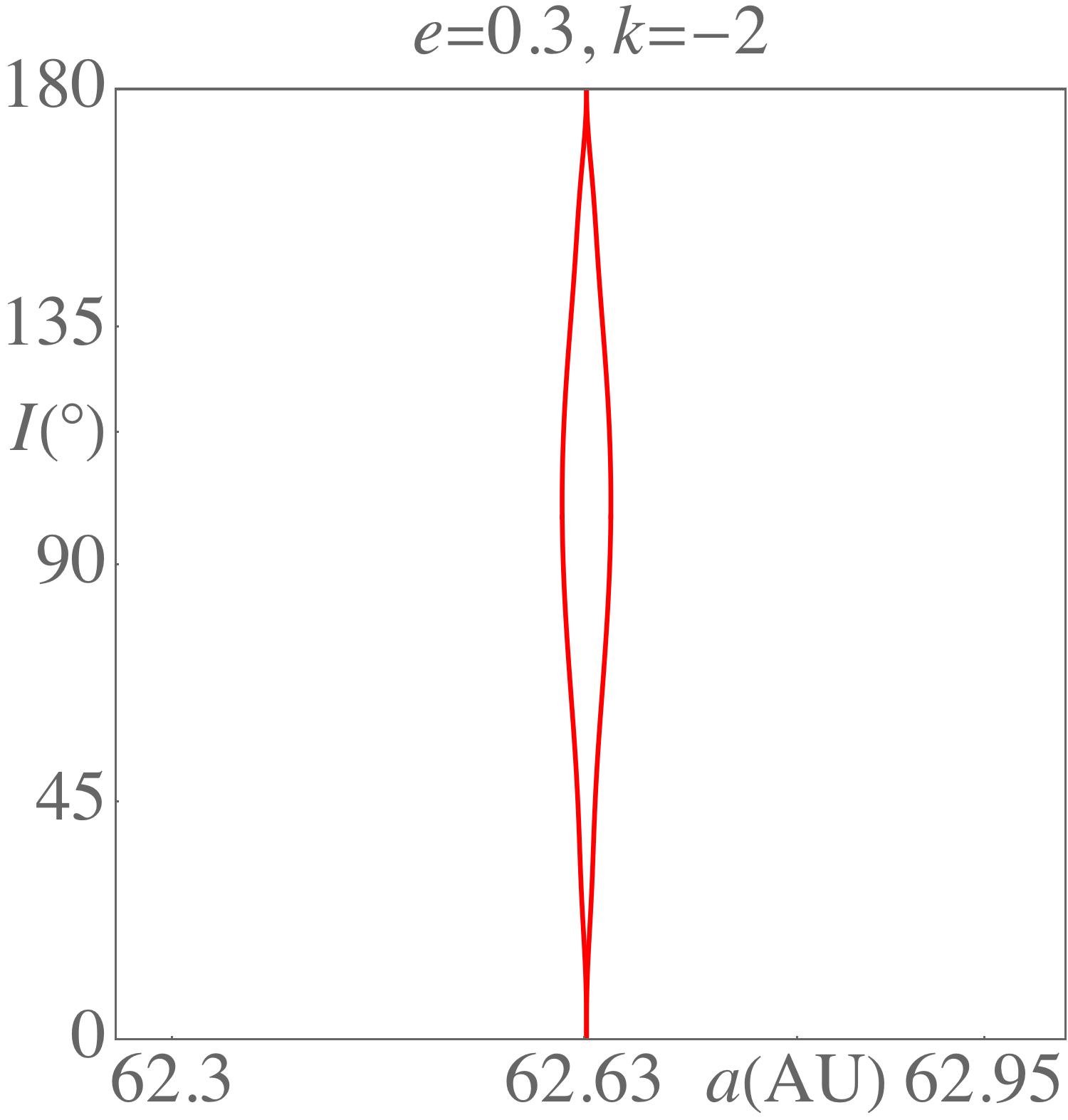}\hspace{8mm}
\includegraphics[width=36mm]{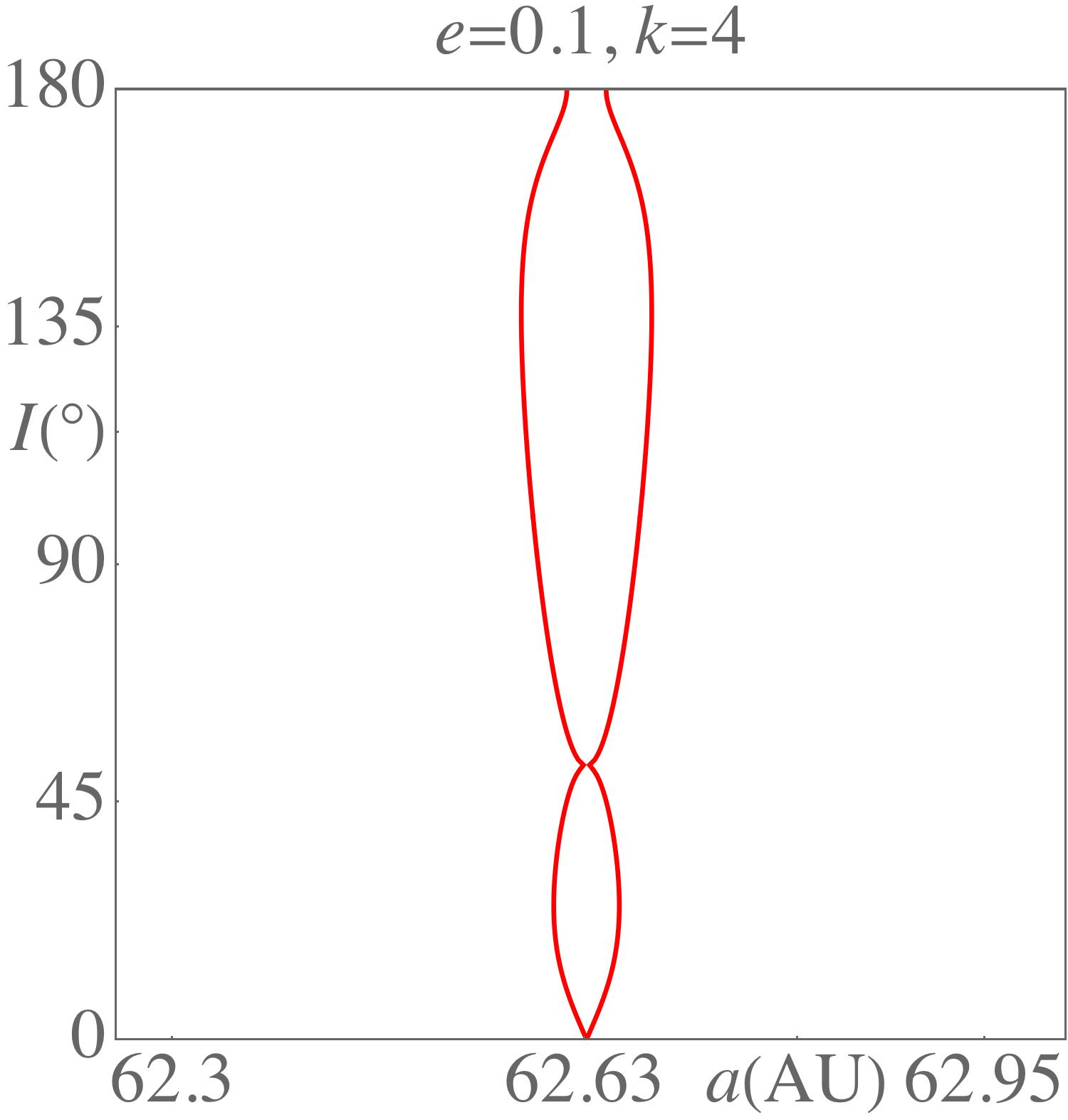}\\
\includegraphics[width=42mm]{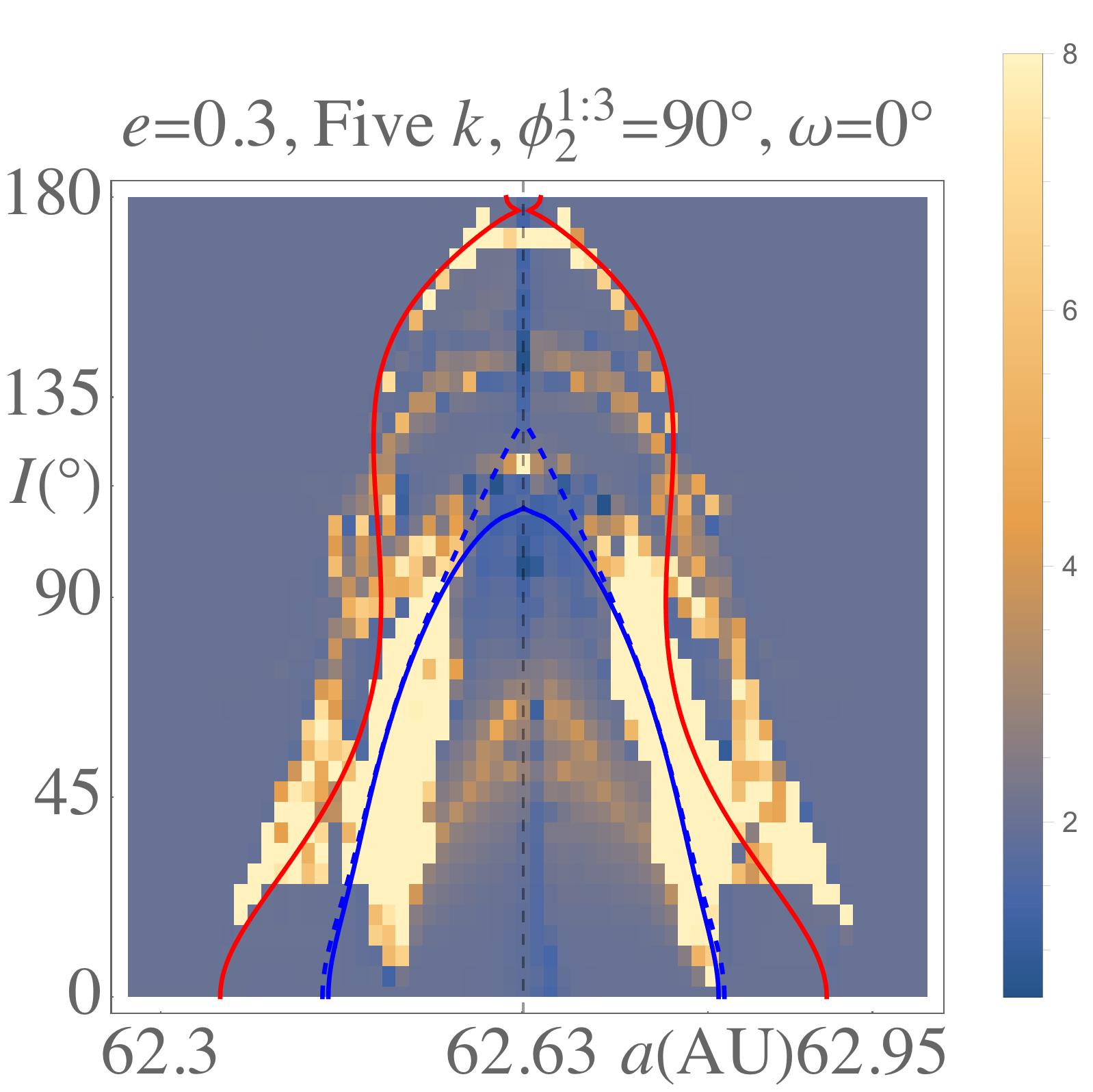}\hspace{2mm}
\includegraphics[width=42mm]{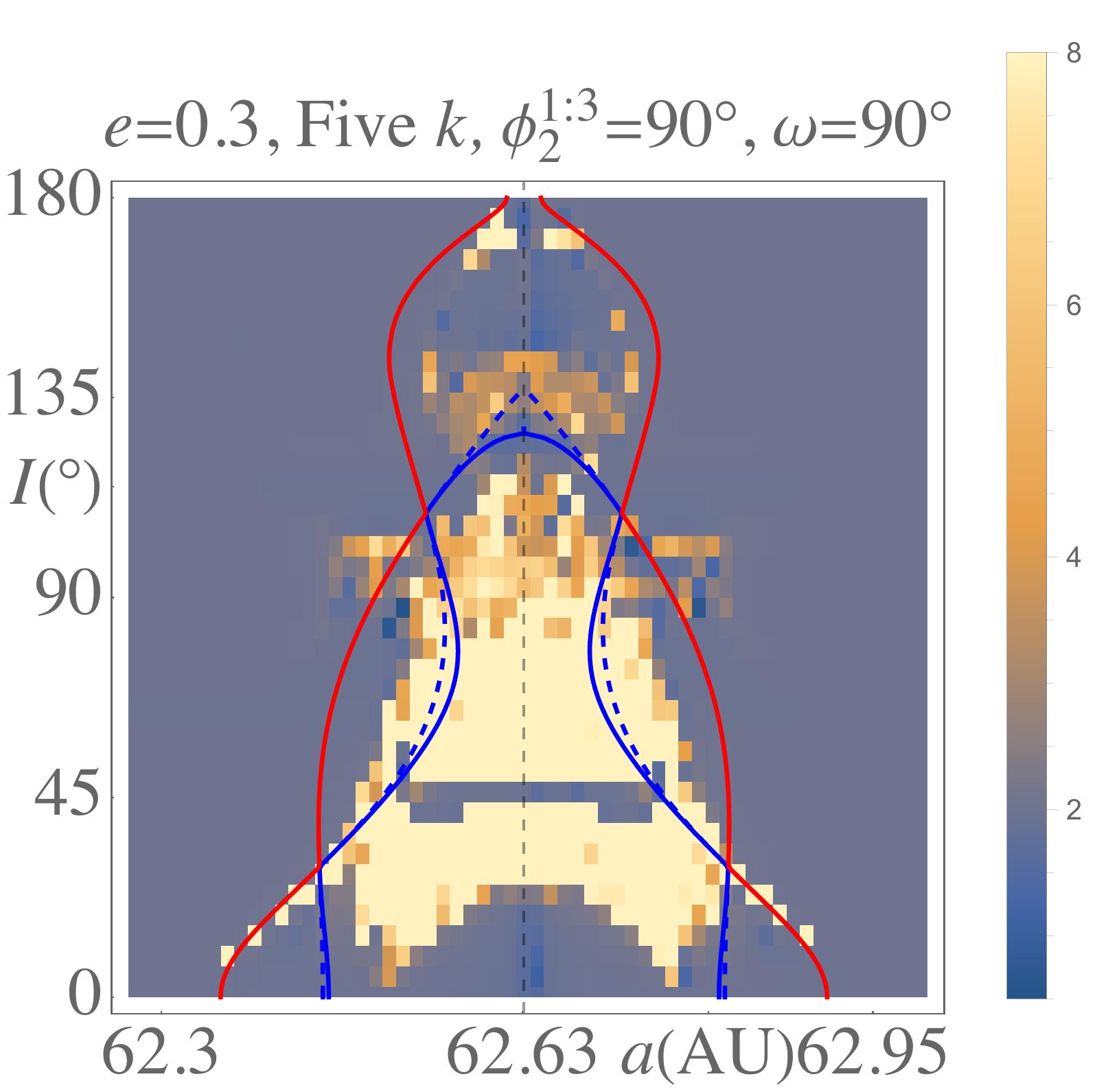}\hspace{1.5mm}
\includegraphics[width=42.5mm]{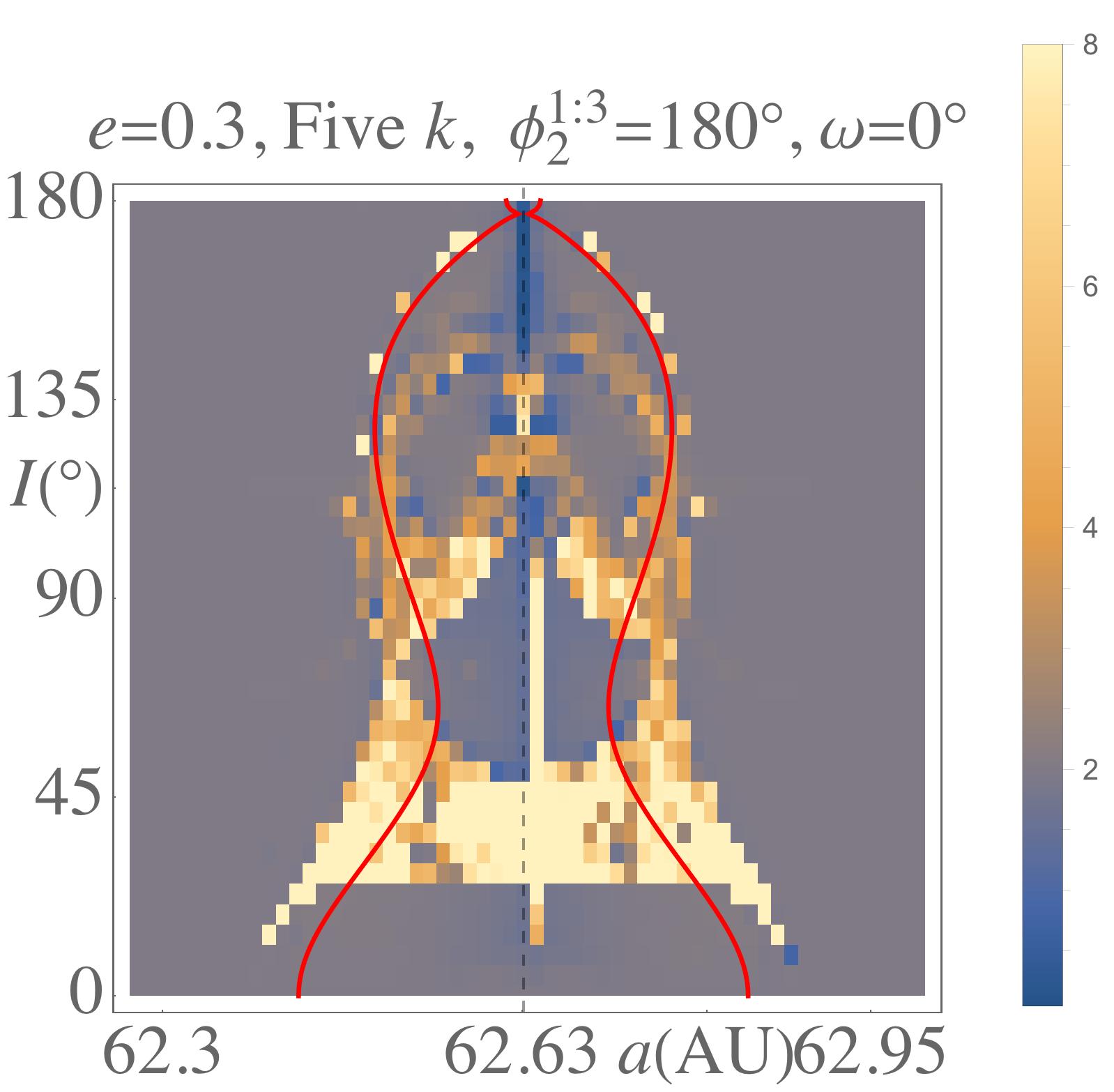}\hspace{1.5mm}
\includegraphics[width=42.5mm]{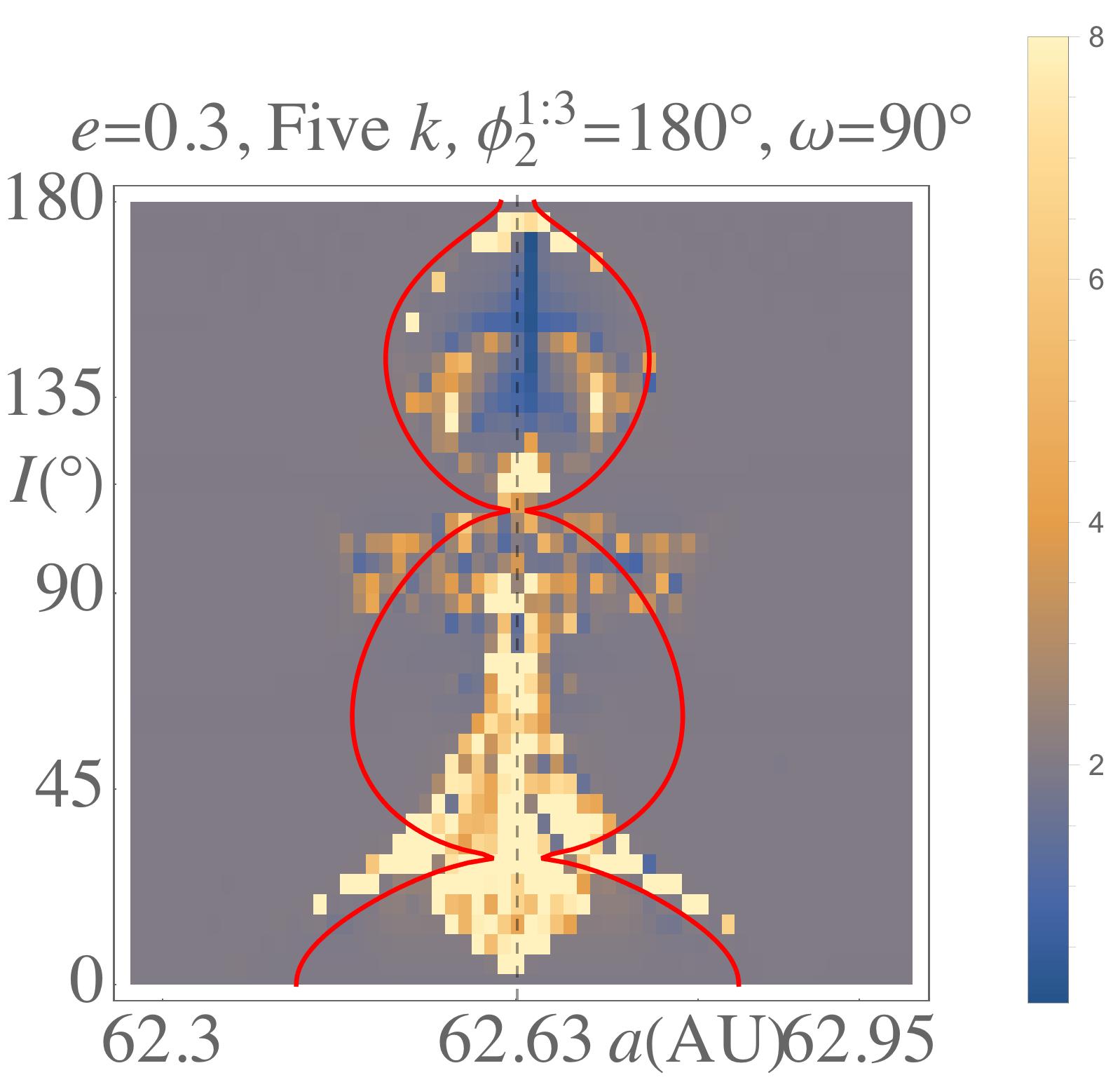}\\
\hspace*{-3mm}
\includegraphics[width=40mm]{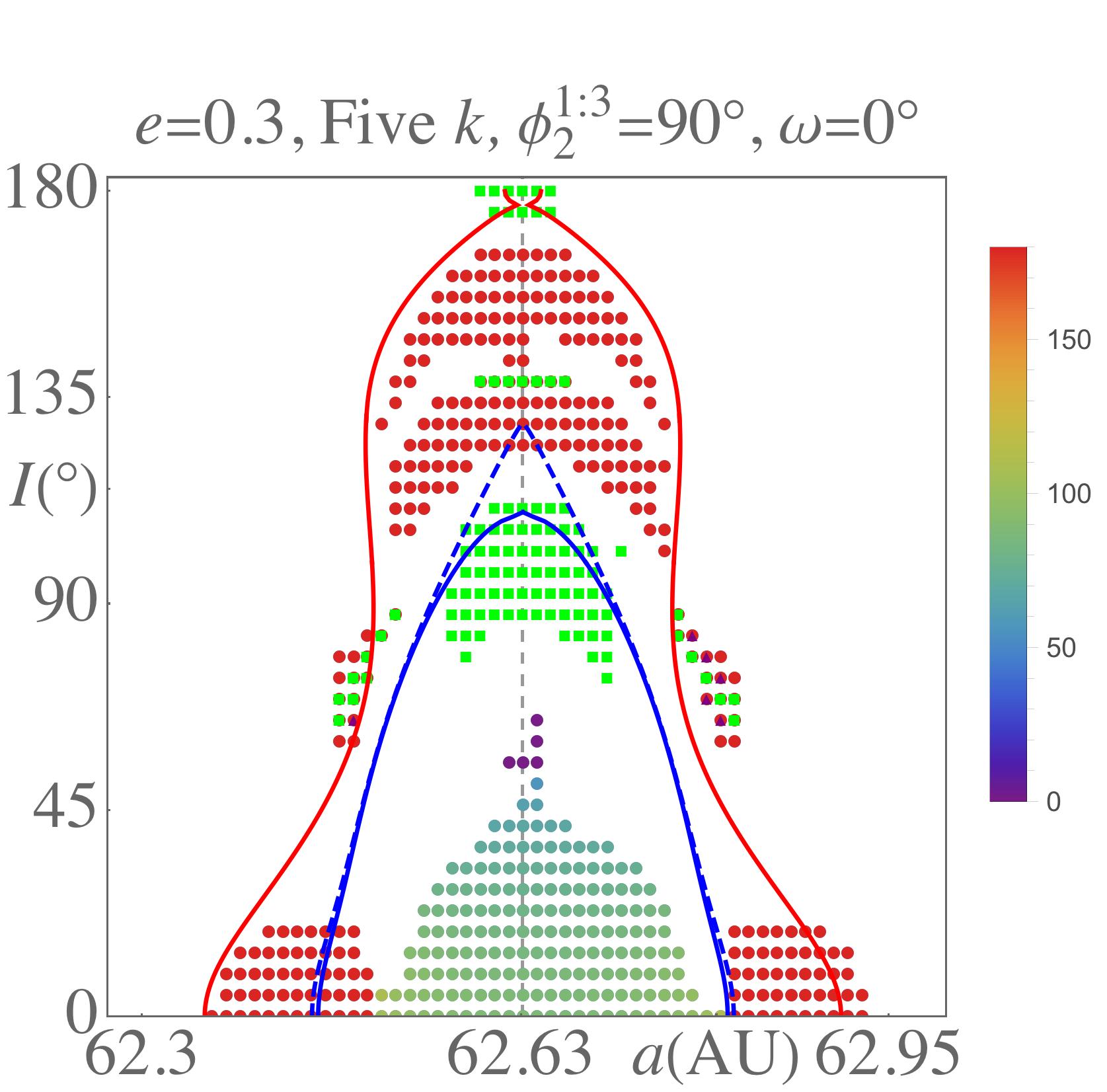}\hspace{3mm}
\includegraphics[width=40mm]{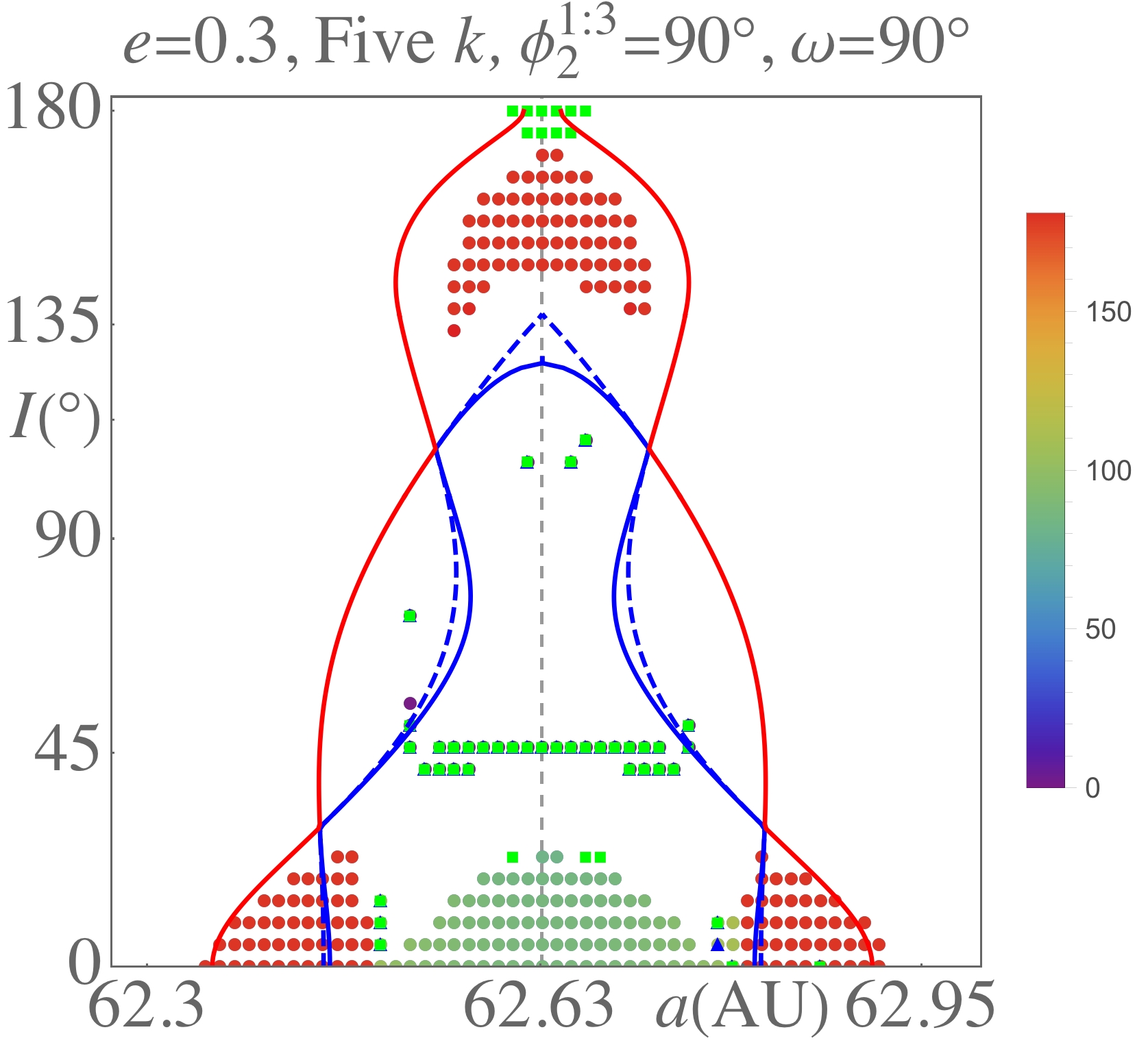}\hspace{3mm}
\includegraphics[width=35mm]{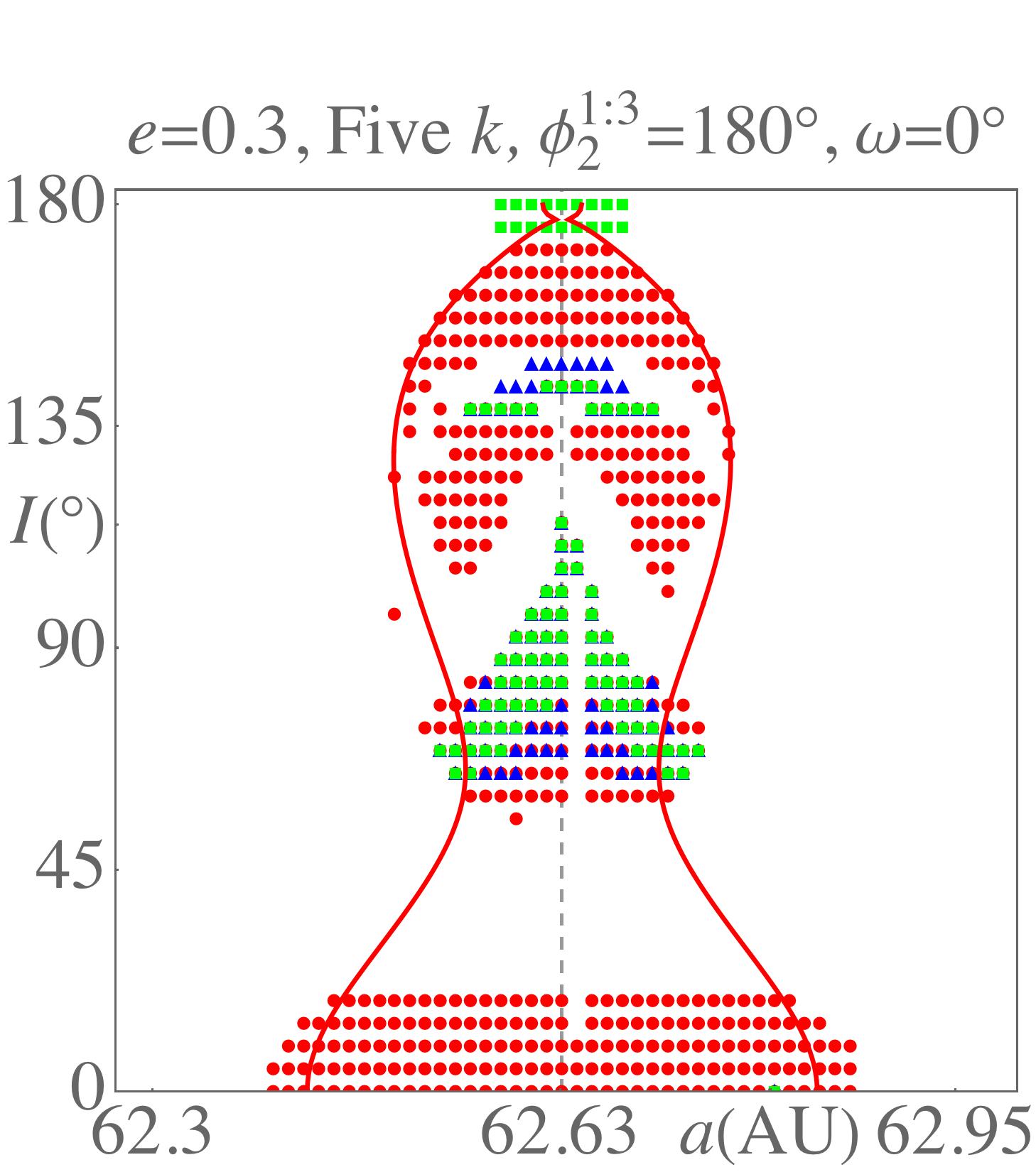}\hspace{8mm}
\includegraphics[width=35mm]{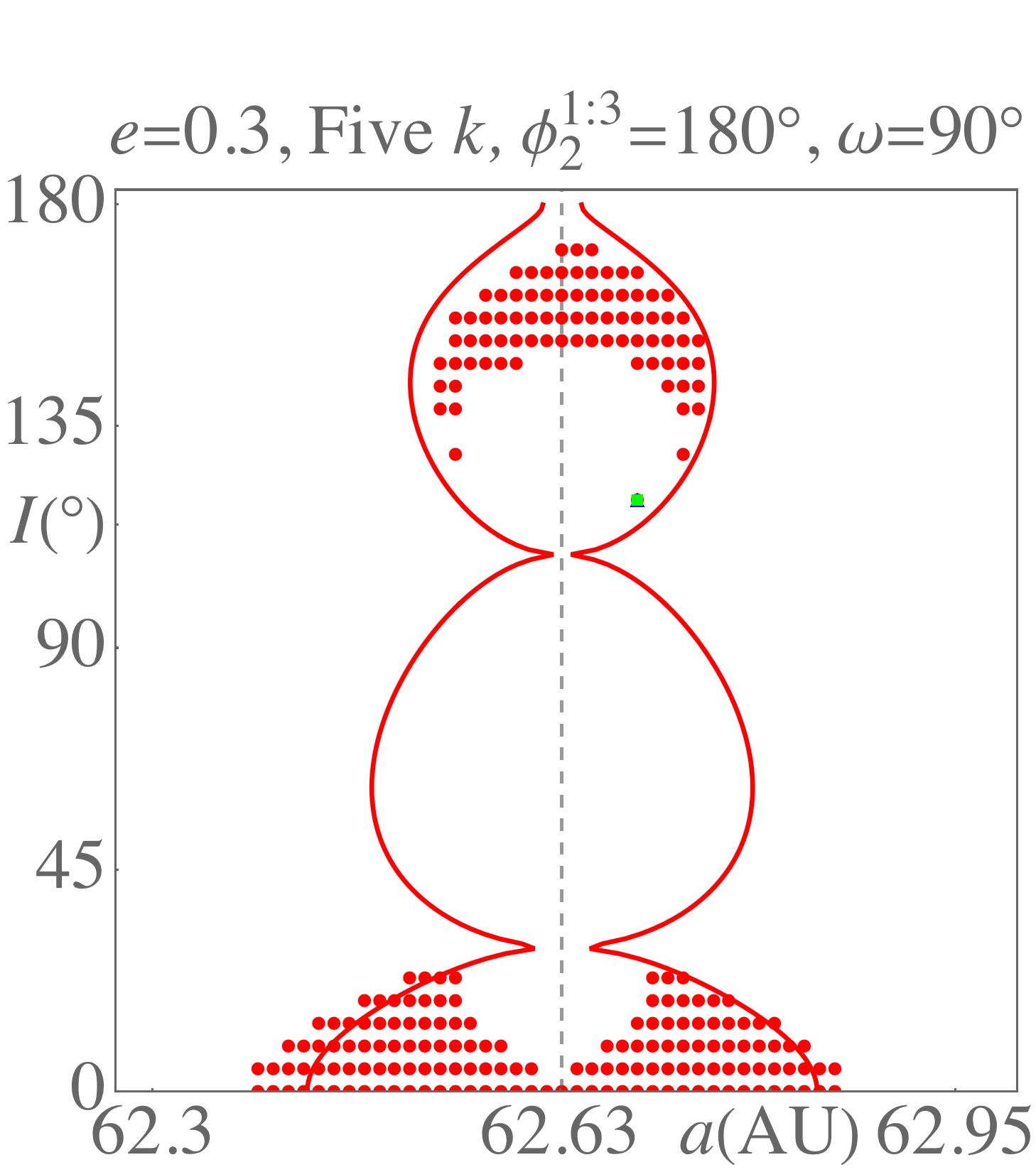}
}
\caption{Outer 1:3 Neptune resonance with $e=0.3$. Initial conditions are shown on top of each panel. Below each  {\sc megno} portrait is the corresponding libration center portrait. Filled circles denote the location of the  libration center of $\phi^{1:3}_2$ indicated on the color scale. Green filled  squares denote the librations of the retrograde mode $k=4$. Blue filled triangles denote the mode $k=0$. `Five $k$' indicates the set $\{k: 2,-2,4,-4,0\}$.}\label{fN1t3e3}
\end{center}
\end{figure*}

Outer mean motion resonances are known to exhibit asymmetric librations: that is librations around values other than $0^\circ$ and $180^\circ$. For this reason, we have developed a first analytical width model in Paper I based on the use of the second harmonic of the perturbation known to be at the origin of {libration} asymmetry. The resonance width of the pure eccentricity resonance $k=1$ ($\phi^{1:2}_1=2\lambda-\lambda^\prime-\varpi$) derived in Paper I from the disturbing function for arbitrary inclinations using the two-harmonics pendulum model is shown in Figure \ref{fN1t2e1} for $e=0.1$ and in Figure \ref{fN1t2e3} for $e=0.3$ (leftmost panel in the first row).  In the {numerical}  integrations of the outer Neptune resonances, we do not measure the absolute resonance width from the position of the asymmetric equilibrium point (red line in the leftmost panel of the bottom row of Figure \ref{fd}). Without a priori  knowledge of the exact position of the asymmetric point, we fix the initial value of $\phi^{1:2}_1$ at $90^\circ$ to detect asymmetric librations as well as librations around both asymmetric centers. In Section 4.3.2, we derived the corresponding widths 
$\Delta p_{\rm asym,90^\circ}$ (\ref{reswidasR21}) and $\Delta p_{\rm both,90^\circ}$   (\ref{reswidasR22}). We also use the initial value $\phi^{1:2}_1=180^\circ$ to detect more librations around both asymmetric points. The corresponding width is given by  $\Delta p_{180^\circ}$ (\ref{reswidGRg21}). The semimajor axis widths (\ref{areswidth}) are obtained with $\delta=0^\circ$ and  shown as solid lines in Figures \ref{fN1t2e1} to \ref{fN1t3e3} as they correspond to our initial conditions. The widths calculated in Paper I (red line in the leftmost panel of the bottom row of Figure \ref{fd}) are shown using dashed lines in Figures \ref{fN1t2e1} to \ref{fN1t3e3} to distinguish them from those corresponding to the specific choice of the initial value of $\phi^{1:2}_1=90^\circ$. The analytical width of the $k=1$ mode for an initial $\phi^{1:2}_1=90^\circ$ is found not to differ significantly from the full widths (dashed lines) except at the upper inclination limit of asymmetric librations that is not sampled by our initial conditions. The analytical width for an initial $\phi^{1:2}_1=180^\circ$ was not derived in Paper I and will be used shortly.


Before seeking the effect of additional modes, we first examine the case of $e=0.1$ and compare the analytical prediction of the mode $k=1$ derived in Paper I to the  {\sc megno} and libration portraits with $\phi^{1:2}_1=90^\circ$ and $\omega=0^\circ$ (Figure \ref{fN1t2e1}, {first two rows, third panels from left}). The  blue lines denotes the width of the asymmetric libration domain and the red lines that of librations around $180^\circ$. The analytical width reproduces accurately the resonance topology of the asymmetric and symmetric libration domains. However, it underestimates the extent of asymmetric librations. We therefore require the effect of additional modes to widen the resonance without changing resonance topology. A good candidate is the pure inclination $k=0$ mode as its resonance order is even ($\phi^{2:4}_0=4\lambda-2\lambda^\prime-2\Omega$) and therefore does not have a first harmonic that could influence the 1:2 resonance. Its width is shown in Figure \ref{fN1t2e1} along with that of the inclination mode $k=-1$ and the pure eccentricity retrograde mode $k=3$ obtained from the classical pendulum width (\ref{reswidth0}) and Table A3. We used the classical pendulum width for these modes as unlike the $k=1$ mode, their $\beta$ values (\ref{beta}) for $e=0.1$ are smaller than unity. The width of $k=0$ is important but is not indicative of how the mode interacts with $k=1$ as  it does not have a first harmonic. Mode $k=1$ is next in terms of size and may be significant mainly for retrograde inclinations. The width of the retrograde mode $k=3$ is modest.  That of $k=-3$ is not shown as it is much smaller than that of $k=3$. 

The resonance width of the simultaneous librations of the two modes $k=1$ and $k=0$ (on the resonant {timescale}) is obtained from the expressions (\ref{reswidasR21},\ref{reswidasR22}, \ref{areswidth}) and shown in the rightmost panels of the first two rows of Figure \ref{fN1t2e1} along with the  {\sc megno} and libration portraits. It can be seen that the analytical width follows accurately the numerically determined separatrices. In particular, the overall shapes of the asymmetric and symmetric domain is reproduced precisely.  

Adding the remaining modes $k=-1,$ 3 and $-3$, makes the resonance width somewhat larger as can be seen in the bottom two rows of Figure  \ref{fN1t2e1}. For the two sets of initial conditions ($\phi^{1:2}_1=90^\circ$, $\omega=0^\circ$) and  ($\phi^{1:2}_1=90^\circ$, $\omega=90^\circ$), the analytical resonance widths describe accurately the resonance domains particularly that of asymmetric libration which is dependent on the value of $\omega$ for $\phi^{1:2}_1=90^\circ$. For instance, when $\omega=90^\circ$,  two disconnected regions appear, a stable one at low inclination  and another from moderate to high inclination where motion is unstable {whose numerically-determined shapes agree with the analytical estimates. }

For the initial condition  $\phi^{1:2}_1=180^\circ$ that samples librations around both asymmetric points without crossing the asymmetric libration domain, we show the analytical width using the five fundamental modes of the 1:2 resonance $k=1,$ $-1$, 3, $-3$ and 0. The five mode combination (\ref{reswidGRg21},\ref{areswidth}) predicts correctly the resonance width depending on the value of $\omega$ as seen from the differing proportions of the two resonant islands centered around $60^\circ$ where motion is unstable and $140^\circ$ where motion is stable.  Apart from mode $k=1$, the only librations that were detected are those of the  pure eccentricity retrograde mode $k=3$. Their extent is predictably smaller for nearly coplanar motion. Elsewhere $k=3$ librations occur simultaneously with those of $k=1$ where $\phi^{1:2}_1$ oscillates around $180^\circ$. 

Increasing eccentricity to $e=0.3$ increases the widths of all single modes as shown in the top row of Figure \ref{fN1t2e3}. In particular the pure inclination mode exhibits a small region of asymmetric librations but in practice  that mode is dwarfed by the pure eccentricity one at the location of the 1:2 resonance because it does not contribute a first harmonic to the resonance. Additionally, the asymmetric libration island of $k=0$ is generated by $\phi^{2:4}_0$ and its second harmonic $\phi^{4:8}_0$ that has little effect on the 1:2 resonance.  The agreement between the analytical width and the numerical simulation is satisfactory except for low inclinations where the analytical models predict a larger (smaller)  width for an initial  $\phi^{1:2}_1=90^\circ$ ($\phi^{1:2}_1=180^\circ$). This mismatch is curious as the analytical model predicts correctly the resonance width for inclinations larger than $45^\circ$. In particular, librations in the pure eccentricity retrograde mode $k=3$ occur precisely inside the analytically-derived width. We also note that librations in the retrograde pure eccentricity mode  away from $I=180^\circ$ occur simultaneously with librations in the pure eccentricity mode.

\subsection{Neptune's outer 1:3 resonance}
The five fundamental modes of the even order 1:3 resonance are the prograde pure eccentricity mode $k=2$,  the pure inclination mode $k=0$, the retrograde pure eccentricity mode $k=4$, and the inclination modes $k=-2$ and $k=-4$. They are shown for $e=0.3$ in the top row of Figure \ref{fN1t3e3} except $k=-4$ whose width is negligible. After the pure eccentricity mode, the pure inclination mode is the largest but it satisfies $\beta<1$ (\ref{beta}) and therefore has no asymmetric librations. It contributes to the resonance width through its first and second harmonics ($\phi^{1:3}_0$ and $\phi^{2:6}_0$) unlike the case of the 1:2 resonance. The next mode in terms of width is $k=4$. When combined together the five modes agree with the {\sc megno} and libration portraits. The problem of the larger (smaller) analytical widths for low inclinations  encountered in {the previous Section} for the 1:2 resonance with $e=0.3$ disappears for $\phi^{1:3}_1=90^\circ$ ($\phi^{1:3}_1=180^\circ$). An important difference between the 1:3 and the 1:2 portraits is the presence of large chaotic regions inside the resonance width. This difference was encountered earlier between the inner 2:1 and 3:1 Jupiter resonances. The origin of these internal chaotic regions is likely related to the presence of internal separatrices for high order resonances that the pendulum model does not possess. For instance, the modeling of low inclination inner resonances can be done using the pendulum model but more accurately by the Poincar\'e hamiltonian model. The former describes only one stable and one unstable points of the latter, those that persist at larger eccentricity and have the largest width. Other critical points give rise to separatrices that occur inside the larger resonance width \citep{ssdbook}. For outer resonances the Andoyer hamiltonian model may be used  to describe the effect of second harmonics in a similar way as the Poincar\'e model does. However, so far no analytical development  exists for the case of large inclinations that could help us explain  the presence of the internal chaotic regions {at arbitrary inclination}. The role of the secular Kozai-Lidov potential in mode selection is likely to interfere with the internal separatrices to create unstable regions.  An indication of this is the presence of various modes in the libration portraits. For instance, for the initial conditions  $\phi^{1:3}_1=90^\circ$ and $\omega=0^\circ$ there is a significant domain centered around $I=90^\circ$ where the retrograde mode $k=4$ suppresses the asymmetric librations of $k=2$.  For the initial conditions  $\phi^{1:3}_1=180^\circ$ and $\omega=0^\circ$, the three modes $k=1$, 4 and 0 librate simultaneously in a region centered around $I=90^\circ$ indicating that the argument of perihelion is stationary. 
The role of the secular potential in mode selection is beyond the scope of this paper and will be addressed in future work. 

\section{Comparison with previous works}
Before the derivation of the disturbing function  in Paper I, semi-analytical methods were developed in the past to study asteroid dynamics at large inclination. For instance, the dynamics of the co-orbital resonance at large inclinations was investigated using the numerical averaging of the disturbing function $R$ over the fastest longitude \citep{Namouni99,Namounietal99,MoraisNamouni16} and led to the identification of new types of co-orbital motion at large inclination.  The semi-analytical method was applied in the context of asteroid motion in the solar system to a wider variety of mean motion resonances \citep{gallardo06} to produce an atlas of resonance strength for the solar system's planets. Whereas semi-analytical methods give a global description of the resonance under study, they cannot by construction access the inner workings of the resonance. For instance, there is so far no simple and precise  way to define resonance width with semi-analytical methods but only  approximations based on the pendulum model \citep{gallardo20}. The situation is even more complex for outer resonances where asymmetric librations occupy a large domain within the resonance as we have shown in the previous section. As far as we are aware, no study semi-analytical or numerical has examined in detail asymmetric libration in outer resonances before this work. 

Recently, \cite{gallardo19a} studied the resonance strength and stability of mean motion resonance at arbitrary inclination. His study was done using the semi-analytical method as well as the so-called `dynamical map' method \citep{gallardo06}. The latter is based on the direct integration of asteroid orbits inside a resonance using the full equations of motion over a few libration periods and the numerical averaging of the fastest oscillation at resonance. That way resonant libration amplitudes of the semimajor axis, eccentricity and inclination may be determined numerically. In essence, the dynamical map method mimics the semi-analytical method using the direct integration of the full equations of motion. \cite{gallardo19a} produced a dynamical map of Jupiter's inner 3:1 resonance and compared it to our analytical widths of the prograde and retrograde pure eccentricity modes $k=1$ and $k=3$ of Paper I. Upon finding that the width obtained with the `dynamical map' is larger than those of either mode, \cite{gallardo19a} concluded that our disturbing function is inefficient at estimating resonance width and claimed that ``[t]his confirms the necessity to consider the whole disturbing function whenever we depart from the central value of the expansion used'' thereby casting doubt on the usefulness of any expansion of the perturbing potential.  The developments in this work allow us to refute these claims. 

First, we note that the chosen value of the argument of perihelion  $\omega=0^\circ$ is that which gives the largest width when the various modes are combined. Had that {study} examined the case of $\omega=90^\circ$, it would have found smaller widths showing that the dynamics is more complex than just summing up single mode widths. This fact was later recognized in a more recent article \citep{gallardo20}. Second, our analytical disturbing function as well as our multiple-argument pendulum models not only give precise resonance widths for each mode, their combinations, and   their variations as functions of $\omega$, they also agree with the  chaos indicator. So our approach is not just to compare the analytical widths to the integration of the full equations of motion like it is done in the `dynamical map'. {Instead}, we seek the  numerical separatrices obtained from measuring precisely  how two nearby orbits diverge. In this sense, it is worth reminding the reader that the  {\sc megno} chaos indicator is one of the most precise diagnostic tools in resonance dynamics. In conclusion, the fact that the disturbing function is an expansion with respect to eccentricity (as in this work we used the second interpretation of the disturbing function without expanding with respect to inclination like in section 5.3 of Paper I) does not mean that it cannot probe the dynamics accurately.  For instance, we could even explain for the first time the width of the chaotic separatrix layer of the inner 2:1 resonance as well as predict accurately  the extent of the asymmetric libration domains for outer resonances.  Therefore the conclusion of \citep{gallardo19a} regarding the importance of the disturbing function at arbitrary inclinations and the analytical pendulum models is unjustified. 

While in the process of developing our multi-argument two-harmonics pendulum model to derive resonance widths for outer as well as inner resonances, we learned of the work of \cite{lei19}. The author tested the conclusion of \citep{gallardo19a} that the expansion of the disturbing function  cannot model the 3:1 resonance accurately and developed what they termed a `multi-harmonic pendulum model' and applied it to the disturbing function of Paper I. The author says that the disturbing function they use is somewhat different but it is not as they choose from the outset not to expand the perturbing potential with respect to inclination. That possibility is already present in  Paper I and was termed  the second interpretation of the disturbing function in  Section 4.3 of Paper I. It was already used to estimate resonance widths in Section 5.3 of Paper I and was reprised  in  the present paper. 

What \cite{lei19} does differently is to rewrite the fully derived disturbing function ($\ref{RY}$)  by trigonometrically {splitting} the resonant argument $\phi^{p:q}_k=q\lambda-p\lambda^\prime-(p-q) \Omega-k\omega$ with respect to $\phi^{p:q}_0$ as follows:
\begin{eqnarray}
\bar R_d&=& \sum C_{p,q} \cos \phi^{p:q}_0 +S_{p,q} \sin\phi^{p:q}_0,\\
C_{p,q}&=&\sum C^R_{p,q,k} \cos k \omega,\\
S_{p,q}&=& \sum S^R_{p,q,k} \sin k \omega.
\end{eqnarray}
The details of the indices in the above sums are found in Lei's paper and are not essential to our discussion.
The author then selects from this new expression the `harmonics' of $\phi^{p:q}_0$ in order to model the 3:1 resonance with the disturbing terms:
\begin{eqnarray}
R_{p,q}&=&C_{p,q} \cos\phi^{p:q}_0 + S_{p,q} \sin\phi^{p:q}_0 + \nonumber\\
&&+ C_{2p,2q} \cos\phi^{2p:2q}_0 + S_{2p,2q} \sin\phi^{2p:2q}_0+\nonumber\\
&&+ C_{3p,3q} \cos\phi^{3p:3q}_0 + S_{3p,3q} \sin\phi^{3p:3q}_0+\nonumber...
\end{eqnarray}
We use the quotes to refer to harmonics in Lei's formulation in order not to confuse them with the physical harmonics of $\phi^{p:q}_0$ in the disturbing function ($\ref{RY}$)   whose force amplitudes are independent of the argument of perihelion $\omega$. 

The author then assumes that the {classical} pendulum model width applies to the presence of higher harmonics and calculates the width from the previous potential at equilibrium using the first two `harmonics' therefore covering the modes $k=0$ as well as 2, $-2$, 4 and $-4$. \cite{lei19} thus derives curves similar to those in the bottom two rows of Figure \ref{fJ3t1} which they compare satisfactorily to the dynamical maps of Gallardo for the Jupiter 3:1 {inner} resonance. After finding agreement,  the author assigns a special role to the mode $k=0$ stated in their abstract as follows ``For a $p$:$q$ resonance at an arbitrary inclination, we define the characteristic resonance argument as $[\phi_0^{p:q}]$ whose amplitude in the disturbing function is a good indicator of representing the total resonance strength.'' For clarity, we substituted our notation for the author's regarding the resonant argument of $k=0$.  

Whereas the author proved successfully that our disturbing function explains the width of the 3:1 resonance estimated by \cite{gallardo19a} there are a number of issues regarding their methodology that obscure the physics of resonance dynamics and can lead to erroneous conclusions when applied to resonances other than the 3:1 as we explain in the following. First, it is mathematically 
acceptable to rewrite the disturbing function for arbitrary inclination of Paper I the way \cite{lei19} did. However the terminology is misleading because when they refer in their article to the `harmonics' of $\phi^{p:q}_0$, one would understand those terms of the disturbing function for arbitrary inclination ($\ref{RY}$)  corresponding to $\phi^{mp:mq}_0=m\phi^{p:q}_0$. However these have completely different force amplitudes from those of \cite{lei19}. In particular, the {physical harmonics of the disturbing function } are independent $\omega$ and are compatible with the harmonics of $\phi^{p:q}_0$ found in the classical disturbing function \citep{ssdbook} (see Paper I). Second, the use of the classical one-harmonic pendulum width  works out for Lei's analysis of the 3:1 resonance, despite the presence of two `harmonics', only because there is in reality one harmonic for each of the four modes they study as we demonstrated in Sections 4 and 5.  Bona fide multiple harmonics pendulum models give rise to asymmetric librations whose analytical modeling can be quite involved as we have shown with the multiple-argument two-harmonics pendulum model of Section 4.2. Third, elevating the mode $k=0$ as the fundamental mode of all outer and inner resonances of even and odd orders alike that represents the strength of any resonance is unwarranted.  In fact, after the successful determination of the 3:1 resonance width, \cite{lei19} calculated the width of the Jupiter 2:1 {inner}  resonance using  $\phi^{2:1}_0$ as a reference. Whereas it is mathematically possible to refer to that mode by artificially splitting the angles of the disturbing function, that argument does not physically exist  {by itself} in the disturbing function ($\ref{RY}$) because the corresponding force amplitude  is exactly zero since the resonance has an odd order. We have shown that the 2:1 resonance width depends on the odd modes $k=-1,\ 1,\ 3,$ and $-3$ and agrees precisely with the  {\sc megno} and libration portraits with no mention of $k=0$. Lei's derived 2:1 width for $\omega=90^\circ$  differs from the  dynamical portraits in Figure \ref{fJ2t1}  and our analytical widths. The mode $k=0$ exists for the inner 2:1 resonance but only as a second harmonic $\phi^{4:2}_0$ whose amplitude is too small to affect the resonance. For the outer 1:2 resonance, mode $k=0$ helps expand the asymmetric libration domain but again only as a second harmonic since outer resonances require the second harmonic of any useful mode  --even if they do not have a first harmonic like $k=0$. Therefore the conclusions about the importance of $k=0$ for all resonances is actually only rooted in a different way of mathematically rewriting the disturbing function that obscures the importance of the physical modes $k$ at the expense of the abstract $k=0$ that contributes to the width of  the 3:1 resonance  but is not the crucial mode of resonances in general. A further proof that the mode $k=0$ is not of particular importance in general  is the resonant librations we determined on the secular timescale. In Figures \ref{fJ2t1} to \ref{fN1t3e3}, it is clear that librations in the mode $k=0$ are not significant whereas librations in the pure eccentricity modes are the most common as these modes define the general shape of the resonance in phase space. 
\section{Conclusion}
In this work, we examined whether the disturbing function derived in Paper I is able to give reliable analytical estimates of resonance width and libration domains. To do so we employed simple pendulum models that were improved with respect  to those of Paper I by the addition of the more realistic situation of simultaneous argument librations on the resonant timescale first encountered in Paper II (see Section 4.2). Although the pendulum models are known to be simple first approximations of resonance dynamics in the context of the classical disturbing function  \citep{ssdbook}, we find that they reproduce accurately most of the features related to resonance width and librations especially the asymmetric type that occurs in outer resonances. This was done by comparison with the resonance separatrices obtained form the accurate  {\sc megno} chaos indicator that measures precisely how two nearby orbits may diverge from one another. Further extensions of the pendulum models such as the Andoyer Hamiltonian models are likely to improve on the present analytical {estimates by accounting for the inner separatrices that are absent in the pendulum models.} However,  whereas resonance width can be explained by the simultaneous contributions of various modes at nominal resonance, mode selection on the secular time scale is  determined by the combination of the resonant terms and the secular Kozai-Lidov potential. The latter was ignored in our analysis as the pendulum approximation is valid on timescales longer than the resonant timescale but shorter than the secular timescale. In this respect, secular precession or possible secular resonances are not found in the pendulum model. Understanding mode selection on the secular timescale is important to characterize the dynamical states that Centaurs and transneptunian objects assume in  their evolution. For instance, understanding mode selection is likely to yield clues about the evolution of the nearly polar transneptunian object (471325) that is currently librating in the $k=4$ mode of the 7:9 resonance with Neptune \citep{MoraisNamounipolar}, and map the pathways followed by high inclination Centaurs in the outer planets domain \citep{MoraisNamouni13b}. Further applications of our findings include the dynamics of irregular satellites of the solar system planets \citep{hinse10} and improving resonance width measurements of semi-analytical methods \citep{gallardo20} especially for outer resonances where asymmetric librations in the pure eccentricity modes dominate parameter space.

\appendix
\section{Force amplitudes \mbox{$f_k^{p:q}$}}
In this Appendix, we list the force amplitudes used in the resonance width formulas in Section {5}. For a resonance $p$:$q$, the resonant angle is written as $\phi_k^{p:q}=q\lambda-p\lambda^\prime -(q-p) \Omega -k \omega$. The perturbing potential is given as $f_k^{p:q}\cos\phi_k^{p:q}$ and the corresponding force amplitude,  $f_k^{p:q}$,  includes both the direct and indirect part of the perturbation as well as their dependence on eccentricity and inclination. The secular part of the perturbation is irrelevant in the pendulum models of resonance. The expressions of $f_k^{p:q}$ make use of  the function $A_{i,j,k,l}=\alpha^l D^l  b_{i+1/2}^{jk}$   where $D^l$ is the $l$th derivative with respect to the semi-major axis ratio $\alpha$ and $b_{i+1/2}^{jk}(\alpha,I_r)$ is the two-dimensional Laplace coefficient defined in Paper I. In this work as well as Paper I, we use the second interpretation of the disturbing function by setting the inclination variable $s\equiv 0$. This implies that the reference inclination $I_r\equiv I$ the asteroid's inclination. In the following formulas, the explicit presence of inclination indicates that the corresponding term comes from the indirect part of the perturbation as the direct part's comes only from the functions $A_{i,j,k,l}$. We list without derivation each force amplitude because that was done in Paper I for all four resonances 2:1, 3:1, 1:2 and 1:3 with the disturbing function of order $N=4$. The corresponding terms may also be found in the following tables. For the outer resonances, we expand the disturbing function to order $N=8$ and retain order 6 terms for the 1:2 resonance and order 8 terms for the 1:3 in order to model retrograde resonances that require second harmonics with $k=6$ and $k=8$ respectively. Lastly we note that to obtain the numerical values of the Laplace coefficients and their derivatives, we use their expansion in terms of $\alpha$ given in Paper I with $N_\alpha=20$. The $\alpha$-expansion of   $ b_{i+1/2}^{jk}$ is the most accurate and computationally-fastest way to evaluate these functions.  
\begin{table}
\caption{Inner 2:1 resonance $\phi^{2:1}_{k}=\lambda-2\lambda^\prime+\Omega-k\omega$. }
\begin{tabular}{cl}
\hline
\hline
Mode $k$& Force amplitude to order $N=4$ \\
\hline
$-1$ & $-\frac{e}{4} (4 A_{0, 2, -2, 0} +  A_{0, 2, -2, 1})+\frac{e^3}{32} (28  A_{0, 2, -2, 0} $\\&$ + 
   5 A_{0, 2, -2, 1}-6  A_{0, 2, -2, 2}-  A_{0, 2, -2, 3})$,
\\   \hline       
$3$ & $\frac{e^3}{96} (4 A_{0, 2, -2, 0}- 3 A_{0, 2, -2, 1} -6 A_{0, 2,-2, 2} $\\&$ - A_{0, 2, -2, 3}),$            \\ \hline 
$1$ & $-\frac{e}{4}  A_{0, 2, 0, 1} +  \frac{e^3}{32} (3 A_{, 2, 0, 1} - 2 A_{0, 2, 0, 2} $\\&$ - A_{0, 2, 0, 3}),$ \\ \hline 
 $-3$ & $ -\frac{e^3}{96}  (136 A_{0, 2, -4, 0} + 93 A_{0, 2, -4, 1} +18 A_{0, 2, -4, 2} $\\&$+ A_{0, 2, -4, 3}).$ \\ 
 \hline\hline
\end{tabular}
 \end{table}

 \begin{table}
\caption{Inner 3:1 resonance $\phi^{3:1}_{k}=\lambda-3\lambda^\prime+2\Omega-k\omega$. }
\begin{tabular}{cl}
\hline
\hline
Mode $k$& Force amplitude to order $N=4$ \\
\hline 
$-2$ & $\frac{e^2}{16}  (21 A_{0, 3, -3, 0} + 10 A_{0, 3, -3, 1} + A_{0, 3, -3, 2})$\\&$ - 
 \frac{e^4}{192}  (186 A_{0, 3, -3, 0} + 122 A_{0, 3, -3, 1} - 
    15 A_{0, 3, -3, 2} $\\&$-12 A_{0, 3, -3, 3} - A_{0, 3, -3, 4})$
\\           \hline 
$0$ &  $\frac{1}{2} A_{0, 3, 1, 0} + 
 \frac{e^2}{8}  (-4 A_{0, 3, 1, 0} + 2 A_{0, 3, 1, 1} + $\\&$A_{0, 3, 1, 2}) + 
 \frac{e^4}{128} (7 A_{0, 3, 1, 0} - 8 A_{0, 3, 1, 1} - 8 A_{0, 3, 1, 2} + 
 $\\&$   4 A_{0, 3, 1, 3} + A_{0, 3, 1, 4})$       
\\ \hline 
$4$ & $\frac{e^4}{768} (-15 A_{0, 3, -3, 0} - 4 A_{0, 3, -3, 1} + 
   30 A_{0, 3, -3, 2} + $\\&$12 A_{0, 3, -3, 3} + A_{0, 3, -3, 4})$
\\ \hline 
$2$ & $\frac{e^2}{16}  (-A_{0, 3, -1, 0} + 2 A_{0, 3, -1, 1} + A_{0, 3, -1, 2}) + $\\&$
 \frac{e^4}{192}  (2 A_{0, 3, -1, 0} + 2 A_{0, 3, -1, 1} - 
    3 A_{0, 3, -1, 2} + 4 A_{0, 3, -1, 3} $\\&$+ A_{0, 3, -1, 4})$
 \\ \hline 
$-4$ & $\frac{e^4}{768}  (1045 A_{0, 3, -5, 0} + 916 A_{0, 3, -5, 1} + 
   258 A_{0, 3, -5, 2} + $\\&$28 A_{0, 3, -5, 3} + A_{0, 3, -5, 4})$
\\ \hline\hline
\end{tabular}
 \end{table}

\begin{table}
\caption{Outer 1:2 resonance first harmonic $\phi^{1:2}_{k}=2\lambda-\lambda^\prime-\Omega-k\omega$. }
\begin{tabular}{cl}
\hline
\hline
Mode $k$& Force amplitude to order $N=6$ \\
\hline
$1$ & $\frac{e}{4} (2 A_{0, 1, 1, 0} -  A_{0, 1, 1, 1}) + 
 \frac{e^3}{32}  (14 A_{0, 1, 1, 1}$\\&$-20 A_{0, 1, 1, 0}  - A_{0, 1, 1, 3}) + 
 \frac{e^5}{768} (136 A_{0, 1, 1, 0} $\\&$- 116 A_{0, 1, 1, 1} - 
    16 A_{0, 1, 1, 2} + 28 A_{0, 1, 1, 3} $\\&$- 2 A_{0, 1, 1, 4} - 
    A_{0, 1, 1, 5}) + $\\&$
 \frac{\alpha e \cos^2 (I/2)}{96} ( 36 e^2 + 11 e^4-45)$

\\     \hline 
$0$ & 0\\ \hline 

$3$ & $\frac{e^3}{96}  ( 6 A_{0, 1, -1, 1}-4 A_{0, 1, -1, 0}  - A_{0, 1, -1, 3}) +$\\&$
 \frac{e^5}{1536} (16 A_{0, 1, -1, 0} - 8 A_{0, 1, -1, 1} - 16 A_{0, 1, -1, 2} + $\\&$
    16 A_{0, 1, -1, 3} - 2 A_{0, 1, -1, 4} - A_{0, 1, -1, 5}) $\\&$- 
 \frac{\alpha e^3 \sin^2(I/2)}{96} (4 + 17 e^2)$

\\ \hline 
$-1$ & $-6\frac{e}{4} (6 A_{0, 1, -3, 0} + A_{0, 1, -3, 1}) + 
 \frac{e^3}{32} (132 A_{0, 1, -3, 0} + $\\&$18 A_{0, 1, -3, 1} - 
    8 A_{0, 1, -3, 2} - A_{0, 1, -3, 3}) $\\&$
- \frac{e^5}{768} (2136 A_{0, 1, -3, 0} + 500 A_{0, 1, -3, 1} - 
    304 A_{0, 1, -3, 2} $\\&$-36 A_{0, 1, -3, 3} +10 A_{0, 1, -3, 4} +
    A_{0, 1, -3, 5})$

\\ \hline 
 $-3$ & $-\frac{e^3}{96} (380 A_{0, 1, -5, 0} + 174 A_{0, 1, -5, 1} + 
    24 A_{0, 1, -5, 2} $\\&$+ A_{0, 1, -5, 3}) + 
 \frac{e^5}{1536} (11840 A_{0, 1, -5, 0} + 6560 A_{0, 1, -5, 1} $\\&$+ 
    592 A_{0, 1, -5, 2} - 152 A_{0, 1, -5, 3} - 26 A_{0, 1, -5, 4} - $\\&$
    A_{0, 1, -5, 5})$
 
 \\
 \hline\hline
\end{tabular}
 \end{table}    

\begin{table}
\caption{Outer 1:2 resonance second harmonic $\phi^{2:4}_{k}=4\lambda-2\lambda^\prime-2\Omega-k\omega$. }
\begin{tabular}{cl}
\hline
\hline
Mode $k$& Force amplitude to order $N=6$ \\
\hline
$2$ & $\frac{e^2}{16}  (26 A_{0, 2, 2, 0} - 10 A_{0, 2, 2, 1} + A_{0, 2, 2, 2}) -  $\\&$
 \frac{e^4}{192}  (1036 A_{0, 2, 2, 0} - 428 A_{0, 2, 2, 1} + $\\&$
    30 A_{0, 2, 2, 2} + 8 A_{0, 2, 2, 3} - A_{0, 2, 2, 4}) +  $\\&$\frac{
 e^6}{6144} (35776 A_{0, 2, 2, 0} - 14656 A_{0, 2, 2, 1} + 
    760 A_{0, 2, 2, 3}  $\\&$- 86 A_{0, 2, 2, 4} - 6 A_{0, 2, 2, 5} + 
    A_{0, 2, 2, 6})$

\\    \hline  
$0$ & $\frac{1}{2} A_{0, 2, 4, 0} + 
 \frac{e^2}{8}  (-64 A_{0, 2, 4, 0} + 2 A_{0, 2, 4, 1}  $\\&$+ A_{0, 2, 4, 2}) + 
 \frac{e^4}{128}  (3952 A_{0, 2, 4, 0} - 128 A_{0, 2, 4, 1} $\\&$ - 
    128 A_{0, 2, 4, 2} + 4 A_{0, 2, 4, 3} + A_{0, 2, 4, 4}) $\\&$ - 
    \frac{e^6}{4608} (214720 A_{0, 2, 4, 0} + 288 A_{0, 2, 4, 1}  $\\&$- 
    11760 A_{0, 2, 4, 2}  +384 A_{0, 2, 4, 3}  $\\&$+ 192 A_{0, 2, 4, 4} -
    6 A_{0, 2, 4, 5}  + A_{0, 2, 4, 6})$

\\ \hline 
$6$ & $\frac{e^6}{92160} (1024 A_{0, 2, -2, 0} - 384 A_{0, 2, -2, 1} - 
   480 A_{0, 2, -2, 2}  $\\&$+ 280 A_{0, 2, -2, 3} - 30 A_{0, 2, -2, 4} - 
   6 A_{0, 2, -2, 5}  $\\&$+ A_{0, 2, -2, 6})$

\\ \hline 
 $-2$ & $\frac{e^2}{16}  (114 A_{0, 2, -6, 0} + 22 A_{0, 2, -6, 1} + 
    A_{0, 2, -6, 2}) - $\\&$ 
 \frac{e^4}{192}  (9948 A_{0, 2, -6, 0} + 1988 A_{0, 2, -6, 1} - 
    42 A_{0, 2, -6, 2}  $\\&$- 24 A_{0, 2, -6, 3} - A_{0, 2, -6, 4}) + 
    \frac{e^6}{6144} (779232 A_{0, 2, -6, 0}  $\\&$+ 185568 A_{0, 2, -6, 1} - 
    10960 A_{0, 2, -6, 2} - 4136 A_{0, 2, -6, 3}  $\\&$- 30 A_{0, 2, -6, 4} + 
    26 A_{0, 2, -6, 5} + A_{0, 2, -6, 6})$
 
 \\ \hline 
 $-6$ & $\frac{e^6}{92160} (5325280 A_{0, 2, -10, 0} + 2746464 A_{0, 2, -10, 1}  $\\&$+ 
   560880 A_{0, 2, -10, 2}  + 58040 A_{0, 2, -10, 3} +  $\\&$
   3210 A_{0, 2, -10, 4}  + 90 A_{0, 2, -10, 5}  + A_{0, 2, -10, 6})$
 
 \\
 \hline\hline
\end{tabular}
 \end{table}    

\begin{table}
\caption{Outer 1:3  resonance first harmonic $\phi^{1:3}_{k}=3\lambda-\lambda^\prime-2\Omega-k\omega$. }
\begin{tabular}{cl}
\hline
\hline
Mode $k$& Force amplitude to order $N=8$ \\
\hline
$2$ & $\frac{e^2}{16}  (9 A_{0, 1, 1, 0} - 6 A_{0, 1, 1, 1}  + A_{0, 1, 1, 2} ) $\\&$ - 
 \frac{e^4}{192}  (162 A_{0, 1, 1, 0}  - 126 A_{0, 1, 1, 1}  +
    21 A_{0, 1, 1, 2}$\\&$  + 4 A_{0, 1, 1, 3}  - A_{0, 1, 1, 4} ) + 
    \frac{e^6}{6144} (2295 A_{0, 1, 1, 0}  $\\&$- 1962 A_{0, 1, 1, 1}  + 219 A_{0, 1, 1, 2}  + 
    204 A_{0, 1, 1, 3} $\\&$ - 51 A_{0, 1, 1, 4}  - 2 A_{0, 1, 1, 5}  + 
    A_{0, 1, 1, 6}) -$\\&$
    \frac{e^8}{368640} (29322 A_{0, 1, 1, 0} - 25218 A_{0, 1, 1, 1} $\\&$+
    315 A_{0, 1, 1, 2} + 5148 A_{0, 1, 1, 3} - 1125 A_{0, 1, 1, 4}$\\&$ - 
    234 A_{0, 1, 1, 5} + 81 A_{0, 1, 1, 6} + A_{0, 1, 1, 8}) $\\&$- 
    \frac{\alpha e^2\cos^2(I/2)}{245760} (92160 - 92160 e^2 - 180488 e^4 $\\&$+ 41509 e^6) $
\\   \hline       
$0$ & $\frac{1}{2} A_{0, 1, 3, 0} - 
 \frac{e^2}{8} (36 A_{0, 1, 3, 0} - 2 A_{0, 1, 3, 1} $\\&$- A_{0, 1, 3, 2}) + 
 \frac{e^4}{128}  (1215 A_{0, 1, 3, 0} - 72 A_{0, 1, 3, 1}$\\&$ - 
    72 A_{0, 1, 3, 2} + 4 A_{0, 1, 3, 3} + A_{0, 1, 3, 4}) $\\&$- 
    \frac{e^6}{4608} (32328 A_{0, 1, 3, 0} + 162 A_{0, 1, 3, 1} $\\&$- 
    3591 A_{0, 1, 3, 2} + 216 A_{0, 1, 3, 3} +108 A_{0, 1, 3, 4}$\\&$ - 
    6 A_{0, 1, 3, 5} -A_{0, 1, 3, 6}) $\\&$+ 
    \frac{e^8}{294912} (590949 A_{0, 1, 3, 0} + 98208 A_{0, 1, 3, 1} $\\&$- 
    110520 A_{0, 1, 3, 2} - 1080 A_{0, 1, 3, 3} $\\&$+ 7074 A_{0, 1, 3, 4} - 
    432 A_{0, 1, 3, 5} $\\&$- 144 A_{0, 1, 3, 6} + 8 A_{0, 1, 3, 7} + 
    A_{0, 1, 3, 8})$
\\ \hline 
$4$ & $-\frac{e^4}{768}  (27 A_{0, 1, -1, 0} - 36 A_{0, 1, -1, 1} +
    6 A_{0, 1, -1, 2}$\\&$ + 4 A_{0, 1, -1, 3} - A_{0, 1, -1, 4}) + 
    \frac{ e^6}{15360} (216 A_{0, 1, -1, 0}$\\&$ - 198 A_{0, 1, -1, 1} - 75 A_{0, 1, -1, 2} + 
    120 A_{0, 1, -1, 3}$\\&$ - 30 A_{0, 1, -1, 4} - 2 A_{0, 1, -1, 5} + 
    A_{0, 1, -1, 6})$\\&$ -
    \frac{e^8}{737280} (405 A_{0, 1, -1, 0} - 2052 A_{0, 1, -1, 1} $\\&$- 
    630 A_{0, 1, -1, 2} + 1368 A_{0, 1, -1, 3} $\\&$- 180 A_{0, 1, -1, 4} - 
    180 A_{0, 1, -1, 5} $\\&$+ 54 A_{0, 1, -1, 6} -
    A_{0, 1, -1, 8})$\\&$ -
    \frac{  \alpha  e^4\sin^2(I/2)}{245760} (5760 + 207704 e^2 - 45883 e^4) $
\\ \hline 
 $-2$ & $\frac{e^2}{16} (75 A_{0, 1, -5, 0} + 18 A_{0, 1, -5, 1} + A_{0, 1, -5, 2}) $\\&$- 
 \frac{e^4}{192}  (3990 A_{0, 1, -5, 0} + 1014 A_{0, 1, -5, 1}$\\&$ -
    33 A_{0, 1, -5, 2} - 20 A_{0, 1, -5, 3} $\\&$- A_{0, 1, -5, 4}) + 
    \frac{ e^6}{6144} (175545 A_{0, 1, -5, 0} $\\&$+ 57654 A_{0, 1, -5, 1} - 
    4305 A_{0, 1, -5, 2}$\\&$ - 2124 A_{0, 1, -5, 3} - 9 A_{0, 1, -5, 4} + 
    22 A_{0, 1, -5, 5} $\\&$+ A_{0, 1, -5, 6})$\\&$ - 
 \frac{e^8}{368640}  (5366790 A_{0, 1, -5, 0} + $\\&$2759778 A_{0, 1, -5, 1} - 
     132867 A_{0, 1, -5, 2} $\\&$- 159228 A_{0, 1, -5, 3} +
     1035 A_{0, 1, -5, 4} $\\&$+ 3330 A_{0, 1, -5, 5} + 51 A_{0, 1, -5, 6} $\\&$- 
     24 A_{0, 1, -5, 7} - A_{0, 1, -5, 8})$
 \\
 \hline 
 $-4$ &  $\frac{e^4}{768} (9681 A_{0, 1, -7, 0} + 4332 A_{0, 1, -7, 1} $\\&$+ 
    678 A_{0, 1, -7, 2} + 44 A_{0, 1, -7, 3} + A_{0, 1, -7, 4})$\\&$ -
    \frac{e^6}{15360} (646632 A_{0, 1, -7, 0} + 331902 A_{0, 1, -7, 1} $\\&$+ 
    47655 A_{0, 1, -7, 2} - 600 A_{0, 1, -7, 3} $\\&$- 630 A_{0, 1, -7, 4} -
    46 A_{0, 1, -7, 5} - A_{0, 1, -7, 6})$\\&$ + 
 \frac{e^8}{737280} (32245479 A_{0, 1, -7, 0} + 21444156 A_{0, 1, -7, 1}$\\&$ + 
     3520314 A_{0, 1, -7, 2} - 239832 A_{0, 1, -7, 3}$\\&$ - 
     93960 A_{0, 1, -7, 4} - 3252 A_{0, 1, -7, 5} $\\&$+ 
     582 A_{0, 1, -7, 6} + 48 A_{0, 1, -7, 7} + A_{0, 1, -7, 8})$
 \\ 
 \hline\hline
\end{tabular}
 \end{table}

\begin{table}
\caption{Outer 1:3  resonance second harmonic $\phi^{2:6}_{k}=6\lambda-2\lambda^\prime-4\Omega-k\omega$. }
\begin{tabular}{cl}
\hline
\hline
Mode $k$& Force amplitude to order $N=8$ \\
\hline
$4$ & $\frac{e^4}{768}  (2760 A_{0, 2, 2, 0} - 1464 A_{0, 2, 2, 1}$\\&$ + 
    300 A_{0, 2, 2, 2} - 28 A_{0, 2, 2, 3} + A_{0, 2, 2, 4}) $\\&$- 
    \frac{e^6}{15360} (196704 A_{0, 2, 2, 0} -105984 A_{0, 2, 2, 1} $\\&$+
    20160 A_{0, 2, 2, 2} - 960 A_{0, 2, 2, 3} $\\&$- 180 A_{0, 2, 2, 4} + 
    26 A_{0, 2, 2, 5} - A_{0, 2, 2, 6}) $\\&$+ 
 \frac{e^8}{1/737280}  (12611808 A_{0, 2, 2, 0} - 6717600 A_{0, 2, 2, 1} $\\&$+ 
     1083024 A_{0, 2, 2, 2} + 47664 A_{0, 2, 2, 3}$\\&$ - 
     34020 A_{0, 2, 2, 4} + 3192 A_{0, 2, 2, 5} $\\&$+ 60 A_{0, 2, 2, 6} - 
     24 A_{0, 2, 2, 7} + A_{0, 2, 2, 8})$
\\   \hline       
$0$ & $\frac{1}{2} A_{0, 2, 6, 0} -
 \frac{e^2}{8}  (144 A_{0, 2, 6, 0} - 2 A_{0, 2, 6, 1} $\\&$- A_{0, 2, 6, 2}) + 
 \frac{e^4}{128}  (20412 A_{0, 2, 6, 0} - 288 A_{0, 2, 6, 1}$\\&$ - 
    288 A_{0, 2, 6, 2} + 4 A_{0, 2, 6, 3} + A_{0, 2, 6, 4}) $\\&$- 
    \frac{e^6}{4608} (2738160 A_{0, 2, 6, 0} + 648 A_{0, 2, 6, 1}$\\&$ - 
    61020 A_{0, 2, 6, 2} +864 A_{0, 2, 6, 3} $\\&$+ 432 A_{0, 2, 6, 4} - 
    6 A_{0, 2, 6, 5} - A_{0, 2, 6, 6})+ $\\&$
 \frac{e^8}{294912}  (333832320 A_{0, 2, 6, 0} + 10548288 A_{0, 2, 6, 1}$\\&$ - 
     10597536 A_{0, 2, 6, 2} - 4320 A_{0, 2, 6, 3} $\\&$+ 
     121608 A_{0, 2, 6, 4} - 1728 A_{0, 2, 6, 5}$\\&$ - 576 A_{0, 2, 6, 6} + 
     8 A_{0, 2, 6, 7} + A_{0, 2, 6, 8})$
\\ \hline 
$8$ & $\frac{e^8}{20643840} (186624 A_{0, 2, -2, 0} - 62208 A_{0, 2, -2, 1}$\\&$ - 
    72576 A_{0, 2, -2, 2} + 48384 A_{0, 2, -2, 3} $\\&$- 
    10080 A_{0, 2, -2, 4} + 336 A_{0, 2, -2, 5} $\\&$+ 168 A_{0, 2, -2, 6} - 
    24 A_{0, 2, -2, 7} + A_{0, 2, -2, 8})$
\\ \hline 
 $-4$ & $\frac{e^4}{768}  (66240 A_{0, 2, -10, 0} + 17496 A_{0, 2, -10, 1} $\\&$+ 
    1668 A_{0, 2, -10, 2} + 68 A_{0, 2, -10, 3} + A_{0, 2, -10, 4}) $\\&$-
    \frac{e^6}{15360} (14022720 A_{0, 2, -10, 0} + 3927024 A_{0, 2, -10, 1}$\\&$ +
    322680 A_{0, 2, -10, 2} - 1920 A_{0, 2, -10, 3} $\\&$- 
    1500 A_{0, 2, -10, 4} - 70 A_{0, 2, -10, 5} - 
    A_{0, 2, -10, 6}) + $\\&$
 \frac{e^8}{737280} (2749749120 A_{0, 2, -10, 0} + 
     853461792 A_{0, 2, -10, 1} $\\&$+ 66413232 A_{0, 2, -10, 2} - 
     3526704 A_{0, 2, -10, 3} $\\&$- 649980 A_{0, 2, -10, 4} - 
     14040 A_{0, 2, -10, 5} $\\&$+ 1332 A_{0, 2, -10, 6} + 
     72 A_{0, 2, -10, 7} + A_{0, 2, -10, 8})$
 \\
 \hline 
 $-8$ &  $\frac{e^8}{20643840} (14277949056 A_{0, 2, -14, 0} $\\&$+ 
    6937481088 A_{0, 2, -14, 1} + 1424904768 A_{0, 2, -14, 2}$\\&$ + 
    161534016 A_{0, 2, -14, 3} + 11052720 A_{0, 2, -14, 4} $\\&$+ 
    467376 A_{0, 2, -14, 5} + 11928 A_{0, 2, -14, 6} $\\&$+ 
    168 A_{0, 2, -14, 7} + A_{0, 2, -14, 8})$
 \\ 
 \hline\hline
\end{tabular}
 \end{table}

\section*{Acknowledgments}
 M.H.M. Morais research had financial support from S\~ao Paulo Research Foundation (FAPESP/2018/08620-1) and CNPq-Brazil
(Pq2/304037/2018-4). This research was supported in part by FINEP and FAPESP through the computational resources provided by the Center for Scientific Computing (NCC/GridUNESP) of the S\~ao Paulo State University (UNESP).
 
 \bibliographystyle{mnras}

\bibliography{ms}

\end{document}